\documentclass[journal]{vgtc}               

\usepackage{xspace}
\usepackage{multirow}
\usepackage{comment}
\usepackage{colortbl}
\usepackage[svgnames]{xcolor}
\usepackage{soul}
\usepackage{setspace}
\usepackage{amsfonts}
\usepackage[export]{adjustbox}
\usepackage{numprint}

\renewcommand{\paragraph}[1]{\textbf{#1.}}

\makeatletter
\let\orgdescriptionlabel\descriptionlabel
\renewcommand*{\descriptionlabel}[1]{%
  \let\orglabel\label
  \let\label\@gobble
  \phantomsection
  \edef\@currentlabel{#1\unskip}
  \let\label\orglabel
  \orgdescriptionlabel{#1}%
}
\makeatother

\newcommand {\mm}[1] {\ifmmode{#1}\else{\mbox{\(#1\)}}\fi}

\newcommand{\Rspace}        {\mm{{\mathbb R}}}
\newcommand{\Ucal}{\mathcal{U}}

\renewcommand{\phi}{\varphi}

\newcommand{\mog}  {\textit{mapper on graphs}\xspace} 

\newcommand{\moag}  {{\emph{mapper graph}}\xspace}

\newcommand{\AGD} {\mm{AGD}\xspace}

 \usepackage{color}

\graphicspath{{./figs/}}

\definecolor{mygray}{rgb}{0.8,0.8,0.8}
\definecolor{mylightgray}{rgb}{0.95,0.95,0.95}

\onlineid{1453}

\vgtccategory{Research}

\vgtcpapertype{Data Transformations}

\title{Homology-Preserving Multi-Scale Graph \texorpdfstring{\\}{} Skeletonization Using Mapper on Graphs}

\author{Paul Rosen, Mustafa Hajij and Bei Wang}
\authorfooter{
\item
 Paul Rosen is with the University of Utah. 
 E-mail: paul.rosen@utah.edu.
\item
 Mustafa Hajij is with the University of San Francisco. 
 E-mail: mhajij@usfca.edu.
\item
 Bei Wang is with the University of Utah. 
 E-mail: beiwang@sci.utah.edu.
}

\shortauthortitle{Rosen \MakeLowercase{\textit{et al.}}: Graph Skeletonization Using Mapper}

\abstract{Node-link diagrams are a popular method for representing graphs that capture relationships between individuals, businesses, proteins, and telecommunication endpoints. However, node-link diagrams may fail to convey insights regarding graph structures, even for moderately sized data of a few hundred nodes, due to visual clutter. We propose to apply the mapper construction---a popular tool in topological data analysis---to graph visualization, which provides a strong theoretical basis for summarizing the data while preserving their core structures. We develop a variation of the mapper construction targeting weighted, undirected graphs, called {\mog}, which generates homology-preserving skeletons of graphs. We further show how the adjustment of a single parameter enables multi-scale skeletonization of the input graph. We provide a software tool that enables interactive explorations of such skeletons and demonstrate the effectiveness of our method for synthetic and real-world data.} 


\teaser{
  \centering
    \begin{minipage}[b]{0.2\linewidth}
        \subfloat[\textsc{hic\_5k\_net\_1} input graph]{%
        \fbox{\includegraphics[width=0.95\linewidth]{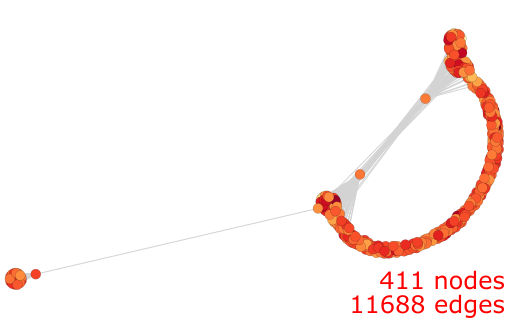}}}
        \vspace{147pt}
    \end{minipage}
    \hspace{-0.23\linewidth}
    \subfloat[PageRank; modularity clustering; $n\!=\!\{40,20,10,6,4\}$; $\epsilon\!=\!0$]{%
    \begin{minipage}[b]{0.49\linewidth}

        \hfill
        \fbox{\includegraphics[width=0.46\linewidth]{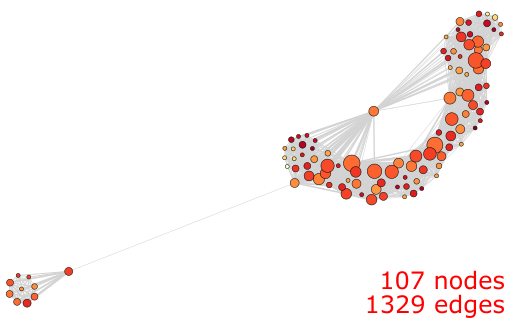}}
        \put(-110,60){
  		\begin{minipage}[t][0pt][t]{0pt}
			\scriptsize
			\mbox{$n\!=\!40$}
		\end{minipage}
		}        

        \vspace{2pt}
        \fbox{\includegraphics[width=0.46\linewidth]{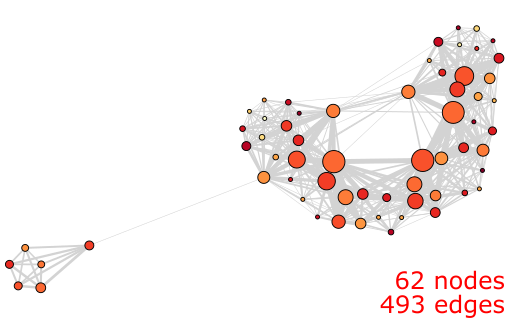}}
        \put(-110,60){
  		\begin{minipage}[t][0pt][t]{0pt}
			\scriptsize
			\mbox{$n\!=\!20$}
		\end{minipage}
		}        
        \hfill
        \fbox{\includegraphics[width=0.46\linewidth]{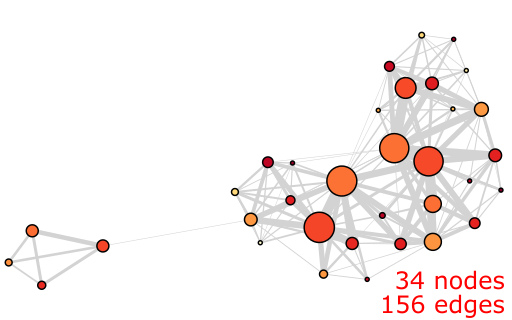}}
        \put(-110,60){
  		\begin{minipage}[t][0pt][t]{0pt}
			\scriptsize
			\mbox{$n\!=\!10$}
		\end{minipage}
		}        

        \vspace{2pt}
        \fbox{\includegraphics[width=0.46\linewidth]{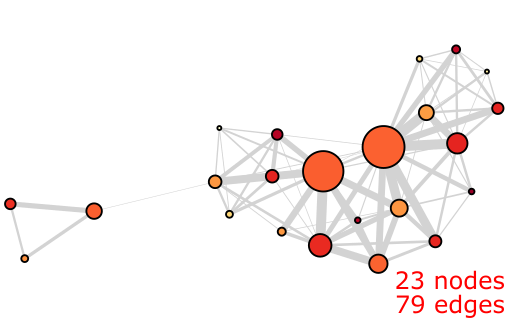}}
        \put(-110,60){
  		\begin{minipage}[t][0pt][t]{0pt}
			\scriptsize
			\mbox{$n\!=\!6$}
		\end{minipage}
		}        
        \hfill
        \fbox{\includegraphics[width=0.46\linewidth]{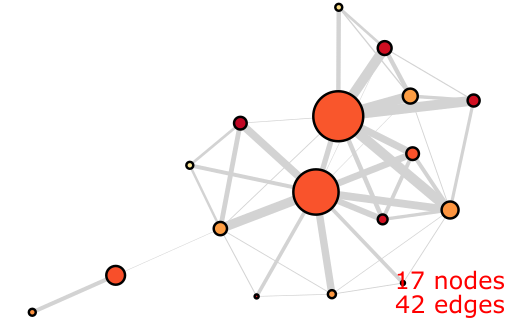}}
        \put(-110,60){
  		\begin{minipage}[t][0pt][t]{0pt}
			\scriptsize
			\mbox{$n\!=\!4$}
		\end{minipage}
		}        
    \end{minipage}}    
    \hspace{20pt}
    \begin{minipage}[b]{0.2\linewidth}
        \subfloat[\textsc{bcsstk20} input graph]{%
        \fbox{\includegraphics[width=0.95\linewidth]{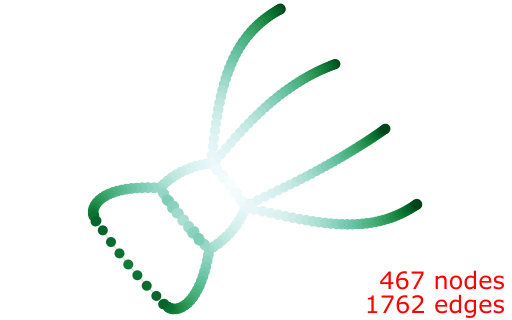}}}
        \vspace{147pt}
    \end{minipage}
    \hspace{-0.23\linewidth}
    \subfloat[Eccentricity (equalized); conn.\ components; $n\!=\!\{40,20,10,6,4\}$; $\epsilon\!=\!0$]{%
    \begin{minipage}[b]{0.49\linewidth}
        \hfill        
        \fbox{\includegraphics[width=0.46\linewidth]{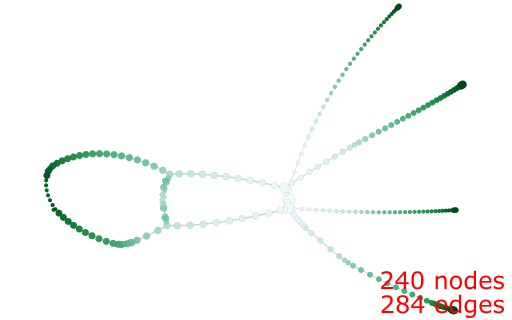}} 
        \put(-110,60){
  		\begin{minipage}[t][0pt][t]{0pt}
			\scriptsize
			\mbox{$n\!=\!40$}
		\end{minipage}
		}        

        \vspace{2pt}
        \fbox{\includegraphics[width=0.46\linewidth]{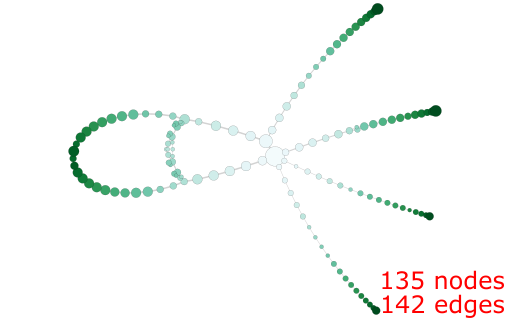}} 
        \put(-110,60){
  		\begin{minipage}[t][0pt][t]{0pt}
			\scriptsize
			\mbox{$n\!=\!20$}
		\end{minipage}
		}        
        \hfill        
        \fbox{\includegraphics[width=0.46\linewidth]{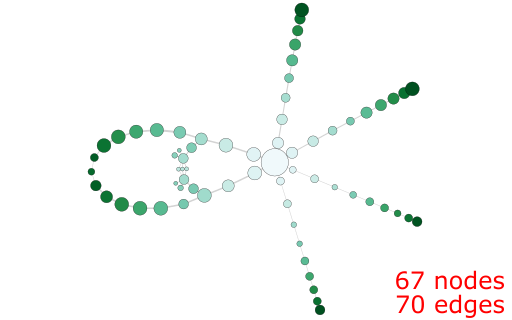}} 
        \put(-110,60){
  		\begin{minipage}[t][0pt][t]{0pt}
			\scriptsize
			\mbox{$n\!=\!10$}
		\end{minipage}
		}        

        \vspace{2pt}
        \fbox{\includegraphics[width=0.46\linewidth]{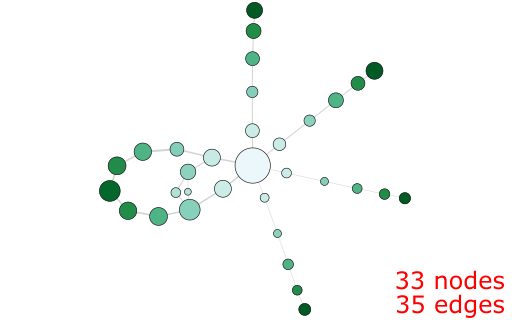}} 
        \put(-110,60){
  		\begin{minipage}[t][0pt][t]{0pt}
			\scriptsize
			\mbox{$n\!=\!6$}
		\end{minipage}
		}        
        \hfill        
        \fbox{\includegraphics[width=0.46\linewidth]{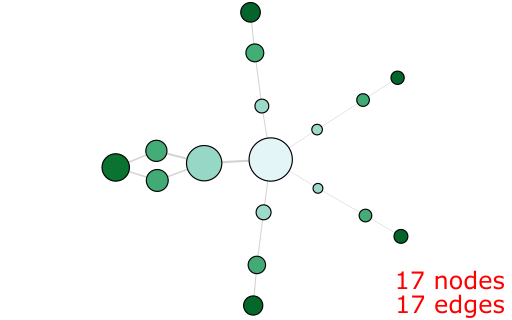}}
        \put(-110,60){
  		\begin{minipage}[t][0pt][t]{0pt}
			\scriptsize
			\mbox{$n\!=\!4$}
		\end{minipage}
		}        
    \end{minipage}}
    
  \caption{Our approach adapts the \textit{mapper} construction---a popular tool from topological data analysis---to the visualization of graphs. It provides multi-scale skeletonizations of the graph from diverse perspectives. First, a topological lens is applied to the graph, based upon a property of interest from the graph, for instance,  (a)~node importance and (b)~eccentricity. Then, by modifying a single input parameter, namely the number of cover elements $n$, our approach provides a multi-scale skeletonization of the input graph that emphasizes the property of interest. In the examples presented here, the visible level-of-detail is reduced as $n$ decreases (within an appropriate range), whereas the homology of the graph---in the form of components and tunnels---remains well persevered.}
  \label{fig:teaser}
}

\begin{document}

\firstsection{Introduction\label{sec:introduction}}

\maketitle

Graphs are often used to model relationships in social, biological, and technological systems. In recent years, our ability to collect and archive such data has far outpaced our ability to understand them. The challenges for graph visualizations are two-fold: how to effectively extract features from such large and complex data; and how to design effective visualizations to communicate these features to the users?

One approach to addressing this problem has been graph aggregation, which creates a skeleton of a graph via node~\cite{Schaeffer2007} and edge clustering~\cite{CuiZhouQu2008, ErsoyHurterPaulovich2011}. Common node clustering techniques include spectral methods~\cite{Dhillon2001,Fiedler1973,KulisBasuDhillon2009,WhiteSmyth2005}, similarity-based aggregation~\cite{TianHankinsPatel2008}, community detection~\cite{Newman2004,Newman2003}, random walks~\cite{JehWidom2002,PonsLatapy2005}, and hierarchical clustering~\cite{BrandesGaertlerWagner2003,BuiChaudhuriLeighton1987}. There are two limitations to these existing approaches. 
First, they make no guarantees about the preservation of homological features (i.e., components and tunnels) in the original graph (see \cref{fig:teaser}). Second, they provide only a single perspective on the aggregation of the graph, driven by the objective of the clustering algorithm. 

We propose to address these challenges by adapting the \emph{mapper} construction~\cite{SinghMemoliCarlsson2007}---a tool from topological data analysis (TDA)---to develop homology-preserving skeletonizations of graphs for visualization. The original mapper algorithm has enjoyed tremendous success in data science, e.g., cancer research~\cite{NicolauLevineCarlsson2011},  
phonemics~\cite{ZhouKamruzzamanSchnable2021,KamruzzamanKalyanaramanKrishnamoorthy2018}, and others~\cite{LumSinghLehman2013, TorresOliveiraTate2016}. 

Our approach works by first applying a \textit{topological lens} to the input graph that captures a certain property of the graph. We consider topological lenses that preserve graph-theoretic properties such as symmetries, density, and centrality. We then perform a series of operations to skeletonize the graph, producing a \emph{topological summary}. The benefits of our approach are three-fold.
First, it provides guarantees on preserving the homology of the graph it is summarizing, assuming certain sampling conditions~\cite{MunchWang2016,BrownBobrowskiMunch2021}. Second, the topological lens allows observing the graph from multiple perspectives, each highlighting different aspects of the graph being studied. Finally, the mapper construction connects naturally with visualization by providing a strong theoretical basis for multiscale simplifications of the input graph while preserving its core homological structures. 
In this paper, we develop and evaluate a variation of the mapper construction targeting weighted undirected graphs; the method is referred to as {\mog}, which produces {\moag} as output. Specifically:
\begin{itemize}[noitemsep]
\item We describe a series of modifications required to make the mapper algorithm effective on graph data;
\item We study the utility of five graph-specific topological lenses;
\item We demonstrate the effectiveness of our method on synthetic and real-world data;
\item We perform a human-subject study showing \mog produces good quality skeletons. 
\end{itemize}
Finally, we provide an open-sourced implementation together with our experimental datasets via GitHub.

\section{Related Work}
\label{sec:related-work}

\subsection{Graph Visualization}

We limit our review to node-link diagrams, which are utilized by many visualization software tools, including Gephi~\cite{BastianHeymannJacomy2009}, GraphViz~\cite{EllsonGansnerKoutsofios2002}, and NodeXL~\cite{HansenShneidermanSmith2010}. For a comprehensive overview of graph visualization techniques, see~\cite{LandesbergerKuijperSchreck2011}.
One of the biggest challenges with node-link diagrams is visual clutter, which has been extensively studied in graph visualization~\cite{EllisDix2007}. It is addressed in several  ways: improved node layouts, edge bundling, lenses, and alternative visual representations.

Tutte~\cite{Tutte1963} provided the earliest graph layout method for node-link diagrams, followed by methods driven by linear programming~\cite{GansnerKoutsofiosNorth1993}, force-directed embedding~\cite{FruchtermanReingold1991,Hu2005}, graph metric embedding~\cite{GansnerKorenNorth2005}, and connectivity structures~\cite{BrandesPich2007,KhouryHuKrishnan2012,Koren2003,KorenCarmenHarel2002}. TopoLayout~\cite{ArchambaultMunznerAuber2007} creates a hybrid layout by decomposing a graph into subgraphs based on their topological features, including trees, complete graphs, bi-connected components, and clusters, which are grouped and laid out as meta-nodes. The difference between TopoLayout and our work is that we use functions defined on the graph to automatically guide decomposition and feature extraction among subgraphs, and multiple functions induce graph skeletonizations, capturing different properties of the graph. 

Edge bundling, which bundles adjacent edges together, is commonly used to reduce visual clutter on dense graphs~\cite{HoltenVanWijk2009}. For massive graphs, hierarchical edge bundling scales to millions of edges~\cite{GansnerHuNorth2011}, while divided edge bundling~\cite{SelassieHellerHeer2011} tends to produce higher-quality visual results. Nevertheless, these approaches only deal with edge clutter, not node clutter, and they only support limited types of analytic tasks~\cite{bach2017towards,mcgee2012empirical}.

The concept of applying lenses to view large and complex data has also been used in graph visualization. An early example is the topologically-driven ``fisheye views'' for visualizing large graphs~\cite{gansner2005topological}. Another example is the TugGraph used to explore neighborhoods and paths extending from the foci of interest within a large graph~\cite{archambault2010tugging, golodetz2020simplifying}.

Finally, alternative visual representations have been used to remove clutter, ranging from variations on node-link diagrams, such as replacing nodes with modules~\cite{DwyerRicheMarriott2013} and motifs~\cite{DunneShneiderman2013}, to abstract representations, such as matrix diagrams~\cite{DinklaWestenbergWijk2012} and graph statistics~\cite{KairamMacLeanSavva2012}.

\subsection{Graph Clustering}
The objective in node clustering (or graph clustering) is to group the nodes of the graph by taking into consideration their edge structure~\cite{Schaeffer2007}. 
Vehlow et al.\ provided a recent survey on the visualization of clustered structures~\cite{vehlow2017visualizing} (see \cref{sec:introduction} for examples).  
Edge clustering has also been studied~\cite{CuiZhouQu2008, ErsoyHurterPaulovich2011}. 
Importantly, in their assessment of graph readability, Archambault et al.~\cite{archambault2010readability} concluded that only clustering could be efficiently performed on large graphs.
Broadly speaking, our approach is a type of graph clustering that simultaneously preserves relationships between clusters to retain the homological features of the graph.

Several approaches have leveraged clustering and hierarchical relationships to improve exploration within graphs. For example, van Ham and van Wijk provided methods for interacting with clusters in graph visualizations~\cite{van2004interactive}. Abello et al.\ built a system, ASK-GraphView, which enables interacting with very large graphs using clustering, among other techniques~\cite{AbelloHamKrishnan2006}. Grouse and GrouseFlocks are systems designed to ease the exploration through a feature topology-based hierarchy of the graph~\cite{ArchambaultMunznerAuber2008}. Batagelj et al.\ created the (x,y)-clustering, with the goal of generating visualizations where intercluster (x) and intracluster (y) graphs retained desired topological properties~\cite{batagelj2010visual}.

\subsection{TDA in Graph Analysis and Visualization}
Persistent homology (the study of topological features across multi-scales) and mapper construction are two of the most widely used tools in TDA. Persistent homology has been used to analyze graphs~\cite{DonatoPetriScolamiero2012, HorakMaleticRajkovic2009, PetriScolamieroDonato2013, PetriScolamieroDonato2013b}, with applications to collaboration networks~\cite{BampasidouGentimis2014, CarstensHoradam2013} and brain networks \cite{CassidyRaeSolo2015, DabaghianMemoliFrank2012, LeeChungKang2011b, LeeChungKang2011, LeeKangChung2012, LeeKangChung2012b, PirinoRiccomagnoMartinoia2015}. In terms of graph visualization, persistent homology has been used in capturing changes in time-varying graphs~\cite{HajijWangScheidegger2018}, as well as supporting interactive force-directed layouts~\cite{doppalapudi2022untangling, SuhHajijWang2019}. 
It has also been used to simplify and visualize hypergraphs~\cite{ZhouRathorePurvine2022}.  

The mapper construction~\cite{SinghMemoliCarlsson2007}, which provides a multiscale skeletonization of point clouds has been widely utilized in TDA for a number of applications~\cite{Carlsson2014, LiuMaljovecWang2017, LumSinghLehman2013, NicolauLevineCarlsson2011, TorresOliveiraTate2016}. It has witnessed recent major theoretical developments (e.g.,~\cite{CarriereOudot2018,DeyMemoliWang2017,MunchWang2016}) that further adjudicate its use in data analysis and provide mathematical foundations for its homology-preserving properties. Recently, Bodnar et al.\ considered learning the lens of a mapper graph using graph neural networks~\cite{BodnarCangeaLio2020}, which is based upon a preprint draft of this work on applying mapper to graphs~\cite{hajij2018mog}.

\begin{figure*}[!t]
\centering
\begin{minipage}[b]{0.015\linewidth}
	    \rotatebox{90}{\footnotesize\hspace{17pt}Mapper Graph\hspace{50pt}Input Graph}
    \end{minipage}
	\subfloat[Average Geodesic Distance\label{fig:lenses:agd}]{
	    \begin{minipage}[b]{0.17\linewidth}
    	    \begin{minipage}[b]{0.95\linewidth}
    	        \fbox{\includegraphics[width=\linewidth]{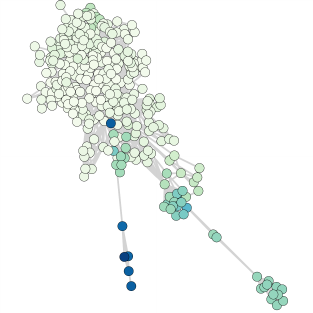}} 
    	    \end{minipage}
    	    \begin{minipage}[b]{0pt}
    	    \hspace{-7pt}
        	    \includegraphics[height=2pt, width=1.7cm, angle=90]{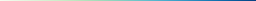}
    	    \vspace{30pt}
    	    \end{minipage}
    	    \fbox{\includegraphics[width=0.95\linewidth]{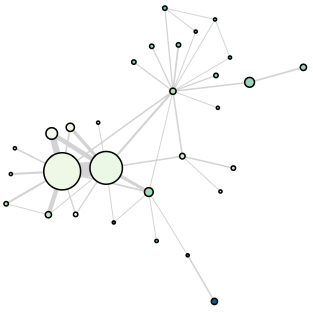}}
	    \end{minipage}
            \vspace{-3pt}
	}
	\subfloat[Density ($\delta\!=\!0.5$)\label{fig:lenses:den}]{
	    \begin{minipage}[b]{0.17\linewidth}
    	    \begin{minipage}[b]{0.95\linewidth}
            	\fbox{\includegraphics[width=\linewidth]{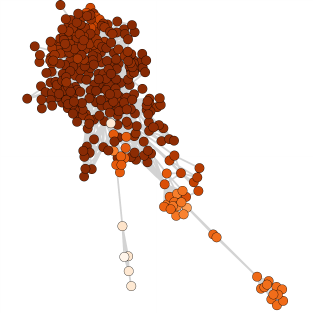}}
    	    \end{minipage}
    	    \begin{minipage}[b]{0pt}
    	    \hspace{-7pt}
            	\includegraphics[height=2pt, width=1.7cm, angle=90]{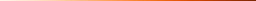} 
    	    \vspace{30pt}
    	    \end{minipage}
        	\fbox{\includegraphics[width=0.95\linewidth]{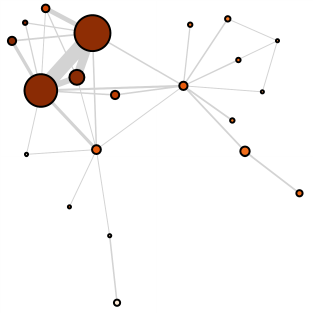}} 
	    \end{minipage}
            \vspace{-3pt}
	}
	\subfloat[Eccentricity\label{fig:lenses:ecc}]{
	    \begin{minipage}[b]{0.17\linewidth}
    	    \begin{minipage}[b]{0.95\linewidth}
            	\fbox{\includegraphics[width=\linewidth]{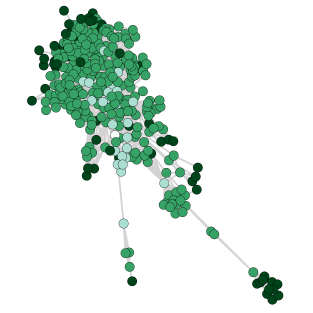}} 
    	    \end{minipage}
    	    \begin{minipage}[b]{0pt}
    	    \hspace{-7pt}
            	\includegraphics[height=2pt, width=1.7cm, angle=90]{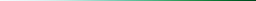}
    	    \vspace{30pt}
    	    \end{minipage}
        	\fbox{\includegraphics[width=0.95\linewidth]{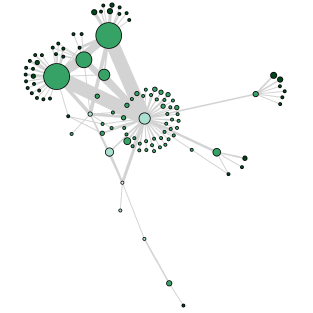}} 
	    \end{minipage}
            \vspace{-3pt}
        }
	\subfloat[Fiedler Vector\label{fig:lenses:fv}]{
	    \begin{minipage}[b]{0.17\linewidth}
    	    \begin{minipage}[b]{0.95\linewidth}
            	\fbox{\includegraphics[width=\linewidth]{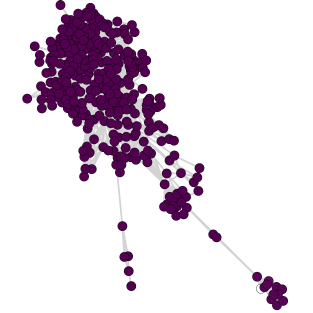}} 
    	    \end{minipage}
    	    \begin{minipage}[b]{0pt}
    	    \hspace{-7pt}
            	\includegraphics[height=2pt, width=1.7cm, angle=90]{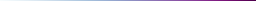} 
    	    \vspace{30pt}
    	    \end{minipage}
        	\fbox{\includegraphics[width=0.95\linewidth]{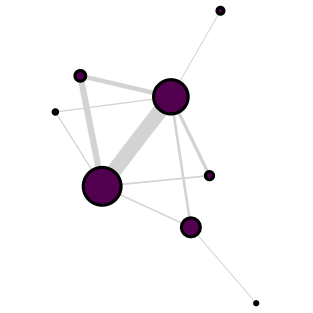}} 
	    \end{minipage}
            \vspace{-3pt}
	}
	\subfloat[PageRank ($d\!=\!0.85$)\label{fig:lenses:pr}]{
	    \begin{minipage}[b]{0.17\linewidth}
    	    \begin{minipage}[b]{0.95\linewidth}
            	\fbox{\includegraphics[width=\linewidth]{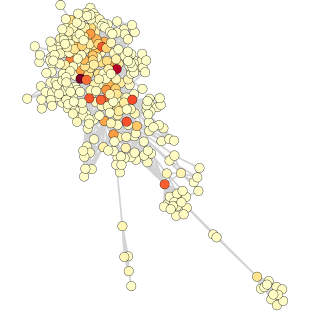}} 
    	    \end{minipage}
    	    \begin{minipage}[b]{0pt}
    	    \hspace{-7pt}
            	\includegraphics[height=2pt, width=1.7cm, angle=90]{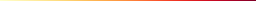} 
    	    \vspace{30pt}
    	    \end{minipage}
        	\fbox{\includegraphics[width=0.95\linewidth]{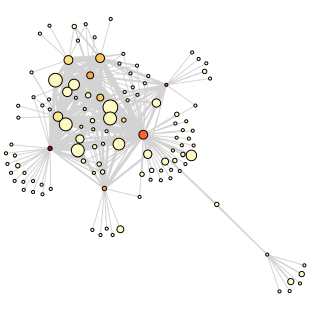}} 
	    \end{minipage}
            \vspace{-3pt}
	}

        \vspace{-5pt}
	\caption{Examples of the topological lenses applied to the \textsc{usair97} input graph (top) and the resulting \moag (bottom) using $n\!=\!8$, $\epsilon\!=\!0$, and modularity clustering. Each of the topological lenses provides a slightly different perspective on the input graph, which is reflected in the output \moag.}
    \label{fig:lenses}
\end{figure*}

\section{Mapper on Graphs Construction}
\label{sec:methods}

Our {\mog} method---a variation of the classic mapper construction---provides a general framework to skeletonize and visualize a graph. 
However, to accomplish this, \textit{every stage of the classic mapper construction required significant modification} (see \cref{fig:mog-pipeline}). This includes the selection of graph-specific topological lenses, and modification of the methods used to calculate the cover and the output {\moag} nodes and edges.

Whereas the input to classic mapper construction is a high-dimensional point cloud $P \subset \Rspace^n$, the input to \mog is a weighted undirected graph $G=(V,E)$ equipped with a positive edge weight $w: E \to \Rspace$ (see \cref{fig:mog-pipeline:function}). If a weight is not provided, several strategies are possible. We assign a uniform edge weight, as seen in other recent works~\cite{SuhHajijWang2019,doppalapudi2022untangling}.

\subsection{Topological Lens \texorpdfstring{(\cref{fig:mog-pipeline:function})}{}}

The first step of classic mapper construction is to apply a real-valued function  
to each point. The function plays the role of a \textit{topological lens} through which we look at the properties of the data. An interesting open problem for the classic mapper construction is how to formulate topological lenses beyond the best practice or a rule of thumb~\cite{BiasottiGiorgiSpagnuolo2008, BiasottiMariniMortara2003}. Nevertheless, prior work has shown that different lenses are important for providing different insights into the data~\cite{BiasottiGiorgiSpagnuolo2008, SinghMemoliCarlsson2007}. 

\begin{table}[!b]
\centering
\caption{Topological lens comp.\ scalability and properties preserved}
\label{tab:props}
\resizebox{0.9\linewidth}{!}{%
\begin{tabular}{l|c|c}
    Method                          & Scalability   & Properties Preserved  \\
    \hline \hline
    Average Geodesic Distance (AGD)  & Low           & Symmetries            \\
    \hline
    Density Estimation               & Low           & Density                      \\
    \hline
    Eccentricity                     & Low           & Centrality                     \\
    \hline
    Eigenfunctions (*Fiedler vector) & Low (*High)   & Spectral \\
    \hline
    PageRank                         & High          & Importance \\
    \hline
\end{tabular}}
\end{table}

For \mog, the \textit{topological lens} (see \cref{fig:mog-pipeline:function}) is a real-valued function defined on the nodes, $f: V \to \Rspace$.  
We focus on five graph-theoretic lenses, including average geodesic distance (AGD), density estimation, eccentricity, eigenfunctions of the graph Laplacian, and PageRank (see~\cref{fig:lenses}), although our framework can be easily extended to include other lenses. 
Each lens is chosen to reflect a specific property of interest that is intrinsic to the structure of a graph, which are outlined in \cref{tab:props}. 
In addition to utilizing the function directly, we provide a non-parametric variation obtained using histogram equalization on the topological lens.

\subsubsection{Average Geodesic Distance (AGD)}

The average geodesic distance (AGD)~\cite{KimLipmanChen2010} is a topological lens that detects symmetries in the graph while being invariant to reflection, rotation, and scaling, and it has been used extensively in shape analysis due to these properties~\cite{KimLipmanChen2010}. 
AGD is calculated by first defining a geodesic distance $d(u, v)$ between any two nodes $u, v \in V$, in our case, utilizing Dijkstra's shortest path algorithm. The AGD is given by
$$\AGD(v)=\frac{1}{|V|}\sum_{u\in V}d(v,u).$$
This definition implies that the nodes near the center of the graph will likely have low function values, while points on the periphery will have high values. 
See \cref{fig:lenses:agd} for an example.

\subsubsection{Density Estimation} 

Density estimation helps to differentiate dense regions from sparse regions and outliers. The density estimation function~\cite{Silverman1986} is given by 
$$D_{\delta}(v)=\sum_{u\in V}\exp(\frac{-d(u,v)^2}{\delta}),$$ 
where $d(u,v)$ is the geodesic distance between two nodes in the graph and $\delta>0$. $D_{\delta}$ tends to differentiate dense regions from sparse regions and outliers, and interestingly, it tends to negatively correlate with the AGD. We set $\delta=0.5$ for all examples. See \cref{fig:lenses:den} for an example.

\begin{figure}[!b]
    \centering

    \includegraphics[width=0.975\linewidth]{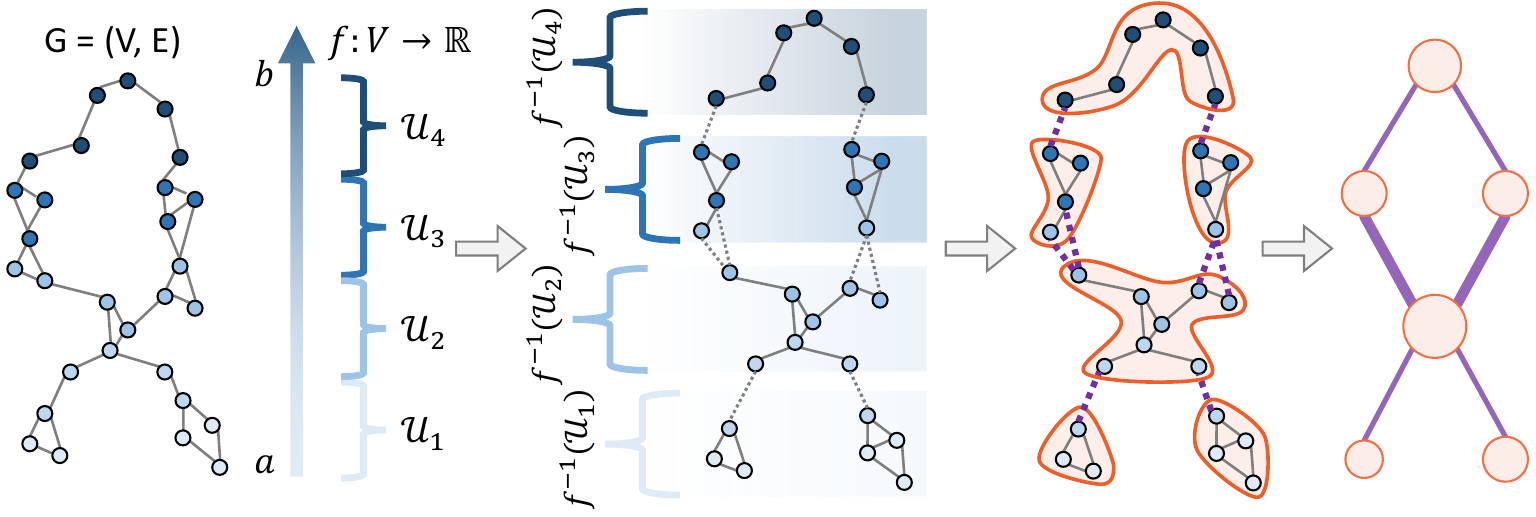} 
    
    \vspace{-17pt}
    \subfloat[\label{fig:mog-pipeline:function}]{\hspace{15pt}}\hspace{0.05\linewidth}
    \hspace{0.01\linewidth}
    \subfloat[\label{fig:mog-pipeline:cover}]{\hspace{15pt}}\hspace{0.125\linewidth}
    \hspace{0.06\linewidth}
    \subfloat[\label{fig:mog-pipeline:inv_img}]{\hspace{15pt}}\hspace{0.175\linewidth}
    \hspace{0.01\linewidth}
    \subfloat[\label{fig:mog-pipeline:cluster}]{\hspace{15pt}}\hspace{0.125\linewidth}
    \subfloat[\label{fig:mog-pipeline:mog}]{\hspace{15pt}}
    
    \vspace{-7pt}
    \caption{An illustration of the {\mog} construction: (a)~A weighted graph $G = (V, E)$ with a topological lens $f:V \to \Rspace$. (b)~A cover $\Ucal$ of the range space is given by intervals $U_{1}$, $U_{2}$, $U_3$, and $U_{4}$ as cover elements. (c-d)~The connected subgraphs induced by $f^{-1}(U_{i})$ form a cover of $G$. (e)~The resulting {\moag} is a $1$-dimensional skeleton whose nodes represent the connected subgraphs (in orange), and edges represent the graph-cut between the subgraphs (in purple). Observe that the output \moag retains the main homological features of the input graph, which is represented by a large cycle and the two components attached at its bottom.}
    \label{fig:mog-pipeline}
\end{figure}

\subsubsection{Eccentricity} 

Eccentricity measures the maximum distance from one node to all others. The eccentricity of a graph node is given by 
$$Ecc(v)=\max_{u\in V}(d(u,v)),$$ 
where $d(u,v)$ is the geodesic distance between two nodes in the graph. Eccentricity measures centrality on the graph, since nodes that are towards the center need to travel a shorter distance to reach all other nodes than those on the periphery. See \cref{fig:lenses:ecc} for an example.

\subsubsection{Eigenfunctions of the Graph Laplacian} 

Eigenfunctions of the graph Laplacian $L$ capture spectral properties of the graph~\cite{LafonLee2006}. 
In particular, the gradient of the eigenfunctions of $L$ tends to follow the overall shape of the data~\cite{Levy2006}, and these functions have been used in applications, such as graph understanding~\cite{ShumanNarangFrossard2013}, segmentation~\cite{ReuterBiasottiGiorgi2009}, spectral clustering~\cite{NgJordanWeiss2002}, and min-cut problems~\cite{Luxburg2007}. 
Let $C(G)$ be the vector space of all functions, $f:V\rightarrow \mathbb{R}$. The unnormalized graph Laplacian of $G$ is the linear operator $L: C(G) \to C(G)$ defined by mapping $f \in C(G)$ to $Lf$, with weight, $w$, where 
$$(L f)(v)=\sum_{u\sim v}w_{u,v}(f(v)-f(u)).$$
Sorting the eigenvectors of $L$ by increasing eigenvalues, we can use eigenvectors of the second or the third smallest eigenvalues of $L$ as the lens, denoted as $l_2$, $l_3$, and so on.  
We focus in particular on $l_2$ (both unnormalized and normalized), commonly referred to as the Fiedler vector~\cite{Levy2006}, since it has desirable geometric properties~\cite{DingHeZha2001}. 
For example, the minimum and maximum of the Fiedler vector tend to occur at nodes with maximum geodesic distances~\cite{ChungSeoAdluru2011}. Efficient computations of the Fiedler vector also exist, making it feasible to use in larger graphs. See \cref{fig:lenses:fv} for an example.

\subsubsection{PageRank}

Except for the Fiedler vector, prior topological lenses are computationally expensive, making them prohibitive to use on large graphs. To address the scalability issue, we utilize \emph{PageRank}~\cite{BrinPage1998} as a computationally efficient lens that measures the importance of nodes.
We use a version of the PageRank algorithm applicable to undirected graphs~\cite{Grolmusz2012}. 
A PageRank vector $R:V \to \Rspace$ is defined for every node $v \in V$,
$$R(v)=\frac{(1-d)}{|V|}+d \sum_{ u\in  N(v)} \frac{R(u)}{ |N(u)|},$$  
where $N(v)$ is the set of neighbors of $v$; $0<d<1$ is the \emph{damping factor}, which is set at $0.85$ for our examples. 
Using this formation of $R(v)$, 
PageRank yields an iterative algorithm that can be computed efficiently in practice~\cite{BrinPage1998,PageBrinMotwani1999}. 
A high PageRank score at $v$ typically means that $v$ is connected to many nodes that also have high PageRank scores. 
PageRank has been shown to be a continuous function that has many properties similar to density~\cite{Pretto2008}.

\subsection{Cover of \texorpdfstring{$f$}{f} \texorpdfstring{(\cref{fig:mog-pipeline:cover})}{}}
\label{subsec:cover}

\begin{figure}[!t]
    \subfloat[$n\!=\!2, \epsilon\!=\!0$\label{fig:cover:n2}]{\fbox{\includegraphics[width=0.220\linewidth]{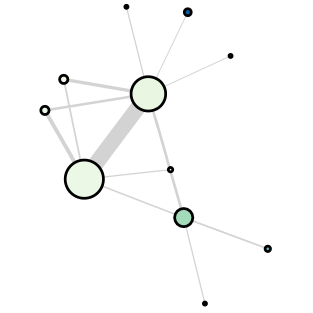}}}\hfill
    \subfloat[$n\!=\!4, \epsilon\!=\!0$\label{fig:cover:n4}]{\fbox{\includegraphics[width=0.220\linewidth]{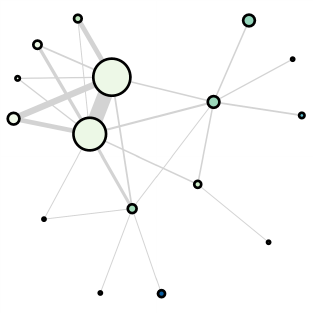}}}\hfill
    \subfloat[$n\!=\!8, \epsilon\!=\!0$\label{fig:cover:n8}]{\fbox{\includegraphics[width=0.220\linewidth]{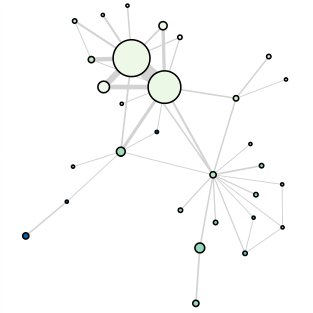}}}\hfill
    \subfloat[$n\!=\!16, \epsilon\!=\!0$\label{fig:cover:n16}]{\fbox{\includegraphics[width=0.220\linewidth]{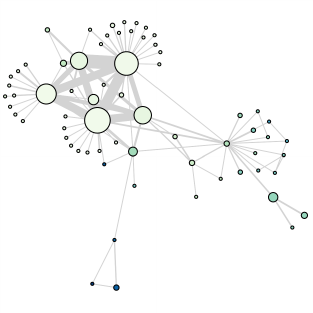}}}
    
    \subfloat[$n\!=\!8, \epsilon\!=\!0$\label{fig:cover:e0}]{\fbox{\includegraphics[width=0.220\linewidth]{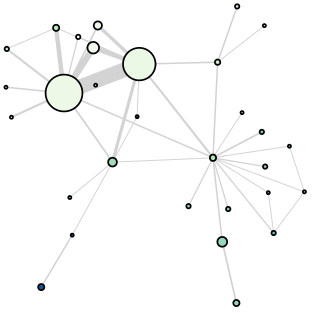}}}\hfill
    \subfloat[$n\!=\!8, \epsilon\!=\!0.1$\label{fig:cover:e1}]{\fbox{\includegraphics[width=0.220\linewidth]{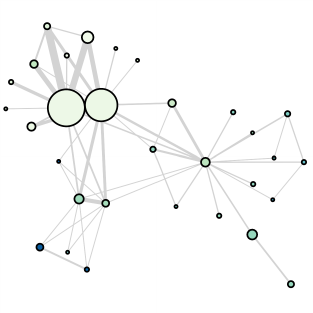}}}\hfill
    \subfloat[$n\!=\!8, \epsilon\!=\!0.2$\label{fig:cover:e2}]{\fbox{\includegraphics[width=0.220\linewidth]{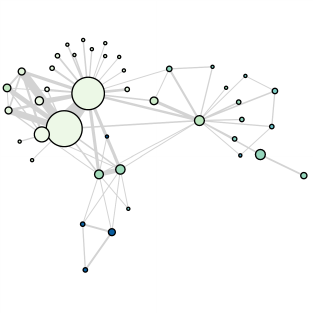}}}\hfill
    \subfloat[$n\!=\!8, \epsilon\!=\!0.4$\label{fig:cover:e4}]{\fbox{\includegraphics[width=0.220\linewidth]{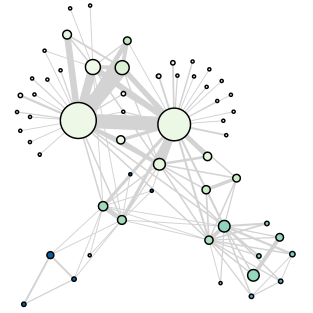}}}

    \caption{An example using AGD while varying $n$ and $\epsilon$ in a cover on the \textsc{usair97} data: (a-d)~fixing $\epsilon\!=\!0$, $n\!=\!\{2,4,8,16\}$, respectively; and (e-h)~fixing $n\!=\!8$, $\epsilon\!=\!\{0,0.1,0.2,0.4\}$, respectively.}
    \label{fig:cover}
\end{figure}

In the second step, a \textit{cover} $\Ucal$ is formed over the range of $f$, $f(V) \subseteq [a,b]$ (see \cref{fig:mog-pipeline:cover}). The cover consists of a finite number of open intervals called cover elements, $\Ucal=\{U_\alpha\}_{\alpha \in A}$. 
To obtain the cover, we use a common strategy, \textit{uniformly sized overlapping} intervals. 
Given a graph $G$ equipped with a topological lens $f: V \to \Rspace$, 
suppose the range of the function $f(V)$ is normalized to be within $[0,1]$. To obtain an initial cover, we split $[0, 1]$ into $n$ (the \emph{resolution} parameter) intervals $[c_1,c_2], \cdots, [c_{n-1},c_{n}]$ with equal length, such that $c_1=0$ and $c_n\!=\!1$.  
Adjusting the resolution parameter increases or decreases the amount of aggregation {\mog} provides. Broadly speaking, smaller $n$ leads to a smaller topological summary of the graph.  
\cref{fig:cover:n2}-\ref{fig:cover:n16} show an example by varying the number of cover elements $n$.

The \emph{overlap} parameter $\epsilon$ is used to obtain the cover $\Ucal$ consisting of cover elements $(c_{i}-\epsilon,c_{i+1}+\epsilon)$ for $1 \leq i \leq n-1$. \cref{fig:cover:e0}-\ref{fig:cover:e4} show an example of increasing $\epsilon$. Increasing the overlap causes the number of output nodes and density of edges in the \moag to increase, which is caused by the duplication of data in the overlapped cover elements. The necessity for the overlapping cover elements in the classic mapper is partially attributable to the lack of connectivity in point clouds. Importantly, \textit{in a significant departure from classic mapper, we utilize the edges of the graph for the connectivity}, as will be discussed shortly. \textit{This modification eliminates the need for overlap.} Therefore, we use $\epsilon\!=\!0$ in all of our experiments.

\subsection{Inverse Image \texorpdfstring{(\cref{fig:mog-pipeline:inv_img})}{}} 
In the third step, the \textit{inverse image of the cover} is calculated, as shown in \cref{fig:mog-pipeline:inv_img}. For each cover element $\Ucal_\alpha$, the inverse image $f^{-1}(\Ucal_\alpha)$ is the subset of nodes $V_{\alpha}\subseteq V$, where $f(V_{\alpha})\in\Ucal_\alpha$. In other words, a  node is associated with a cover element whose interval cover the node's function value. 
In contrast with the classic mapper, for each element of the cover $\Ucal_\alpha$, we find the nodes $V_{\alpha} = f^{-1}(\Ucal_\alpha)$ and extract its induced subgraph $H_\alpha \subseteq G$, where $f(V_\alpha)\in\Ucal_\alpha$. It is important to note that when the overlap is used in cover elements (i.e., $\epsilon\!\neq\!0$), some nodes from the input graph will be replicated across multiple inverse images.

\subsection{Mapper Graph Nodes and Edges \texorpdfstring{(\cref{fig:mog-pipeline:cluster})}{}}
\label{sec:methods:nodes}

The final step of the process is \textit{forming nodes and edges of the output \emph{\moag}}, as shown in \cref{fig:mog-pipeline:cluster}. 

\subsubsection{Forming Mapper Graph Nodes}

To form the nodes of the output, classic mapper construction applies a clustering algorithm to a set of points in the inverse image of a cover element (e.g., agglomerative clustering or DBSCAN~\cite{ester1996density}). 
In our setting, the subgraphs $H_{\alpha}$ extracted with \mog have additional structures coming from the connectivity of the graph. Therefore, to form the nodes of the \moag, we apply a graph clustering algorithm to each subgraph $H_{\alpha}$. 
Every cluster within $H_{\alpha}$ is returned as a node of the output \moag. 

\begin{figure}[!b]
    \centering
    
    {
    \begin{minipage}[b]{0.575\linewidth}
    \centering
    \subfloat[Conn.\ Components\label{fig:node_clustering:cc}]{\fbox{\includegraphics[height=2.25cm]{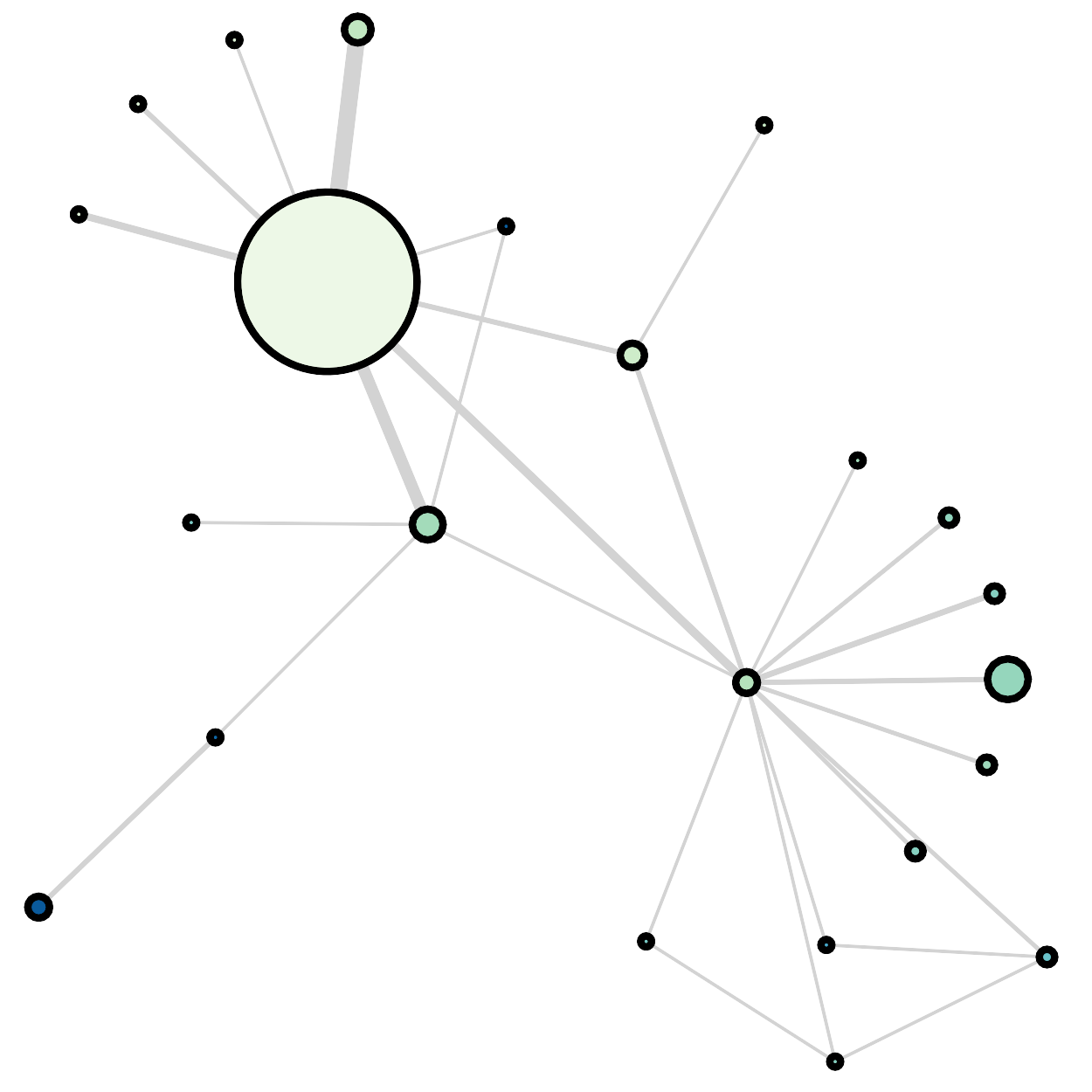}}}\hfill
    \subfloat[Modularity Clustering\label{fig:node_clustering:mc}]{\fbox{\includegraphics[height=2.25cm]{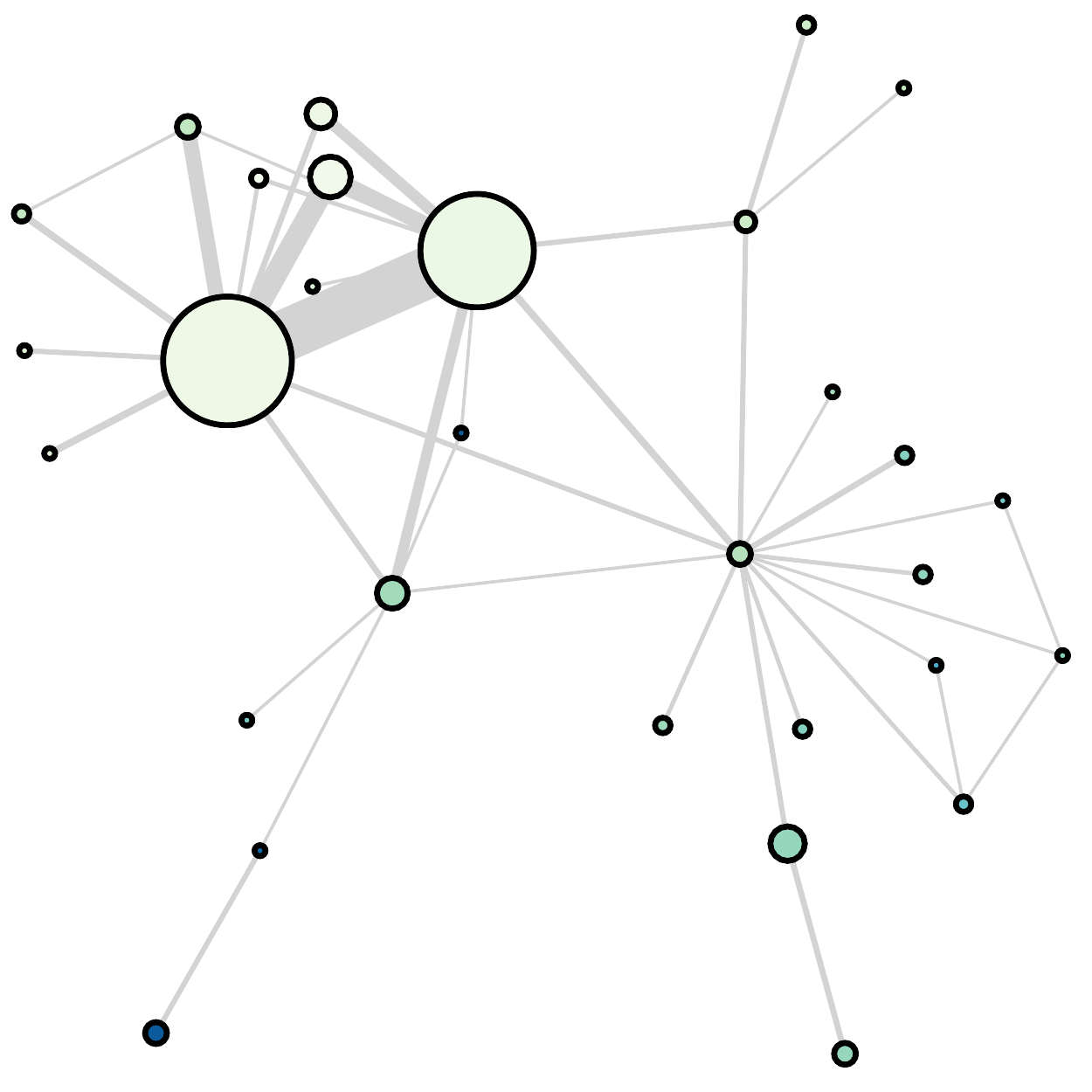}}}\hfill
    \subfloat[Label Propagation\label{fig:node_clustering:lp}]{\fbox{\includegraphics[height=2.25cm]{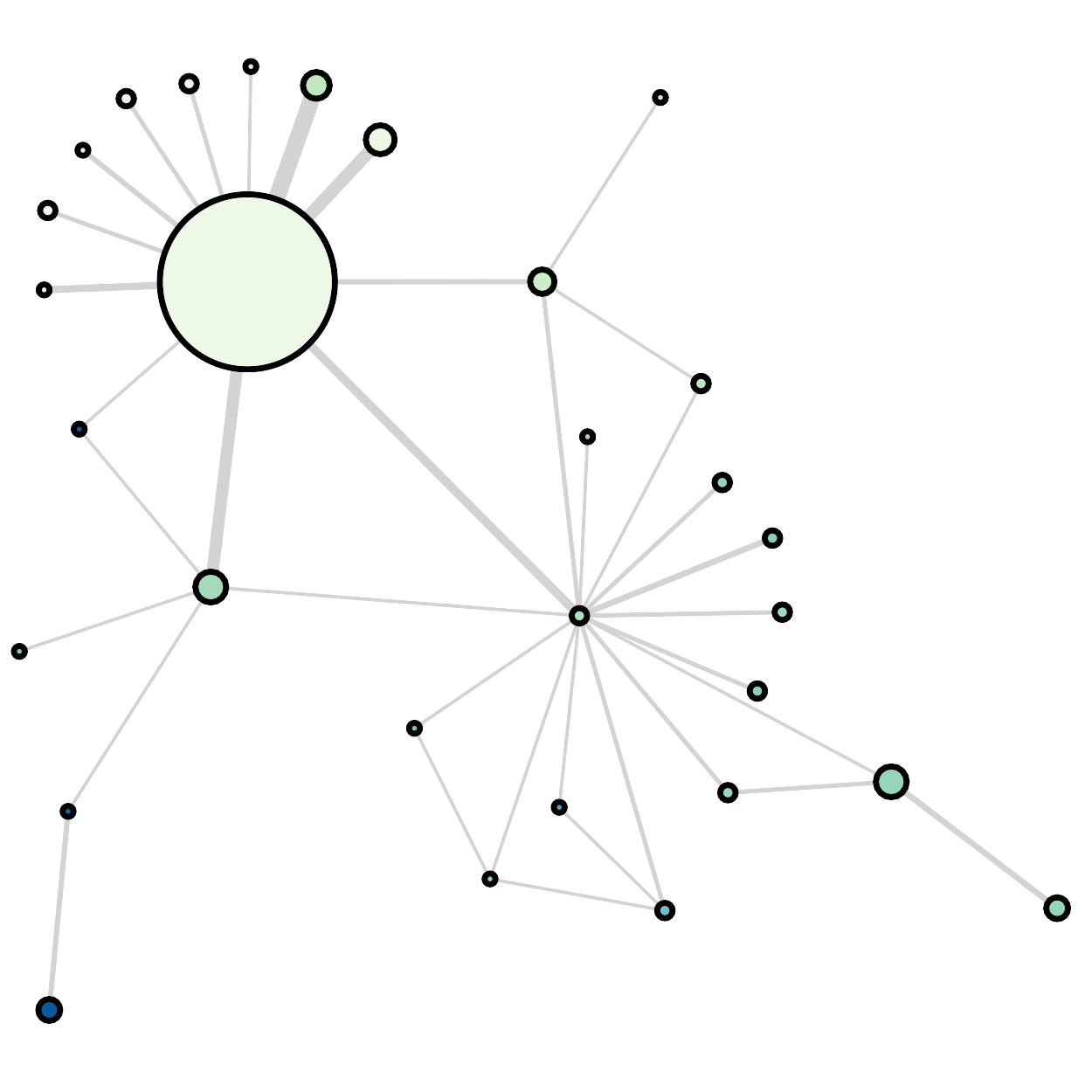}}}
    \end{minipage}
    }
    \hspace{1pt}
    \begin{minipage}[b]{0pt}
    \textcolor{mygray}{\rule{1pt}{4cm}}
    \vspace{10pt}
    \end{minipage}
    \hspace{-8pt}
    {
    \begin{minipage}[b]{0.4\linewidth}
        \centering
        \subfloat[Connectivity using overlap\label{fig:link_methods:conn}]{\hspace{10pt}\fbox{\includegraphics[height=2.25cm]{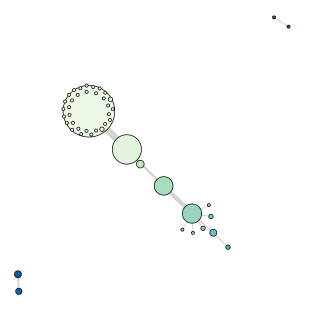}}\hspace{10pt}}
        
        \subfloat[Connectivity using graph-cut\label{fig:link_methods:gc}]{\hspace{10pt}\fbox{\includegraphics[height=2.25cm]{figs/clustering/fig4_mog_agd_8_connected_components.png}}\hspace{10pt}}
    \end{minipage}}
    
    \caption{Left: An example of using (a)~connected components, (b)~modularity clustering, and (c)~asynchronous label propagation as node clustering methods on the \textsc{usair97} data with $n\!=\!8$ and $\epsilon\!=\!0$. Right: An example of using different edge forming methods on the \textsc{usair97} data with $n\!=\!8$, $\epsilon\!=\!0$, and connected components.}
    \label{fig:node_clustering}
\end{figure}

We apply three clustering algorithms that preserve different aspects of the subgraph. 
The first identifies connected components in each subgraph (see \cref{fig:node_clustering:cc}), that is, groups of nodes in $H_{\alpha}$ that are reachable to each other. The second algorithm is modularity clustering~\cite{brandes2007modularity} (see \cref{fig:node_clustering:mc}), which identifies communities with dense connections between nodes and sparse connections between communities. Finally, we consider label propagation (see \cref{fig:node_clustering:lp}), which identifies communities by iteratively assigning node labels and propagating the labels through the graph. We tested two label propagation algorithms (semi-synchronous~\cite{cordasco2010community} and asynchronous~\cite{raghavan2007near}) and obtained similar results (all presented results use asynchronous label propagation). 

\cref{fig:node_clustering} show the results of these clustering algorithms, which preserve similar structures. However, the graph details they retain vary based on the properties of the underlying algorithm.

\subsubsection{Forming Mapper Graph Edges} 
Finally, we form the edges of the \moag. 
Initially, we test a classic (mapper) approach that identifies the overlapping node sets between clusters. For all mapper graph nodes created in the prior step, if they share any node from the original graph, they are connected with a weight equal to the overlap between them. In other words, consider two mapper  nodes $N_i$ and $N_j$ ($i\neq j$), they are connected if $N_i\cap N_j \neq \emptyset$ with $w_{ij}=|N_i\cap N_j|$. Unfortunately, this method does not work as well as we have hoped. As seen in \cref{fig:link_methods:conn}, the method has two undesirable problems. First, the required overlap creates a few dense connections. Second, the method does not guarantee connectivity.  

Instead, we use the connectivity of the input graph to determine the connectivity of the output graph. We utilize the number of edges in the input graph that go between any two \moag nodes $N_i$ and $N_j$ ($i\neq j$). We count the number of edges between the nodes of $N_i$ and $N_j$, which can be efficiently computed using a graph-cut. The nodes are connected if $graph\_cut(N_i,N_j) \neq \emptyset$, with $w_{ij}=|graph\_cut(N_i,N_j)|$. \cref{fig:link_methods:gc} shows an example. Alternative approaches could sum ($w_{ij}=\sum_{q\in graph\_cut(N_i,N_j)} w_q$) or apply a statistic, such as the mean, to the weights of the edges of the graph-cut. The utility of such approaches depends upon the analytical goals of the visualization.
In comparison with the initial approach, the graph-cut-based method better preserves graph details and eliminates the need for enforcing overlaps among cover elements. Unless otherwise specified, all examples in the paper use the graph-cut method to form edges in the \moag.

\begin{figure}[!b]
    \centering
    \fbox{%
    \begin{minipage}[b]{0.325\linewidth}
        \centering
        \includegraphics[height=80pt]{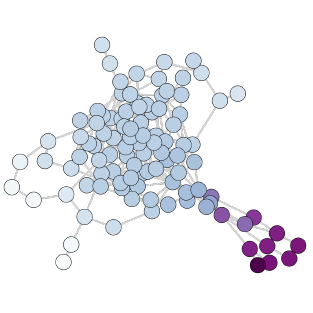}
    \end{minipage}}
    \fbox{%
    \begin{minipage}[b]{0.6\linewidth}
        \centering
        \includegraphics[trim=0 0 55pt 0, clip, height=80pt]{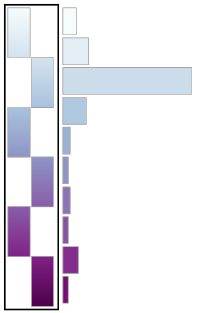}
        \includegraphics[trim= 0 19pt 0 0, clip, height=80pt]{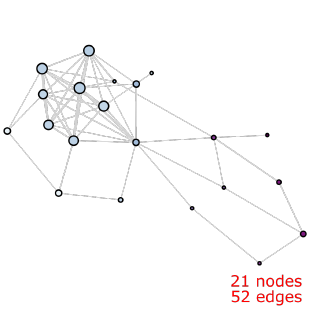}
    \end{minipage}}
     
    \vspace{-10pt}
    {\subfloat[]{\hspace{75pt}}}
    \hspace{8pt}
    {\subfloat[\label{fig:histogram:cover}]{\hspace{12pt}}}
    {\subfloat[\label{fig:histogram:hist}]{\hspace{25pt}}}
    {\subfloat[\label{fig:histogram:mog}]{\hspace{97pt}}}
    \caption{Visualization of the (a)~\textsc{movies} graph includes a visualization of (b)~the cover elements $\Ucal$, (c)~the histogram of the topological lens $f$, and (d)~the \moag ($n\!=\!6$, $\epsilon\!=\!0$, with modularity clustering).}
    \label{fig:histogram}
\end{figure}

\subsection{Visualization \texorpdfstring{(\cref{fig:mog-pipeline:mog})}{}}

The final output \moag (\cref{fig:mog-pipeline:mog}) provides an overview of the homological structure  of the input graph (\cref{fig:mog-pipeline:function}) through multiple perspectives using different topological lenses. The levels of details captured by the \moag may be adjusted by modifying the cover. Its visualization consists of two components (see \cref{fig:histogram}): a visualization of the {\moag}, and a visualization of the cover with a histogram useful for parameter selection.

\subsubsection{Output Mapper Graph Visualization}
The output of \mog is a \moag with additional metadata, which is the primary visualization target, with flexibility in its layout and visual encodings; see \cref{fig:histogram:mog}. 
For the purpose of this paper, we have chosen a simple approach. The \moag is laid out using a standard force-directed layout provided by D3.js. Our approach is ultimately agnostic of the graph layout algorithm, and different layouts (e.g., layered approaches) may improve the presentation under certain scenarios. For illustrative purposes, we also add several (optional) visual encodings to better understand the information captured by \mog. First, a \moag node $N_i$ (i.e., the cluster $N_i \subseteq G$) is colored using the average function value across its members using a topological lens specific colormap (see \cref{fig:lenses}). Next, the sizes of the nodes are proportional to $|N_i|$. Finally, edge thickness is drawn proportional to the edge weight (i.e., $|graph\_cut(N_i,N_j)|$). 
 
\subsubsection{Histogram and Cover Visualization}

To assist with parameter selection, we provide additional information about the input graph, which is particularly useful when the input graph is too large to draw. 

To understand the distribution of a topological lens $f$, a histogram of the function values is laid out vertically, with the minimum value $a$ at the bottom and the maximum value $b$ at the top. \cref{fig:histogram:hist} shows an example of a histogram for the Fiedler lens. The histogram is split into an adjustable number of bins.

The cover (see \cref{fig:histogram:cover}) is visualized using a series of boxes, one per cover element, displayed next to the histogram of the topological lens. Each box is placed and colored based on its corresponding interval. The boxes are laid out horizontally automatically using a greedy packing approach. Observing patterns in the histogram and its alignment with the cover elements can help inform choices for $n$, $\epsilon$, and the equalization of the topological lens.

\section{Evaluation}
\label{sec:results}

We evaluate the computational performance of our approach (see \cref{sec:eval:perf}) and its ability to provide multiscale homology-preserving skeletonizations (see \cref{sec:eval:examp}, \ref{sec:results:community}, and \ref{sec:eval:human}).

\paragraph{Implementation}
Our approach is implemented using Python for the calculation of the \moag and D3.js for visualization. 
The code and datasets are available on GitHub $<${\footnotesize\url{https://github.com/shape-vis/MapperOnGraphs}}$>$. A demo version of the software is available at $<${\footnotesize\url{https://shape-vis.github.io/MapperOnGraphsDemo/}}$>$. 

\paragraph{Datasets}
\label{sec.eval.data}
We evaluate our approach using over 45 graphs (see \cref{tab:datasets}). 
We focus on a subset of small ($|V|\in[100,\numprint{1000})$, $|E|\in[200,\numprint{16651}]$), medium ($|V|\in[\numprint{1000},\numprint{5000})$, $|E|\in[\numprint{1092},\numprint{284924}]$), and large ($|V|\in[\numprint{5000},\numprint{1134890}]$, $|E|\in[\numprint{91342},\numprint{2987624}]$) graphs. 
The datasets are a mix of synthetic and real-world datasets.
Graphs not in the paper are available in the demo.

\begin{table}[!ht]
    \centering
    \caption{Listing of Datasets Used in Evaluation}
    \label{tab:datasets}
    \resizebox{0.95\linewidth}{!}{%
    \begin{tabular}{p{5cm}|@{\hspace{4pt}}c@{\hspace{4pt}}|c|c}
        Dataset	&	$|V|$	&	$|E|$	&	Source	\\
        \hline	
        \arrayrulecolor{mygray}
        \textsc{airport}	&	 2,896 	&	 15,641 	&	\url{Openflights.org}	\\
        \arrayrulecolor{mygray}\hline
        \rowcolor{mylightgray}
        \textsc{amazon0302}	&	 262,111 	&	 899,792 	&	SNAP~\cite{snapnets}	\\
        \arrayrulecolor{mygray}\hline
        \textsc{balanced tree(3,6)}	&	 1,093 	&	 1,092 	&	NetworkX~\cite{hagberg2008exploring}	\\
        \arrayrulecolor{mygray}\hline
        \rowcolor{mylightgray}
        \textsc{barbell(50,20)}	&	 120 	&	 2,471 	&	NetworkX~\cite{hagberg2008exploring}	\\
        \arrayrulecolor{mygray}\hline
        \textsc{bcsstk}	&	 110 	&	 254 	&	UF Sparse Matrix Repo.~\cite{davis2011university}	\\
        \arrayrulecolor{mygray}\hline
        \rowcolor{mylightgray}
        \textsc{bcsstk20}	&	 467 	&	 1,762 	&	UF Sparse Matrix Repo.~\cite{davis2011university}	\\
        \arrayrulecolor{mygray}\hline
        \textsc{bcsstk22}	&	 110 	&	 364 	&	UF Sparse Matrix Repo.~\cite{davis2011university}	\\
        \arrayrulecolor{mygray}\hline
        \rowcolor{mylightgray}
        \textsc{bio-celegans}	&	 453 	&	 2,025 	&	Network Repository~\cite{rossi2015network}	\\
        \arrayrulecolor{mygray}\hline
        \textsc{bio-diseasome}	&	 516 	&	 1,188 	&	Network Repository~\cite{rossi2015network}	\\
        \arrayrulecolor{mygray}\hline
        \rowcolor{mylightgray}
        \textsc{bn-mouse-visual-cortex-2}	&	 193 	&	 214 	&	Network Repository~\cite{rossi2015network}	\\
        \arrayrulecolor{mygray}\hline
        \textsc{ca-condmat}	&	 21,363 	&	 91,342 	&	SNAP~\cite{snapnets}	\\
        \arrayrulecolor{mygray}\hline
        \rowcolor{mylightgray}
        \textsc{caltech}	&	 762 	&	 16,651 	&	Facebook 100	\\
        \arrayrulecolor{mygray}\hline
        \textsc{chch-miner}	&	 1,510 	&	 48,512 	&	BioSNAP~\cite{biosnapnets}	\\
        \arrayrulecolor{mygray}\hline
        \rowcolor{mylightgray}
        \textsc{circular ladder(100)}	&	 200 	&	 300 	&	NetworkX~\cite{hagberg2008exploring}	\\
        \arrayrulecolor{mygray}\hline
        \textsc{collaboration network}	&	 379 	&	 914 	&	\cite{newman2001structure}	\\
        \arrayrulecolor{mygray}\hline
        \rowcolor{mylightgray}
        \textsc{com-amazon}	&	 334,863 	&	 925,872 	&	SNAP~\cite{snapnets}	\\
        \arrayrulecolor{mygray}\hline
        \textsc{com-youtube}	&	 1,134,890 	&	 2,987,624 	&	SNAP~\cite{snapnets}	\\
        \arrayrulecolor{mygray}\hline
        \rowcolor{mylightgray}
        \textsc{connected caveman(15,30)}	&	 450 	&	 6,525 	&	NetworkX~\cite{hagberg2008exploring}	\\
        \arrayrulecolor{mygray}\hline
        \textsc{dch-miner}	&	 7,197 	&	 466,656 	&	BioSNAP~\cite{biosnapnets}	\\
        \arrayrulecolor{mygray}\hline
        \rowcolor{mylightgray}
        \textsc{df-miner}	&	 20,549 	&	 802,760 	&	BioSNAP~\cite{biosnapnets}	\\
        \arrayrulecolor{mygray}\hline
        \textsc{dorogovtsev goltsev mendes(5)}	&	 123 	&	 243 	&	NetworkX~\cite{hagberg2008exploring}	\\
        \arrayrulecolor{mygray}\hline
        \rowcolor{mylightgray}
        \textsc{enron-email}	&	 143 	&	 623 	&	SNAP~\cite{snapnets}	\\
        \arrayrulecolor{mygray}\hline
        \textsc{ff-miner}	&	 46,007 	&	 106,499 	&	BioSNAP~\cite{biosnapnets}	\\
        \arrayrulecolor{mygray}\hline
        \rowcolor{mylightgray}
        \textsc{hic 1k net 1}	&	 5,420 	&	 315,852 	&		\\
        \rowcolor{mylightgray}
        \textsc{hic 1k net 2}	&	 5,096 	&	 333,129 	&		\\
        \rowcolor{mylightgray}
        \textsc{hic 1k net 3}	&	 5,345 	&	 320,659 	& BioSNAP~\cite{biosnapnets}		\\
        \rowcolor{mylightgray}
        \textsc{hic 1k net 4}	&	 5,472 	&	 378,343 	&		\\
        \rowcolor{mylightgray}
        \textsc{hic 1k net 5}	&	 7,015 	&	 434,977 	&		\\
        \rowcolor{mylightgray}
        \textsc{hic 1k net 6}	&	 4,581 	&	 284,924 	&		\\
        \arrayrulecolor{mygray}\hline
        \textsc{hic 5k net 1}	&	 411 	&	 11,688 	&		\\
        \textsc{hic 5k net 2}	&	 308 	&	 9,599 	&		\\
        \textsc{hic 5k net 3}	&	 632 	&	 19,469 	&		\\
        \textsc{hic 5k net 4}	&	 265 	&	 8,402 	&	BioSNAP~\cite{biosnapnets}	\\
        \textsc{hic 5k net 5}	&	 1,032 	&	 32,329 	&		\\
        \textsc{hic 5k net 6}	&	 192 	&	 6,242 	&		\\
        \textsc{hic 5k net 7}	&	 435 	&	 13,428 	&		\\
        \textsc{hic 5k net 8}	&	 1,194 	&	 37,569 	&		\\
        \arrayrulecolor{mygray}\hline
        \rowcolor{mylightgray}
        \textsc{map of science}	&	 554 	&	 2,276 	&	\cite{borner2012design}	\\
        \arrayrulecolor{mygray}\hline
        \textsc{movies}	&	 101 	&	 192 	&	\cite{de2018exploratory}	\\
        \arrayrulecolor{mygray}\hline
        \rowcolor{mylightgray}
        \textsc{random lobster(100,0 6,0 4)}	&	 596 	&	 595 	&	NetworkX~\cite{hagberg2008exploring}	\\
        \arrayrulecolor{mygray}\hline
        \textsc{ring of cliques(6,70)}	&	 420 	&	 14,496 	&	NetworkX~\cite{hagberg2008exploring}	\\
        \arrayrulecolor{mygray}\hline
        \rowcolor{mylightgray}
        \textsc{smith}	&	 2,970 	&	 97,133 	&	Facebook 100	\\
        \arrayrulecolor{mygray}\hline
        \textsc{soc-epinions1}	&	 75,877 	&	 405,739 	&	SNAP~\cite{snapnets}	\\
        \arrayrulecolor{mygray}\hline
        \rowcolor{mylightgray}
        \textsc{usair97}	&	 332 	&	 2,126 	&	Network Repository~\cite{rossi2015network}	\\
        \arrayrulecolor{mygray}\hline
        \textsc{watts strogatz(100,5,0.05)}	&	 100 	&	 200 	&	NetworkX~\cite{hagberg2008exploring}	\\
    \end{tabular}}
\end{table}

\subsection{Performance}
\label{sec:eval:perf}

\subsubsection{Calculating Topological Lenses}

Calculating the topological lens is oftentimes the most expensive part of generating a \moag. 
We recorded the compute time for all lenses on all graphs in our test set. 
In some cases, certain lenses can not be computed in a reasonable amount of time. We terminated those computations after $900$ seconds. 

\cref{fig:ff_perf} shows the (completed) compute time for all lenses. They are laid out by the number of input graph nodes ($|V|$) against the compute time in seconds using a log-log scale. 
A power regression is plotted for each lens. 
The experimental results show that the methods from fastest to slowest are: (1)~PageRank: $O(|V|^{1.07})$, (2)~Fiedler: $O(|V|^{1.17})$, (3)~eccentricity: $O(|V|^{2.45})$, (4)~density: $O(|V|^{2.45})$, and (5)~AGD: $O(|V|^{2.46})$. 
One conclusion is that Fiedler and PageRank perform only slightly worse than linear, making them good candidates for large graphs. 
The other methods---AGD, density, and eccentricity---are worse than quadratic, making them impractical on large graphs.

\subsubsection{Calculating Mapper Graphs}

\begin{figure}[!t]
    \centering   
    
    \subfloat[Compute time against the number of nodes for each topological lens. Dotted lines are power regressions. The results show that PageRank~(yellow) and Fiedler vector~(purple) calculations scale much better than AGD~(teal), density~(orange), and eccentricity~(green).\label{fig:ff_perf}]
    {\begin{minipage}[t]{1\linewidth}
        \centering
        \includegraphics[width=0.8\linewidth]{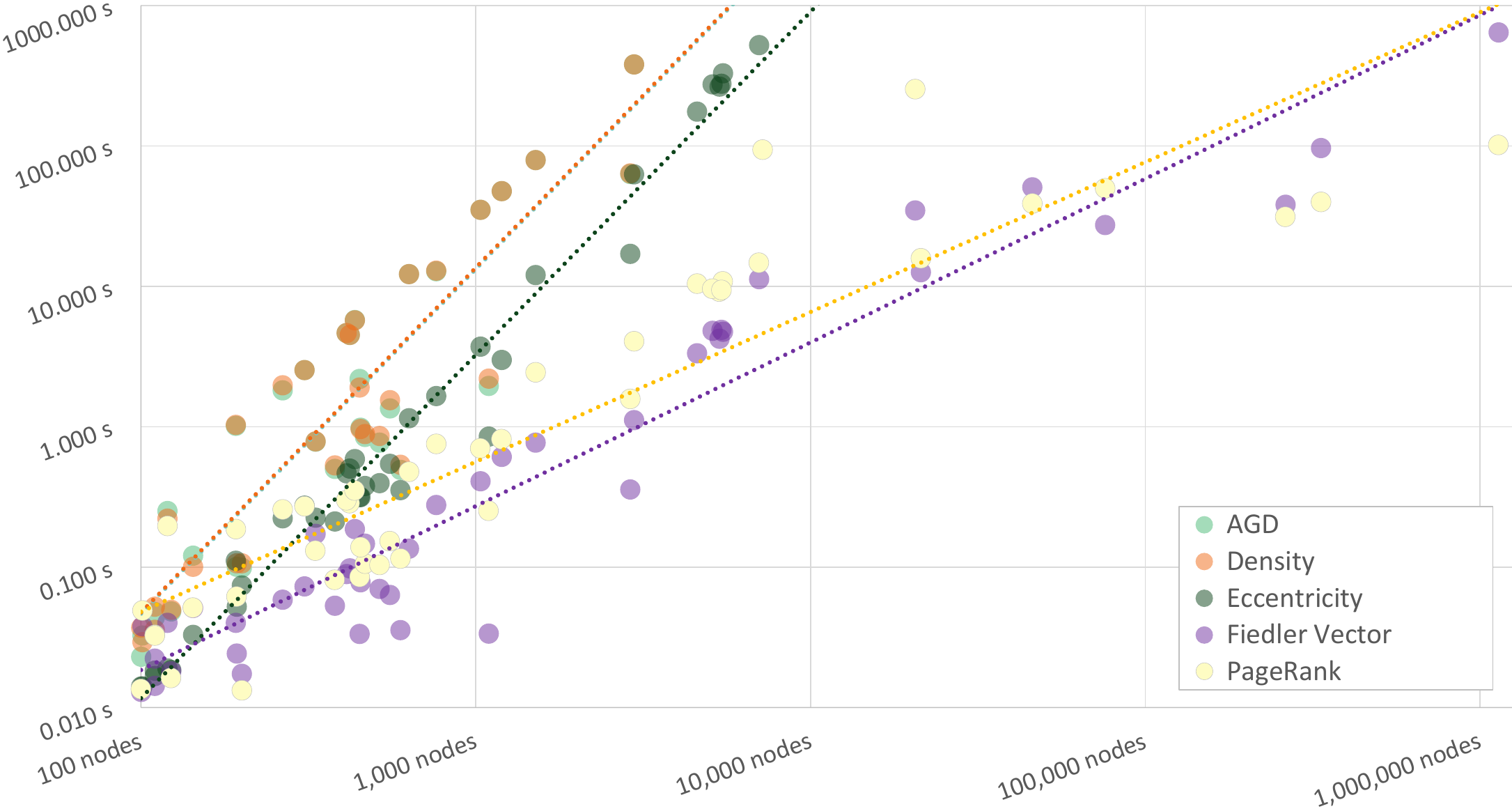}
    \end{minipage}}

    \includegraphics[width=0.65\linewidth]{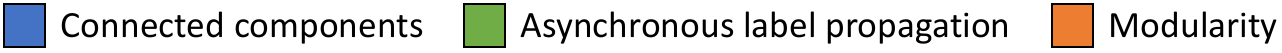}
    
    \subfloat[Compute time for small (\textsc{usair97}), medium (\textsc{smith}), and large (\textsc{hic 1k net 5}) graphs across all topological lenses. The plots show time in seconds vertically, and the number of cover elements horizontally. Instability is associated with variability in clustering time.\label{fig:cluster_time}]
    {\begin{minipage}[t]{1\linewidth}
        \includegraphics[width=0.32\linewidth]{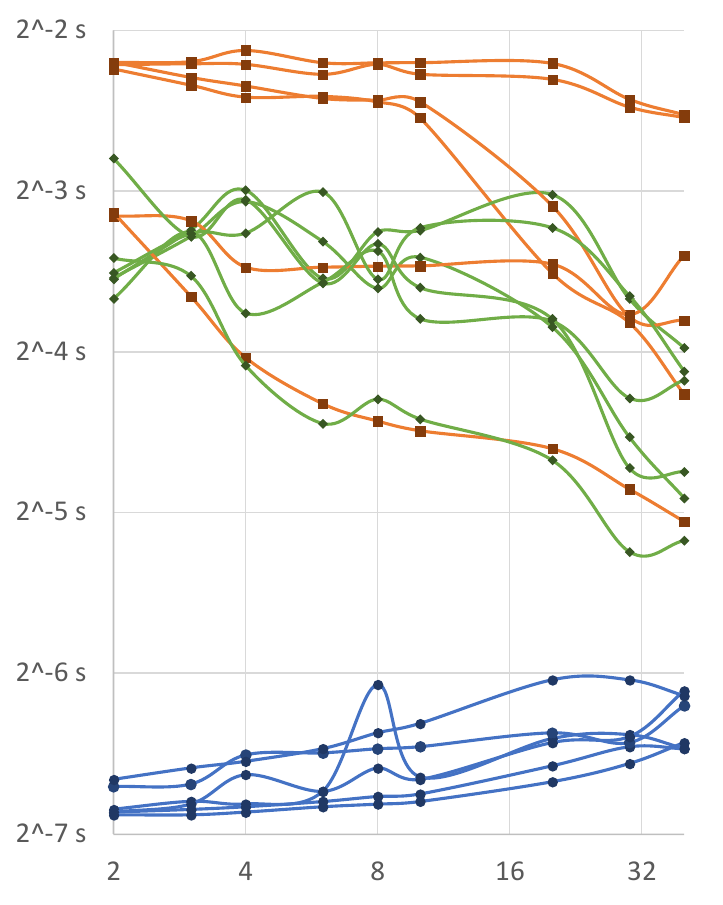}
        \hfill
        \includegraphics[width=0.32\linewidth]{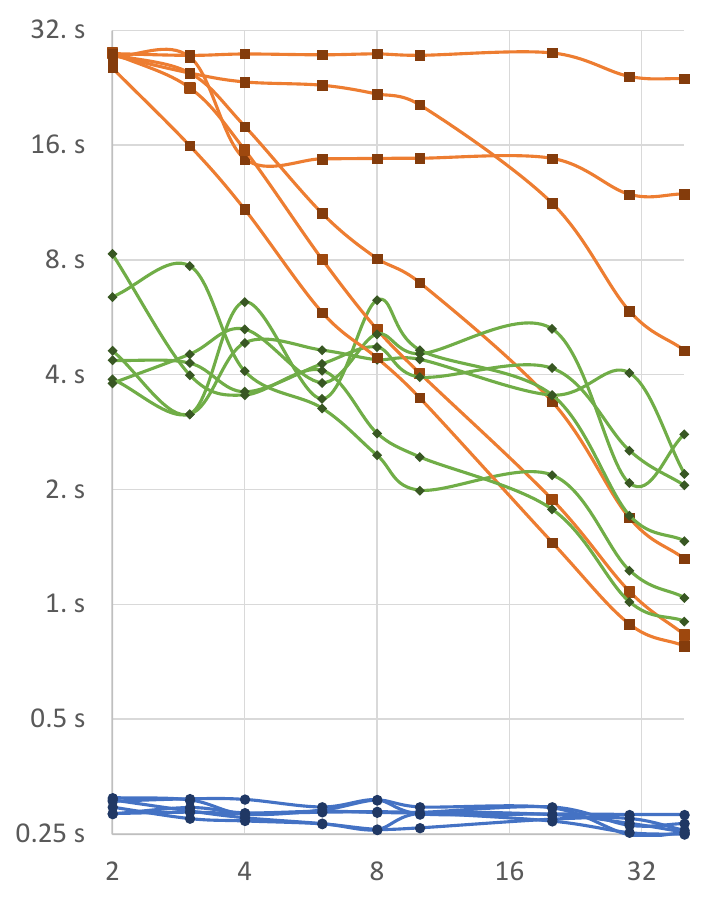}
        \hfill
        \includegraphics[width=0.32\linewidth]{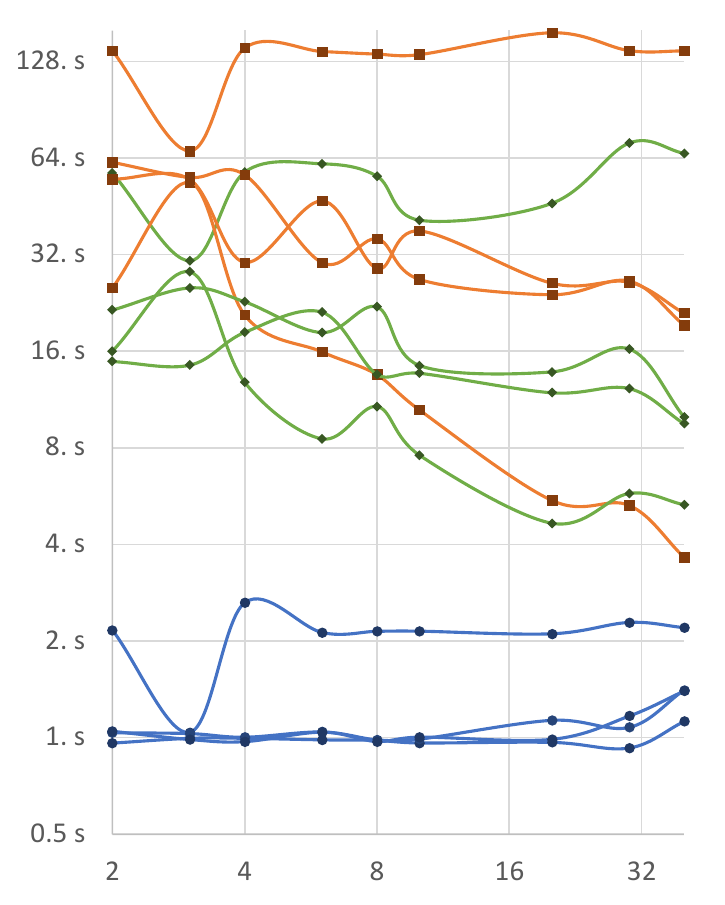}
        
        \vspace{-7pt}
        {\tiny \hspace{32pt}
        \textsc{usair97} \hspace{65pt}
        \textsc{smith} \hspace{63pt}
        \textsc{hic 1k net 5}}
        \vspace{5pt}
    \end{minipage}
    }
    
    \subfloat[Compute time for all graphs and lens functions for (left)~$n\!=\!2$ and (right)~$n\!=\!20$ cover elements. The plot shows the time in seconds vertically, and the number of input graph nodes ($|V|$) horizontally. The power regressions (dashed lines) show that the size of the input graph and the clustering algorithm have an influence on the compute time.\label{fig:all_cluster_time}]
    {\begin{minipage}[t]{1\linewidth}
    \centering
    \includegraphics[width=0.375\linewidth]{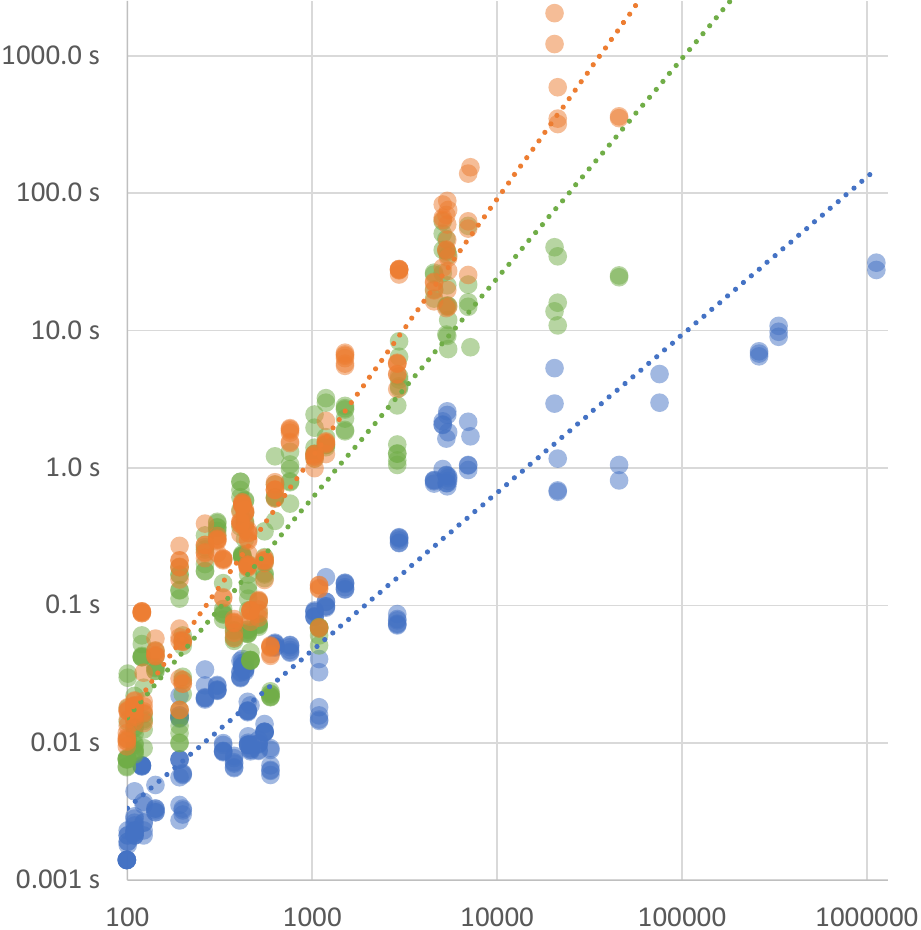} \hspace{5pt}
    \includegraphics[width=0.375\linewidth]{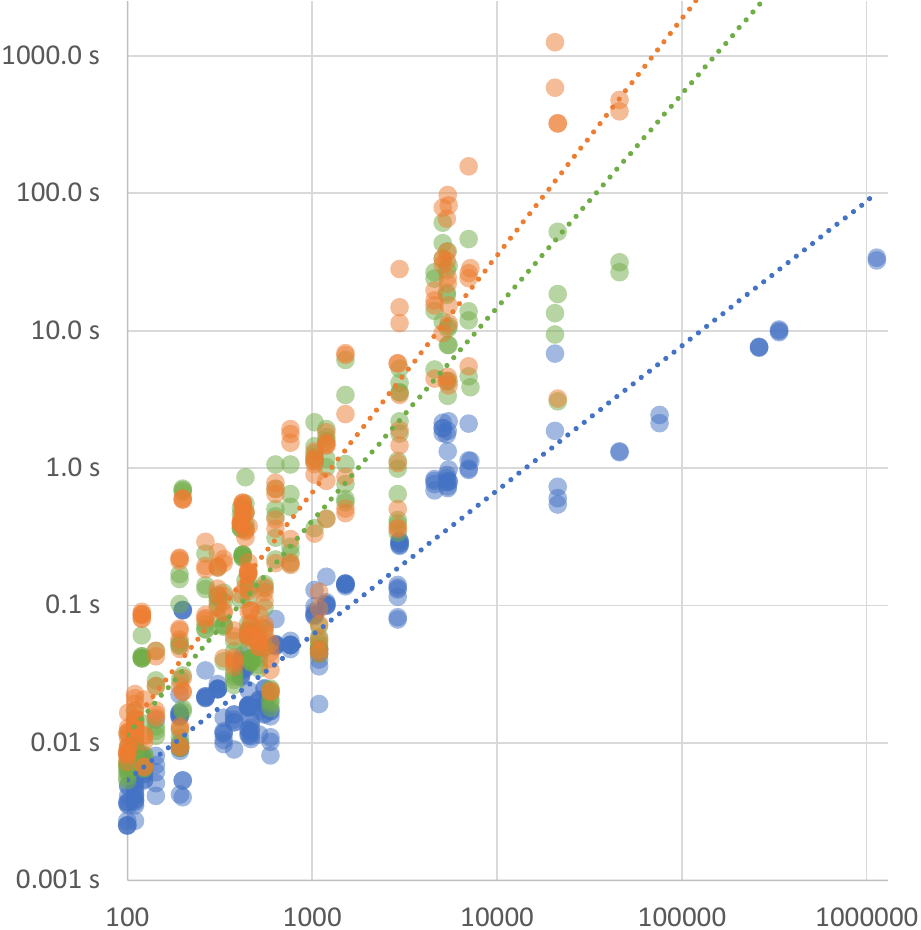}

    \vspace{-4pt}
    {\tiny \hspace{2pt}
        $n\!=\!2$ cover elements \hspace{52pt}
        $n\!=\!20$ cover elements
    }
    \end{minipage}
    }

    \caption{Log-log plots of the compute time of (a)~each topological lens, and \mog for (b)~different sized datasets across all topological lenses and (c)~all graphs and lens functions across different numbers of cover elements. The data are colored by (a)~their associate topological lens and (b-c)~the chosen clustering algorithm. }

\end{figure}

As we investigated the performance of the \mog framework, it became clear that the performance was highly dependent upon the clustering algorithm. 
\cref{fig:cluster_time} shows the time to calculate \moag for three typical datasets, one small, one medium, and one large, across several lens functions and various numbers of cover elements. 
In addition, it differentiates the clustering algorithm by color. The connected components algorithm (in blue) is the most efficient, often by several orders of magnitude. 
We also observed that asynchronous label propagation (in green) is often faster than modularity clustering. In addition, for both asynchronous label propagation and modularity clustering, the time required decreases as the number of cover elements increases. This occurs because as more cover elements are added, the number of input graph nodes that fall into each inverse image decreases. 

\cref{fig:all_cluster_time} shows the time to calculate a \moag (excluding lens calculation) for all datasets and all lenses with $n\!=\!2$ and $n\!=\!20$ cover elements. The plots show the performance trend of different clustering algorithms across a variety of datasets. The power regressions show that from fastest to slowest are: (1)~connected components: $O(|V|^{1.15})$ and $O(|V|^{1.05})$; (2)~asynchronous label propagation: $O(|V|^{1.60})$ and $O(|V|^{1.55})$; and (3)~modularity clustering: $O(|V|^{1.88})$ and $O(|V|^{1.72})$, for $n\!=\!2$ and $n\!=\!20$, respectively. 
In summary, connected component clustering is significantly more scalable than label propagation, which is somewhat more scalable than modularity clustering.

\begin{figure}[!b]
    \centering
  
    \subfloat[\textsc{bio-celegans}: PageRank, modularity clustering]{
    \begin{minipage}[t]{0.234\linewidth} \centering
        \fbox{\includegraphics[width=0.95\linewidth]{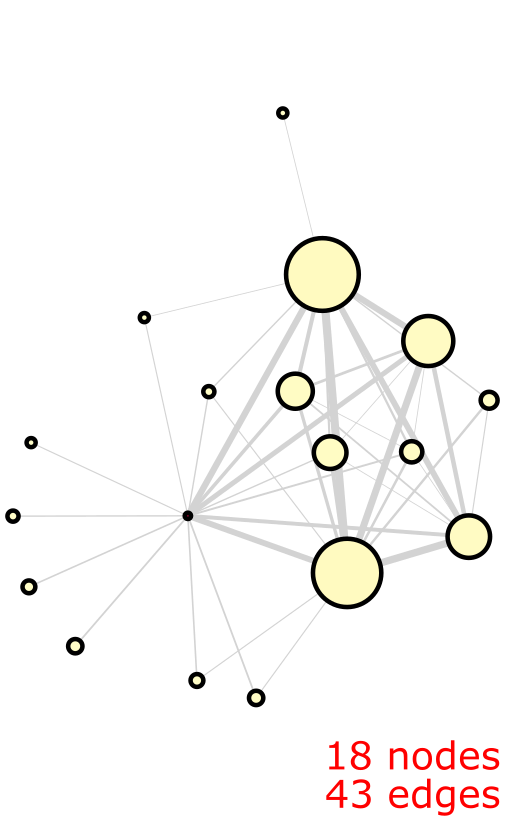}}  {\tiny $n\!=\!2, \epsilon\!=\!0$}
    \end{minipage} \hfill
    \begin{minipage}[t]{0.234\linewidth} \centering        
        \fbox{\includegraphics[width=0.95\linewidth]{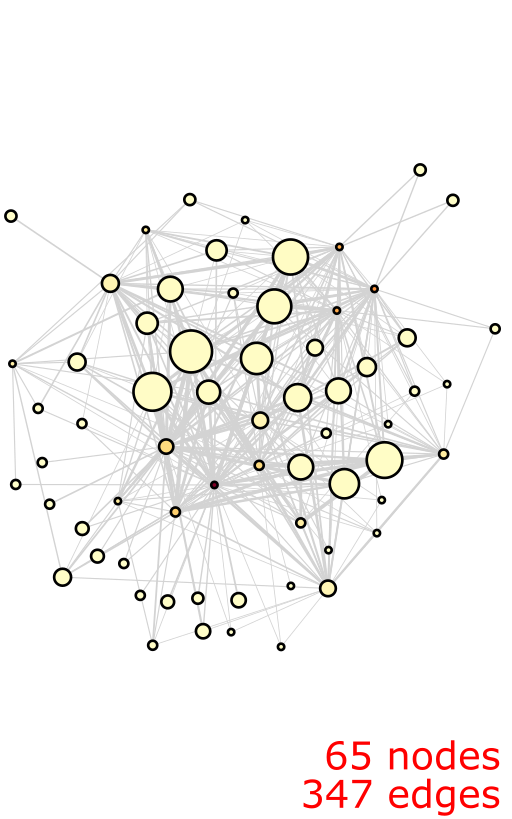}} {\tiny $n\!=\!20, \epsilon\!=\!0$}
    \end{minipage}
    \begin{minipage}[t]{0.234\linewidth} \centering        
        \fbox{\includegraphics[width=0.95\linewidth]{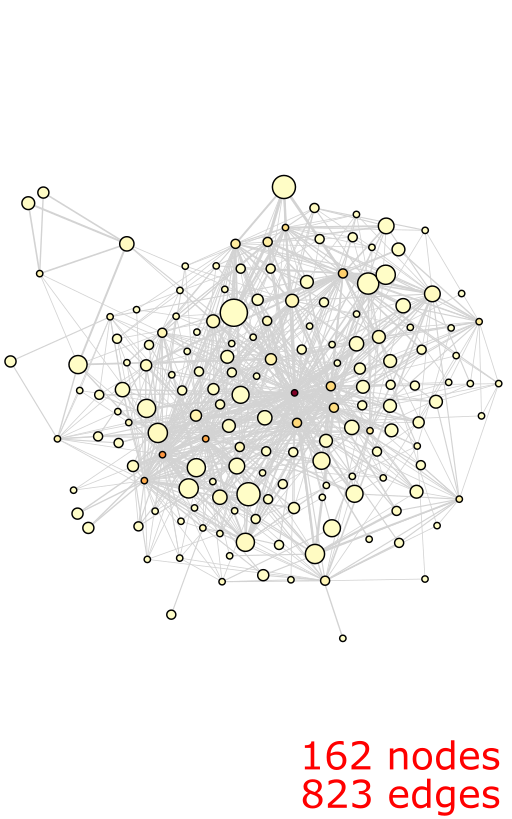}} {\tiny $n\!=\!40, \epsilon\!=\!0$}
    \end{minipage}        
    \begin{minipage}[t]{0.234\linewidth} \centering        
        \fbox{\includegraphics[width=0.95\linewidth]{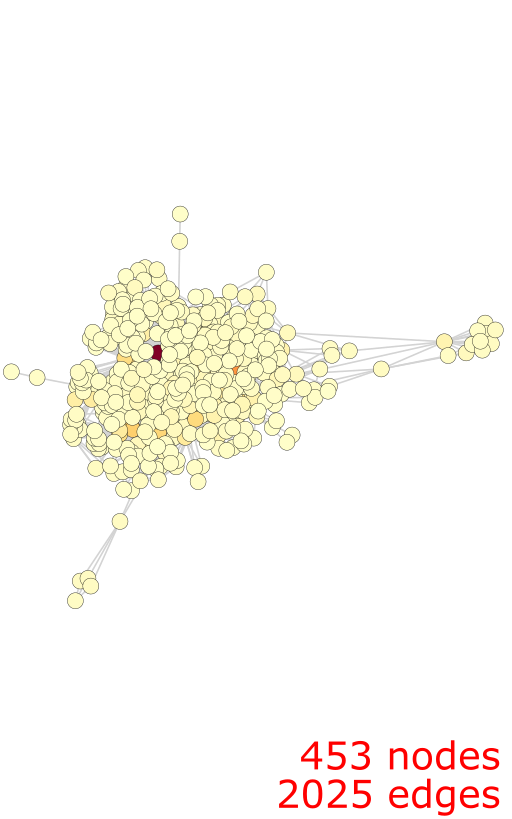}} {\tiny Input Graph}
    \end{minipage}
    }

    \subfloat[\textsc{hic 5k net 7}: density (equalized), connected components]{
    \begin{minipage}[t]{0.234\linewidth} \centering
        \fbox{\includegraphics[width=0.95\linewidth]{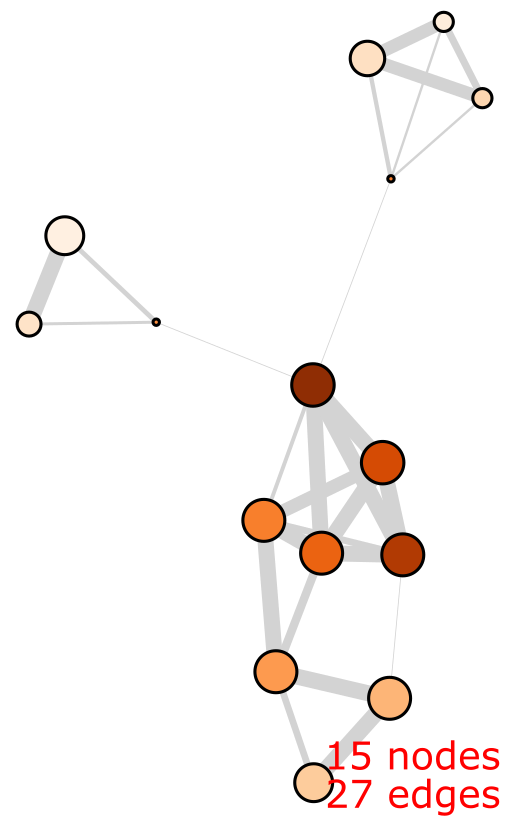}}  {\tiny $n\!=\!10, \epsilon\!=\!0$}
    \end{minipage} \hfill
    \begin{minipage}[t]{0.234\linewidth} \centering
        \fbox{\includegraphics[width=0.95\linewidth]{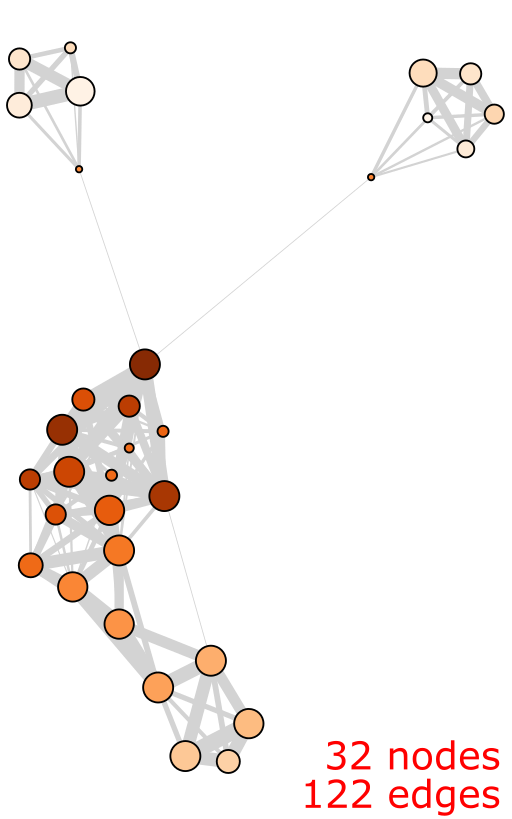}}  {\tiny $n\!=\!20, \epsilon\!=\!0$}
    \end{minipage} \hfill
    \begin{minipage}[t]{0.234\linewidth} \centering
        \fbox{\includegraphics[width=0.95\linewidth]{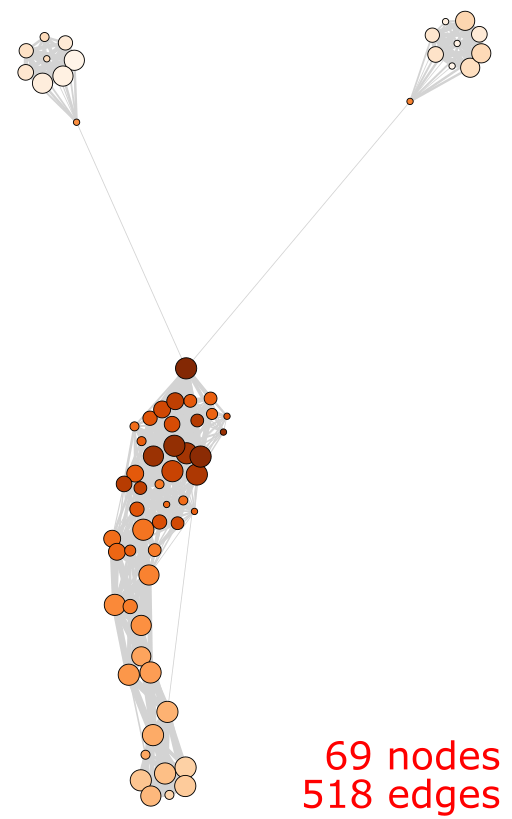}} {\tiny $n\!=\!40, \epsilon\!=\!0$}
    \end{minipage} \hfill
    \begin{minipage}[t]{0.234\linewidth} \centering
        \fbox{\includegraphics[width=0.95\linewidth]{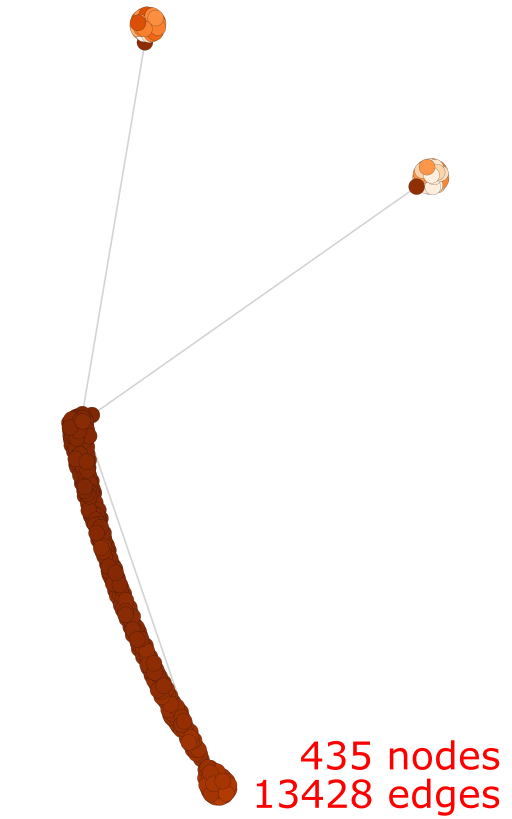}} {\tiny Input Graph}
    \end{minipage}
    }

    \vspace{-5pt}
    \caption{A series of small input graphs demonstrating how \mog retains the prominent homological structure of the input graph as the number of cover elements, $n$, is varied.}
    \label{fig:small_graphs1}
\end{figure}

\begin{figure*}[!t]
    \centering

    \subfloat[\textsc{collaboration network}: eccentricity, connected components\label{fig:small_graphs2:collab}]{
    \begin{minipage}[t]{0.117\linewidth} \centering
        \fbox{\includegraphics[width=0.95\linewidth]{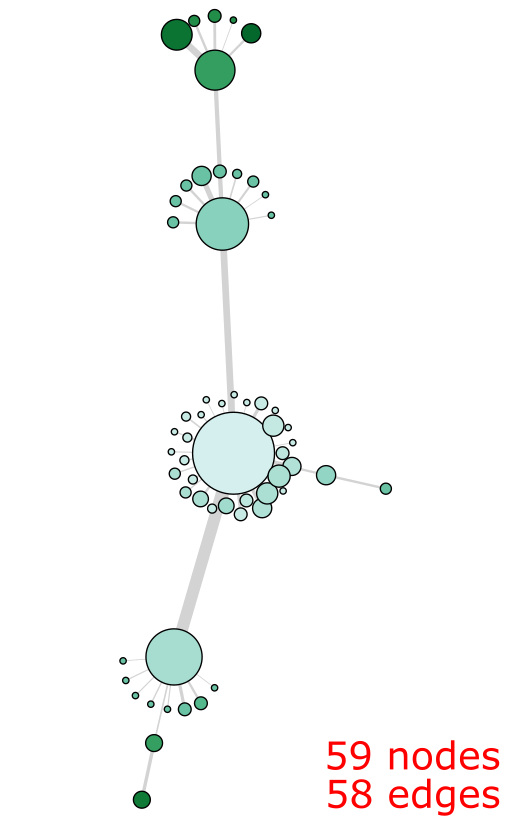}}  {\tiny $n\!=\!4, \epsilon\!=\!0$}
    \end{minipage} \hfill
    \begin{minipage}[t]{0.117\linewidth} \centering    
        \fbox{\includegraphics[width=0.95\linewidth]{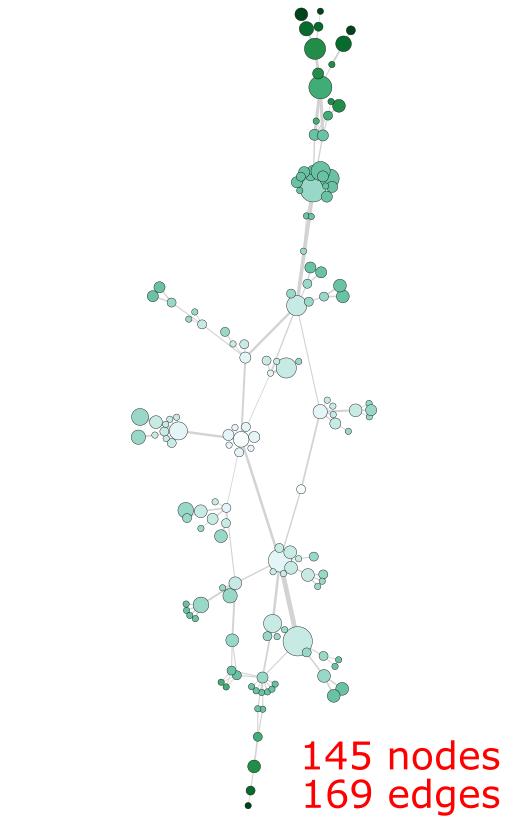}} {\tiny $n\!=\!10, \epsilon\!=\!0$}
    \end{minipage} \hfill
    \begin{minipage}[t]{0.117\linewidth} \centering        
        \fbox{\includegraphics[width=0.95\linewidth]{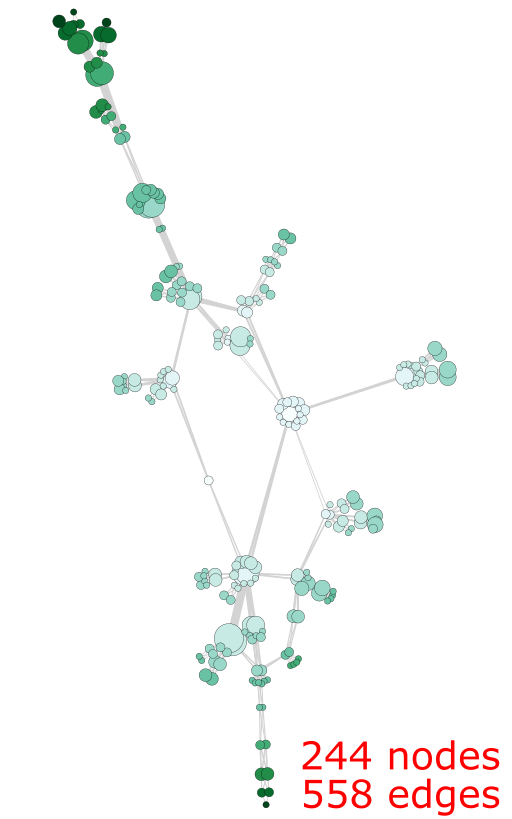}} {\tiny $n\!=\!40, \epsilon\!=\!0$}
    \end{minipage} \hfill
    \begin{minipage}[t]{0.117\linewidth} \centering        
        \fbox{\includegraphics[width=0.95\linewidth]{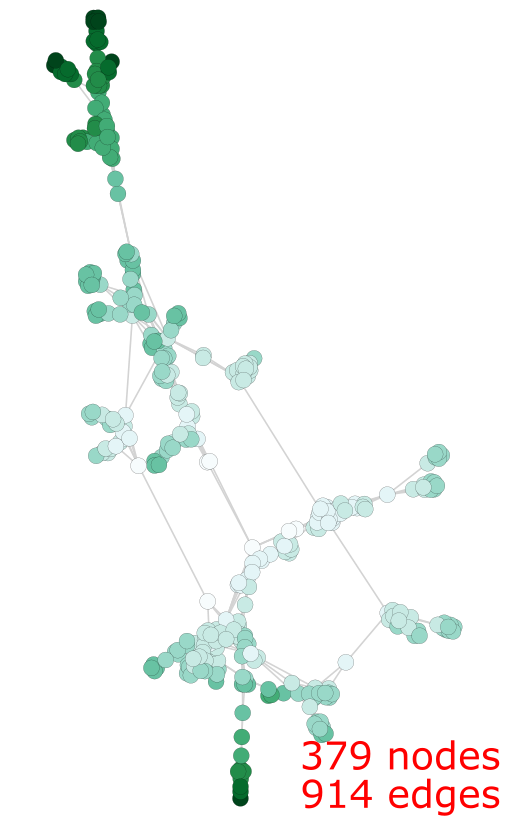}} {\tiny Input Graph}
    \end{minipage}
    }
    \hfill
    \subfloat[\textsc{bio-diseasome}: Fiedler vector, modularity clustering\label{fig:small_graphs2:diseas}]{
    \begin{minipage}[t]{0.117\linewidth} \centering
        \fbox{\includegraphics[width=0.95\linewidth]{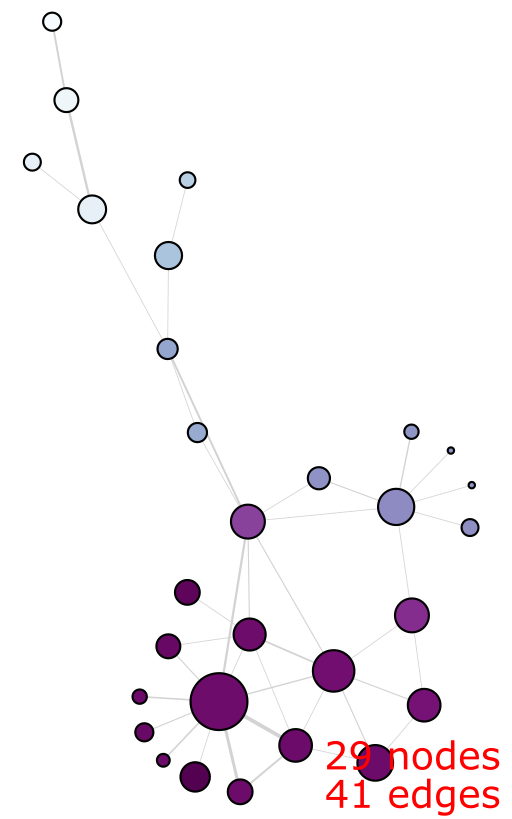}}  {\tiny $\epsilon\!=\!0, n\!=\!2$}
    \end{minipage} \hfill
    \begin{minipage}[t]{0.117\linewidth} \centering        
        \fbox{\includegraphics[width=0.95\linewidth]{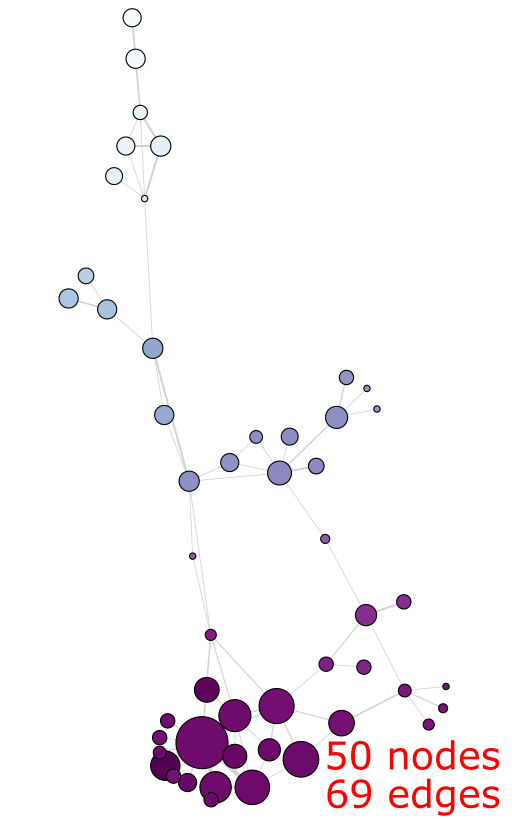}}  {\tiny $\epsilon\!=\!0, n\!=\!8$}
    \end{minipage} \hfill
    \begin{minipage}[t]{0.117\linewidth} \centering
        \fbox{\includegraphics[width=0.95\linewidth]{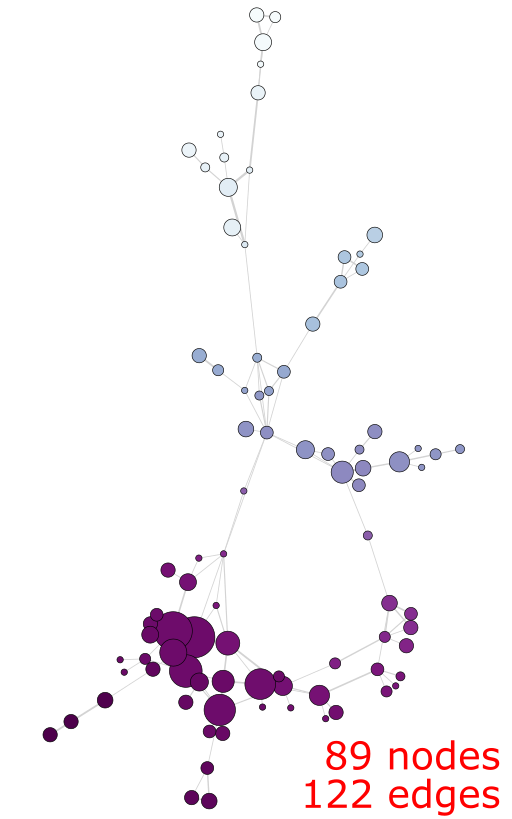}} {\tiny $\epsilon\!=\!0, n\!=\!40$}
    \end{minipage} \hfill
    \begin{minipage}[t]{0.117\linewidth} \centering
        \fbox{\includegraphics[width=0.95\linewidth]{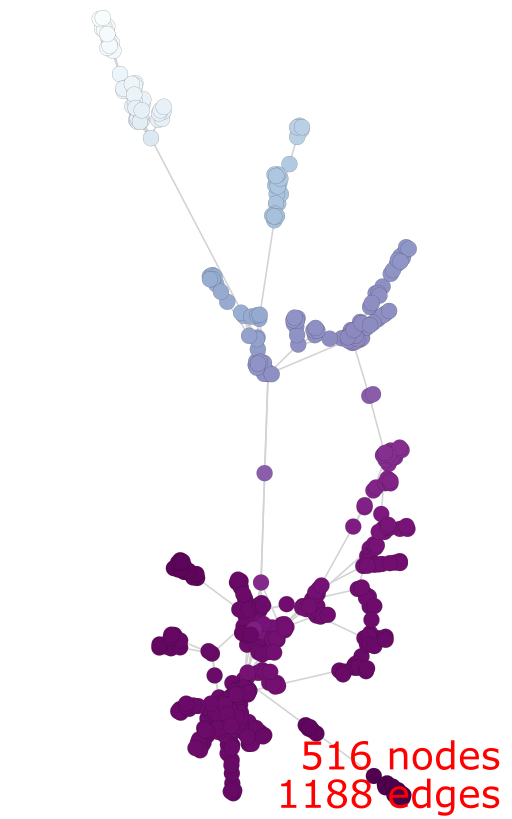}} {\tiny Input Graph}
    \end{minipage}
        
    }

    \subfloat[\textsc{bn-mouse-visual-cortex-2}: Fiedler vec.~(equalized), async.\ label prop.\label{fig:small_graphs2:mouse}]{
    \begin{minipage}[t]{0.117\linewidth} \centering
        \fbox{\includegraphics[width=0.95\linewidth]{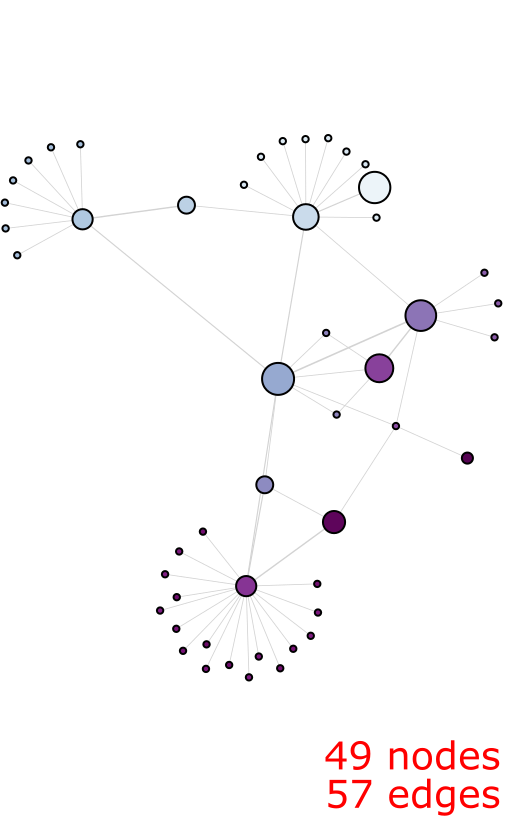}} {\tiny $\epsilon\!=\!0, n\!=\!6$}
    \end{minipage} \hfill
    \begin{minipage}[t]{0.117\linewidth} \centering        
        \fbox{\includegraphics[width=0.95\linewidth]{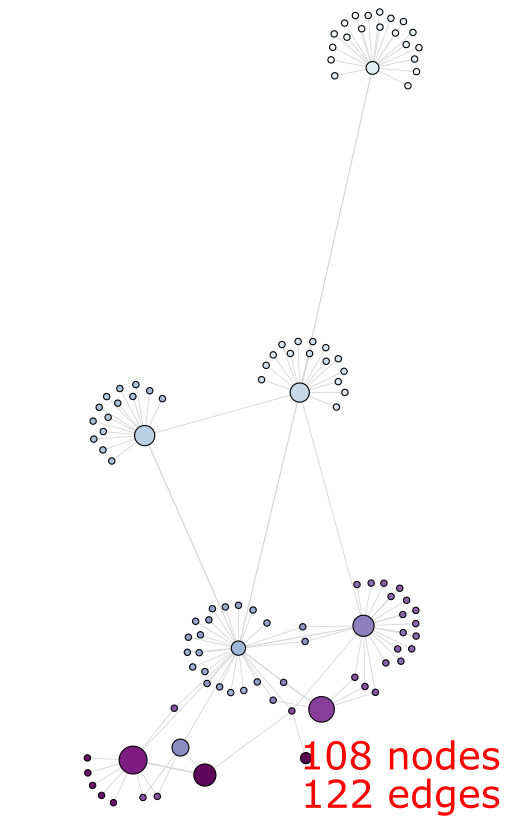}} {\tiny $\epsilon\!=\!0, n\!=\!10$}
    \end{minipage} \hfill
    \begin{minipage}[t]{0.117\linewidth} \centering        
        \fbox{\includegraphics[width=0.95\linewidth]{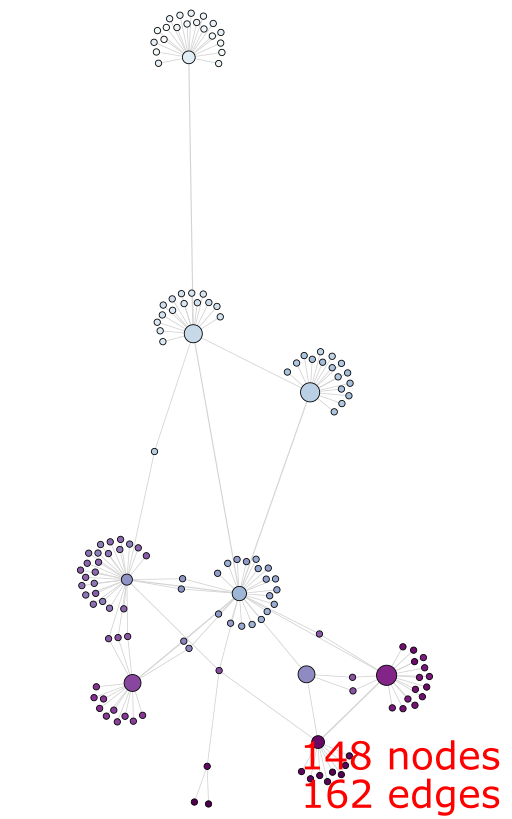}} {\tiny $\epsilon\!=\!0, n\!=\!20$}
    \end{minipage} \hfill
    \begin{minipage}[t]{0.117\linewidth} \centering        
        \fbox{\includegraphics[width=0.95\linewidth]{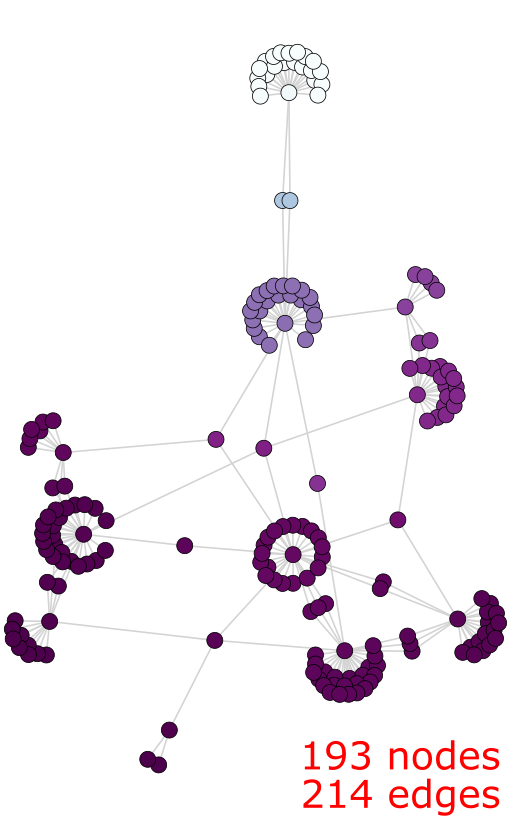}} {\tiny Input Graph}
    \end{minipage}

    }
    \hfill
    \subfloat[\textsc{enron-email}: AGD, asynchronous label propagation\label{fig:small_graphs2:enron}]{
    \begin{minipage}[t]{0.117\linewidth} \centering
        \fbox{\includegraphics[width=0.95\linewidth]{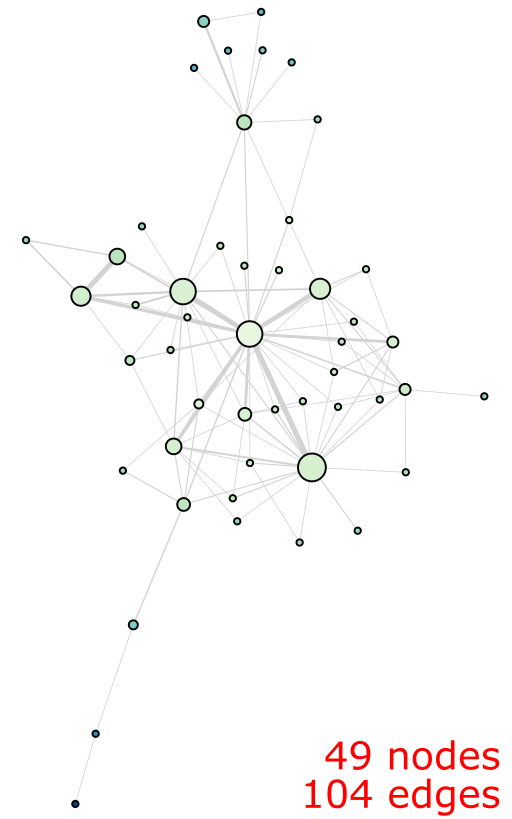}} {\tiny $\epsilon\!=\!0, n\!=\!8$}
    \end{minipage} \hfill
    \begin{minipage}[t]{0.117\linewidth} \centering        
        \fbox{\includegraphics[width=0.95\linewidth]{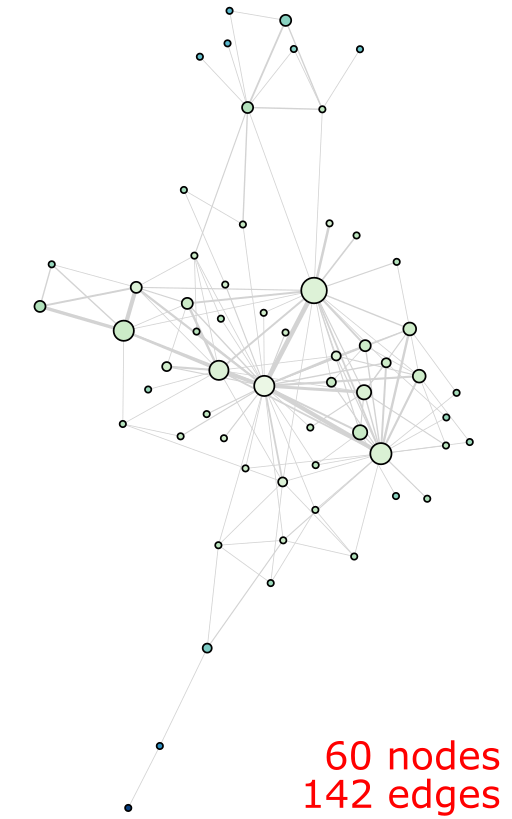}} {\tiny $\epsilon\!=\!0, n\!=\!10$}
    \end{minipage} \hfill
    \begin{minipage}[t]{0.117\linewidth} \centering        
        \fbox{\includegraphics[width=0.95\linewidth]{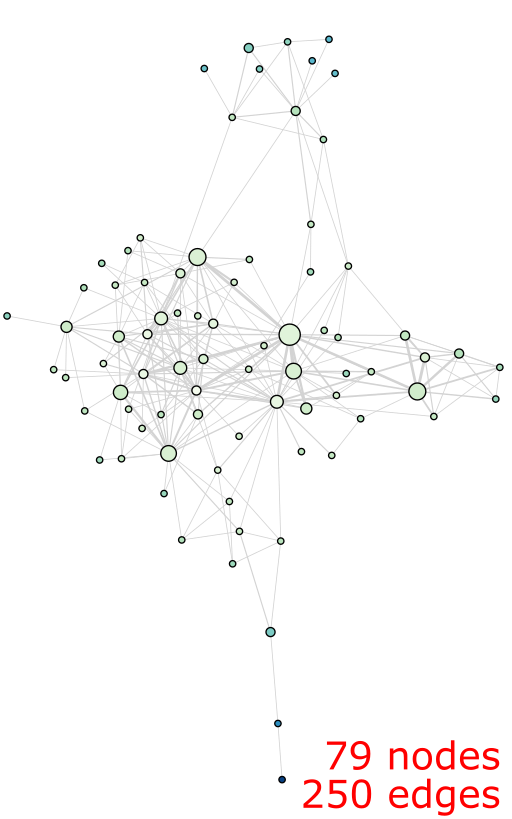}} {\tiny $\epsilon\!=\!0, n\!=\!20$}
    \end{minipage} \hfill
    \begin{minipage}[t]{0.117\linewidth} \centering        
        \fbox{\includegraphics[width=0.95\linewidth]{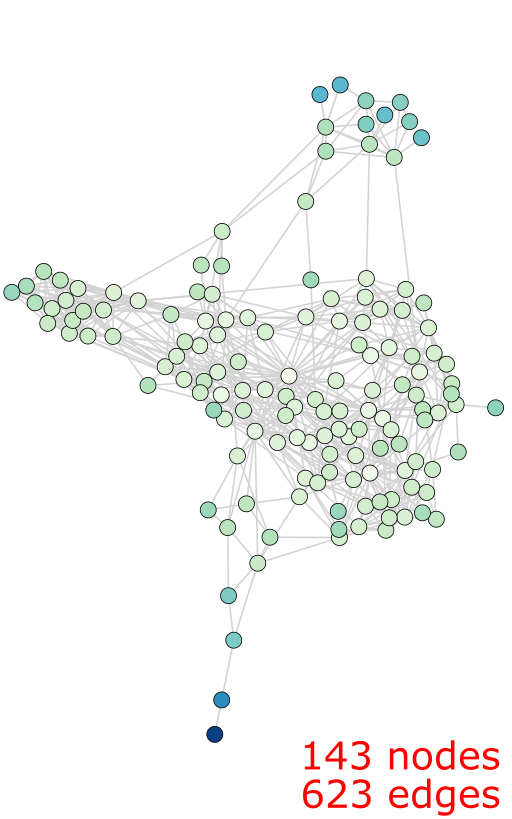}} {\tiny Input Graph}
    \end{minipage}

    }

    \vspace{-5pt}
    \caption{Continuation of \cref{fig:small_graphs1}.}
    \label{fig:small_graphs2}
    \vspace{-10pt}
    
\end{figure*}

\subsection{Examples}
\label{sec:eval:examp}

A major challenge in evaluation is that we were unable to identify appropriate quantitative metrics for comparing skeltonizations in our context~\cite{wills2020metrics}. 
As such, we relied on qualitative evaluations in this section and \cref{sec:results:community} and a human-subject evaluation in \cref{sec:eval:human}.

\subsubsection{Multi-scale Homology Preservation}

To demonstrate the homology preservation properties of our approach, we show a variety of examples. The examples were selected such that the homology of the input graph was clear, and a corresponding \moag was selected that reflected that homology. 
\cref{fig:teaser}, \cref{fig:small_graphs1}, and \cref{fig:small_graphs2} show two, two, and four examples, respectively. These examples were chosen to show diversity among topological lenses and clustering algorithms.
In all examples, as the number of cover elements $n$ increases, the \moag becomes more similar to the input graph.
As $n$ decreases, the \moag contains fewer nodes while still retaining important homological structures. These structures include the preservation of tunnels, which are particularly visible in both graphs of \autoref{fig:teaser} and can also be observed in \autoref{fig:small_graphs2:collab}, \ref{fig:small_graphs2:diseas}, \ref{fig:small_graphs2:mouse}, \ref{fig:small_graphs2:enron}, and others.

\begin{figure*}[!t]
    \centering
    
    \vspace{10pt}
    \subfloat[\textsc{amazon0302} ($|N|=262,111$, $|E|=899,792$), normalized Fiedler vector, connected components]{%
 \begin{minipage}[t]{0.155\linewidth} \centering
    \fbox{\includegraphics[width=0.95\linewidth]{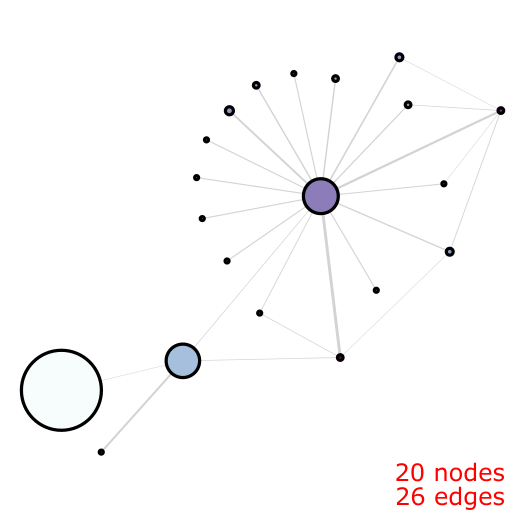}}
 {\tiny $\epsilon\!=\!0, n\!=\!4$}
    \end{minipage} 
    \hspace{2pt}
    \begin{minipage}[t]{0.155\linewidth} \centering
    \fbox{\includegraphics[width=0.95\linewidth]{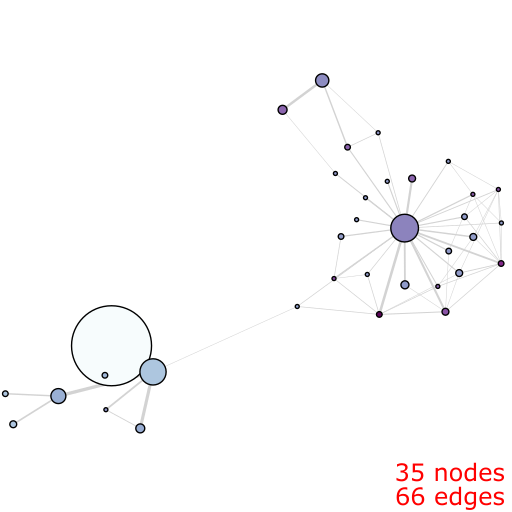}}
     {\tiny $\epsilon\!=\!0, n\!=\!8$}
    \end{minipage} \hspace{2pt}
    \begin{minipage}[t]{0.155\linewidth} \centering
    \fbox{\includegraphics[width=0.95\linewidth]{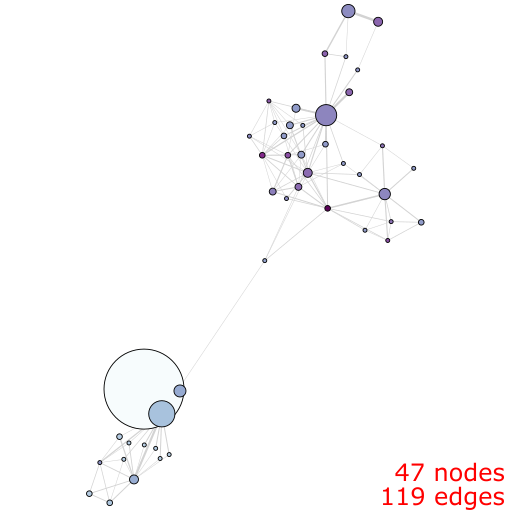}}
     {\tiny $\epsilon\!=\!0, n\!=\!10$}
    \end{minipage} 
    \hspace{2pt}
    \begin{minipage}[t]{0.155\linewidth} \centering
    \fbox{\includegraphics[width=0.95\linewidth]{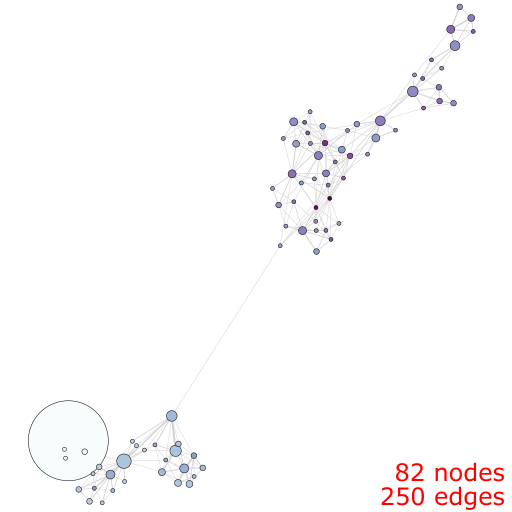}}
     {\tiny $\epsilon\!=\!0, n\!=\!20$}
    \end{minipage} 
    \hspace{2pt}
    \begin{minipage}[t]{0.155\linewidth} \centering
    \fbox{\includegraphics[width=0.95\linewidth]{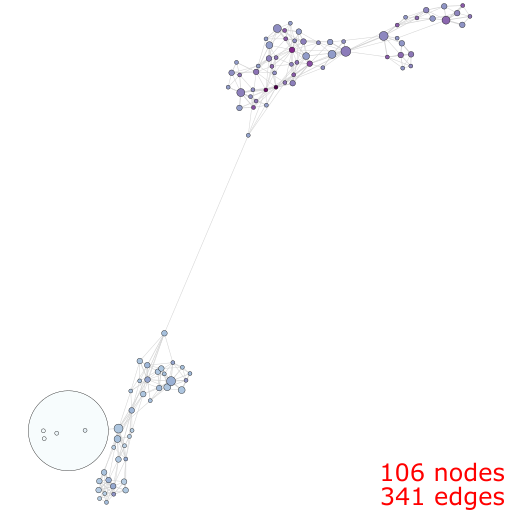}}
     {\tiny $\epsilon\!=\!0, n\!=\!30$}
    \end{minipage} 
    \hspace{2pt}
    \begin{minipage}[t]{0.155\linewidth} \centering
    \fbox{\includegraphics[width=0.95\linewidth]{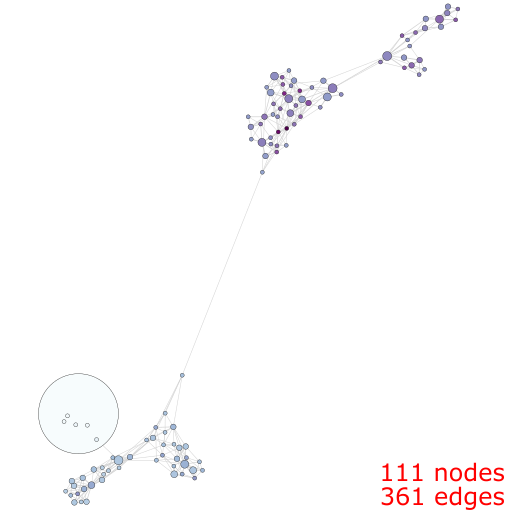}}
     {\tiny $\epsilon\!=\!0, n\!=\!40$}
    \end{minipage} 
    }
    
    \subfloat[\textsc{com-amazon} ($|N|=334,863$, $|E|=925,872$), PageRank, connected components]{%
    \begin{minipage}[t]{0.155\linewidth} \centering    \fbox{\includegraphics[width=0.95\linewidth]{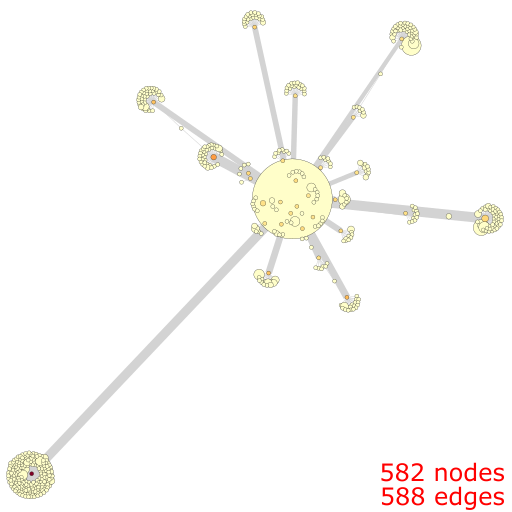}}
 {\tiny $\epsilon\!=\!0, n\!=\!6$}
    \end{minipage} 
    \hspace{2pt}
    \begin{minipage}[t]{0.155\linewidth} \centering    \fbox{\includegraphics[width=0.95\linewidth]{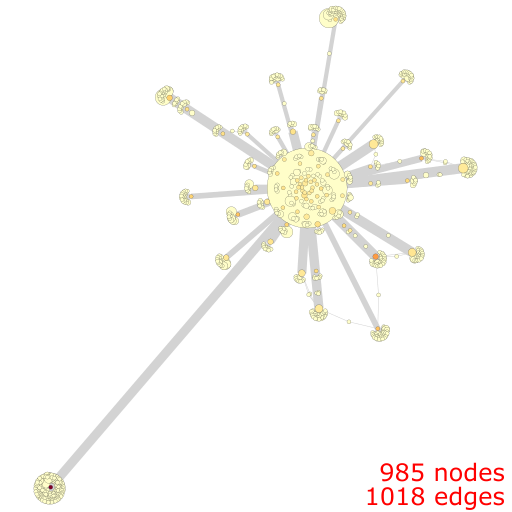}}
 {\tiny $\epsilon\!=\!0, n\!=\!8$}
    \end{minipage} 
    \hspace{2pt}
    \begin{minipage}[t]{0.155\linewidth} \centering    \fbox{\includegraphics[width=0.95\linewidth]{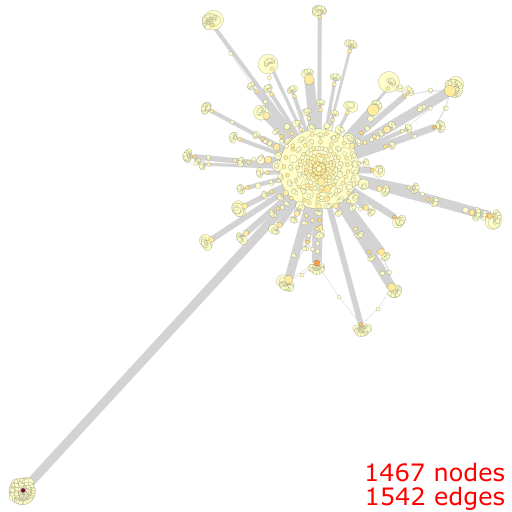}}
 {\tiny $\epsilon\!=\!0, n\!=\!10$}
    \end{minipage} \hspace{2pt}
    \begin{minipage}[t]{0.155\linewidth} \centering    \fbox{\includegraphics[width=0.95\linewidth]{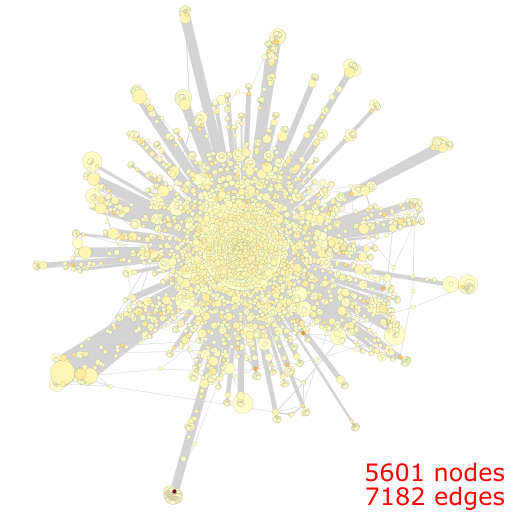}}
 {\tiny $\epsilon\!=\!0, n\!=\!20$}
    \end{minipage} \hspace{2pt}
    \begin{minipage}[t]{0.155\linewidth} \centering    \fbox{\includegraphics[width=0.95\linewidth]{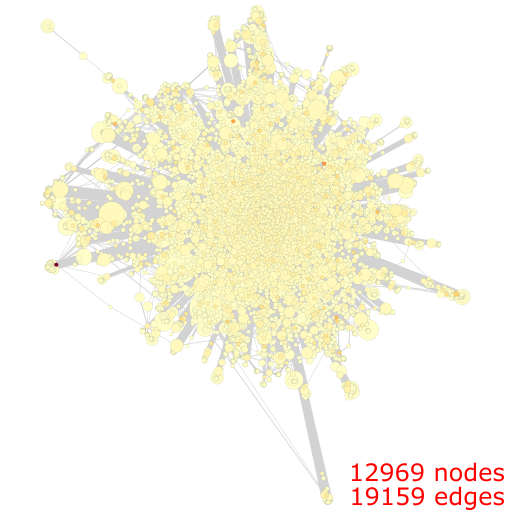}}
 {\tiny $\epsilon\!=\!0, n\!=\!30$}
    \end{minipage} 
    \hspace{2pt}
    \begin{minipage}[t]{0.155\linewidth} \centering    \fbox{\includegraphics[width=0.95\linewidth]{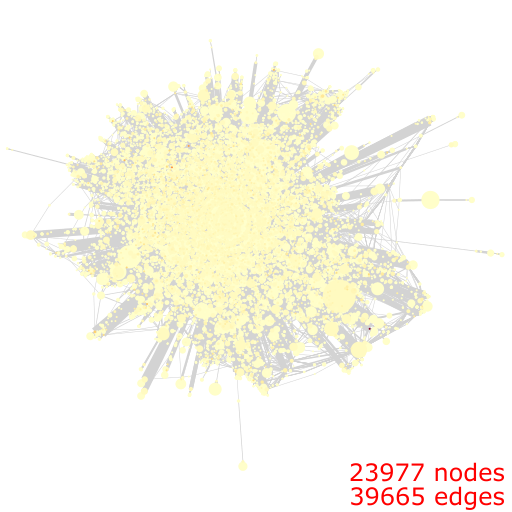}}
 {\tiny $\epsilon\!=\!0, n\!=\!40$}
    \end{minipage}    }

    \subfloat[\textsc{com-youtube} ($|N|=1,134,890$, $|E|=2,987,624$), normalized Fiedler vector, connected components]{%
    \begin{minipage}[t]{0.155\linewidth} \centering\fbox{\includegraphics[width=0.95\linewidth]{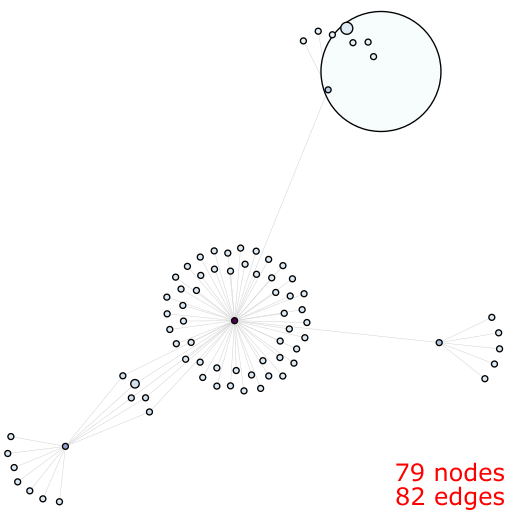}}
 {\tiny $\epsilon\!=\!0, n\!=\!4$}
    \end{minipage} 
    \hspace{2pt}
    \begin{minipage}[t]{0.155\linewidth} \centering    \fbox{\includegraphics[width=0.95\linewidth]{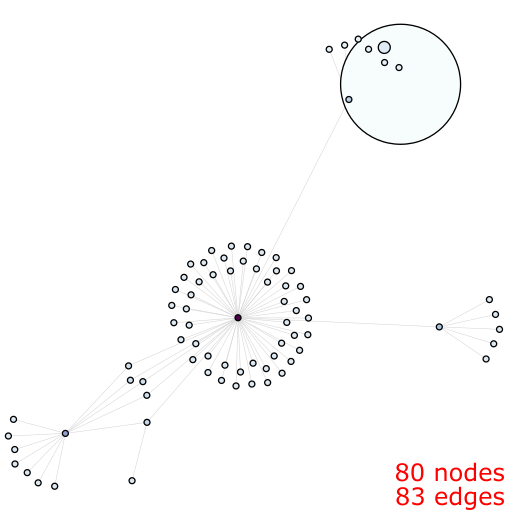}}
 {\tiny $\epsilon\!=\!0, n\!=\!6$}
    \end{minipage} 
    \hspace{2pt}
    \begin{minipage}[t]{0.155\linewidth} \centering    \fbox{\includegraphics[width=0.95\linewidth]{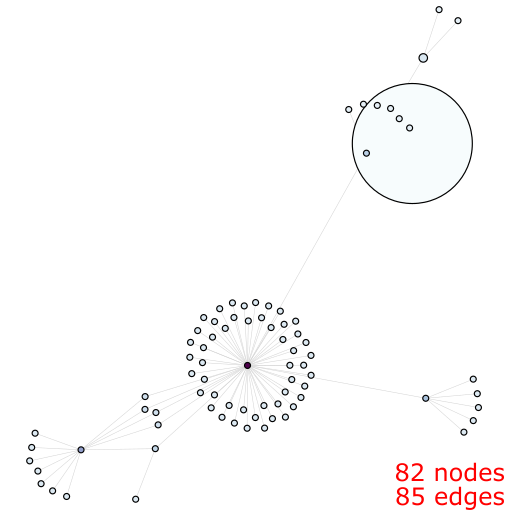}}
 {\tiny $\epsilon\!=\!0, n\!=\!10$}
    \end{minipage} 
    \hspace{2pt}
    \begin{minipage}[t]{0.155\linewidth} \centering    \fbox{\includegraphics[width=0.95\linewidth]{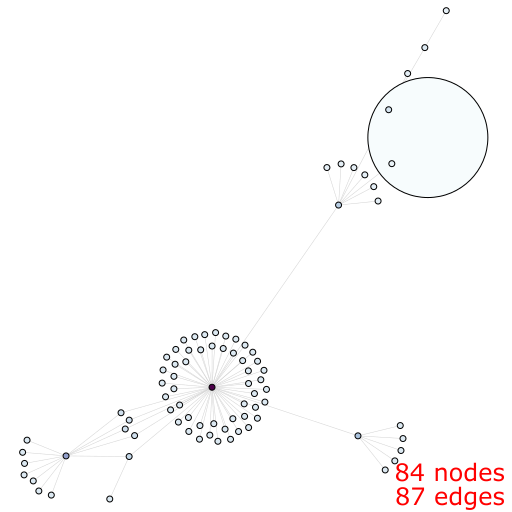}}
 {\tiny $\epsilon\!=\!0, n\!=\!20$}
    \end{minipage} 
    \hspace{2pt}
    \begin{minipage}[t]{0.155\linewidth} \centering    \fbox{\includegraphics[width=0.95\linewidth]{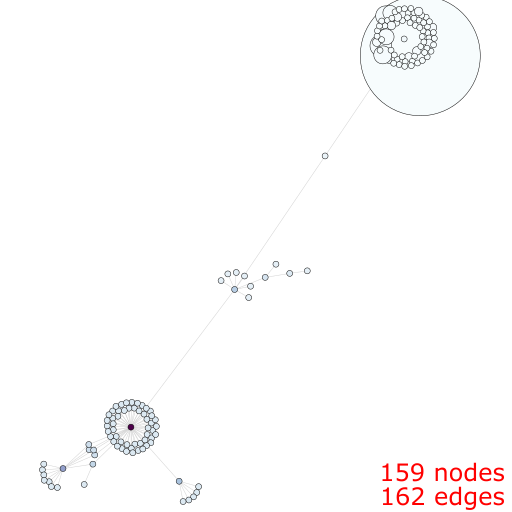}}
 {\tiny $\epsilon\!=\!0, n\!=\!40$}
    \end{minipage}
    \hspace{2pt}
    \begin{minipage}[t]{0.155\linewidth} \centering    \fbox{\includegraphics[width=0.95\linewidth]{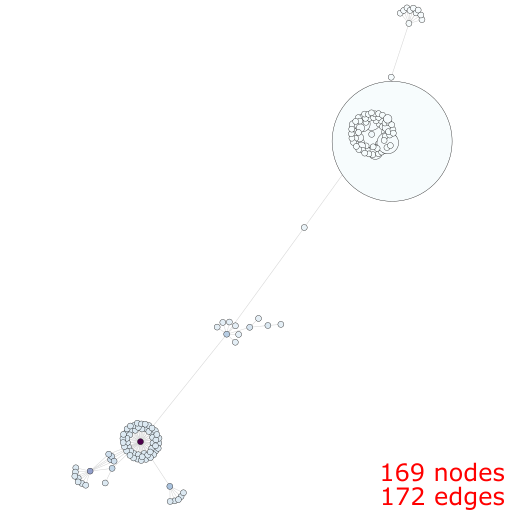}}
 {\tiny $\epsilon\!=\!0, n\!=\!100$}
    \end{minipage}}

    \vspace{-5pt}
    \caption{Examples of large input graphs  demonstrating how a \mog can reveal certain prominent topological structure of graphs so large that meaningful node-link diagrams are difficult to draw.}
    \label{fig:large_graphs}
\end{figure*}

\begin{figure*}[!t]
    \centering

    \begin{minipage}[t]{0.008\linewidth}
        \rotatebox{90}{\tiny \hspace{15pt} \mog \hspace{25pt} Mod.\ (right) / Louvain (left)}
    \end{minipage}
    \subfloat[\textsc{bio-diseasome}\label{fig:community:bio-diseasome}]{
    \begin{minipage}[t]{0.19\linewidth}
        \centering
        \fbox{\includegraphics[width=0.4\linewidth]{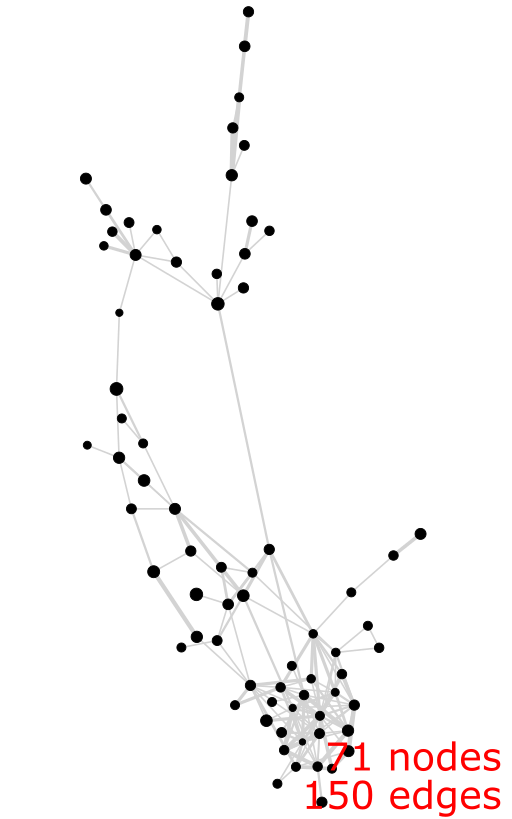}}
        \fbox{\includegraphics[width=0.4\linewidth]{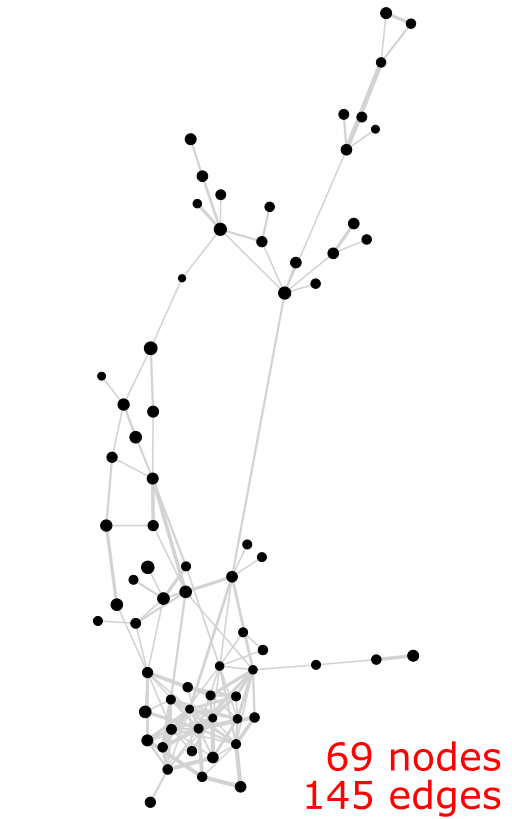}}

        \fbox{\includegraphics[width=0.4\linewidth]{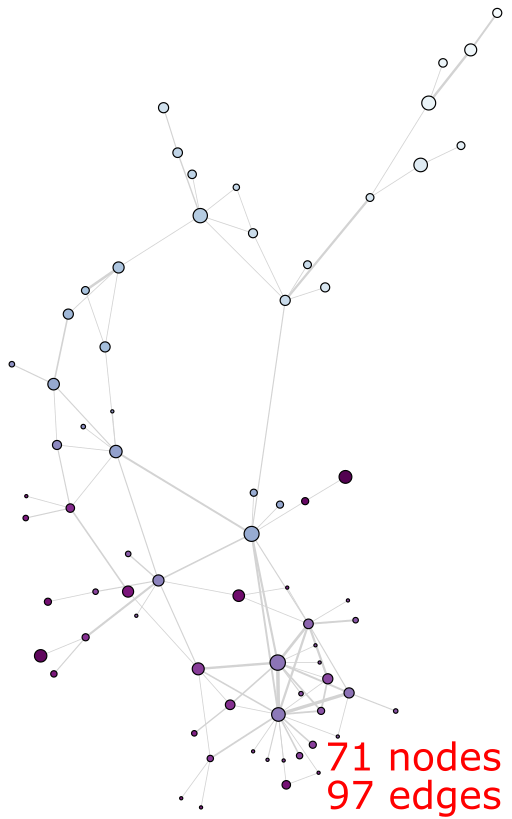}}
        \fbox{\includegraphics[width=0.4\linewidth]{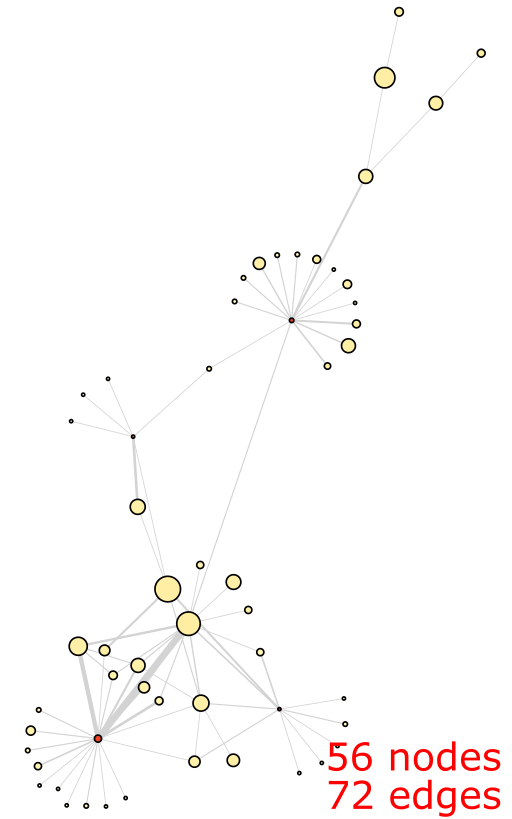}}
    \end{minipage}}
    \subfloat[\textsc{bn-mouse-visual-cortex-2}\label{fig:community:bn-mouse}]{
    \begin{minipage}[t]{0.19\linewidth}
        \centering
        \fbox{\includegraphics[width=0.4\linewidth]{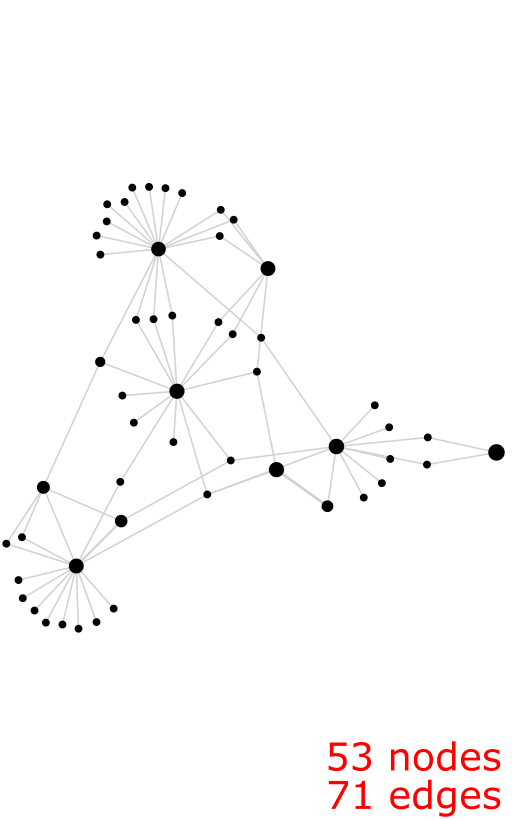}}
        \fbox{\includegraphics[width=0.4\linewidth]{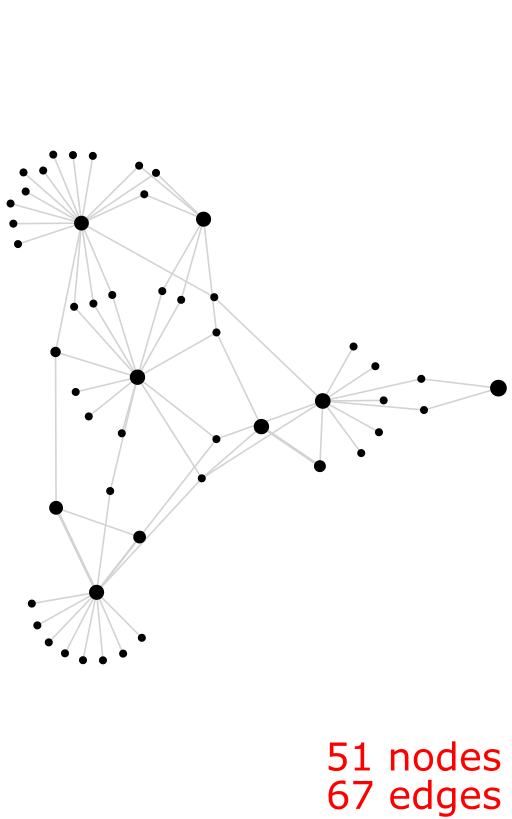}}

        \fbox{\includegraphics[width=0.4\linewidth]{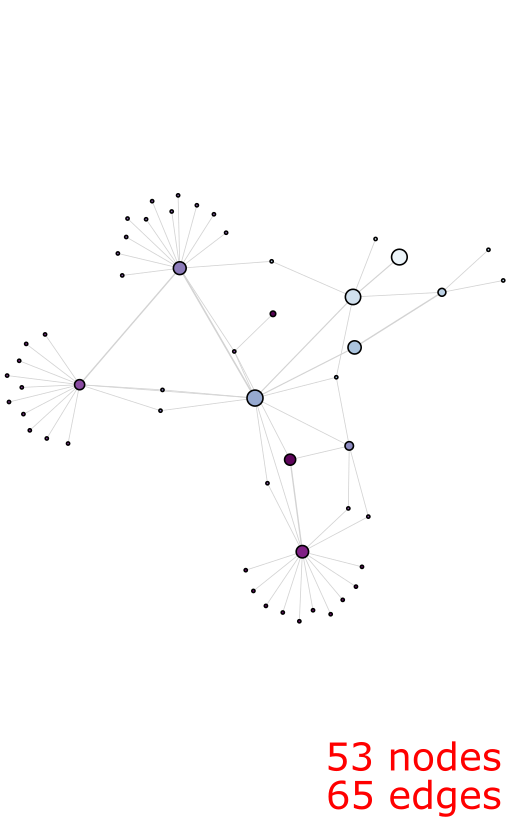}}
        \fbox{\includegraphics[width=0.4\linewidth]{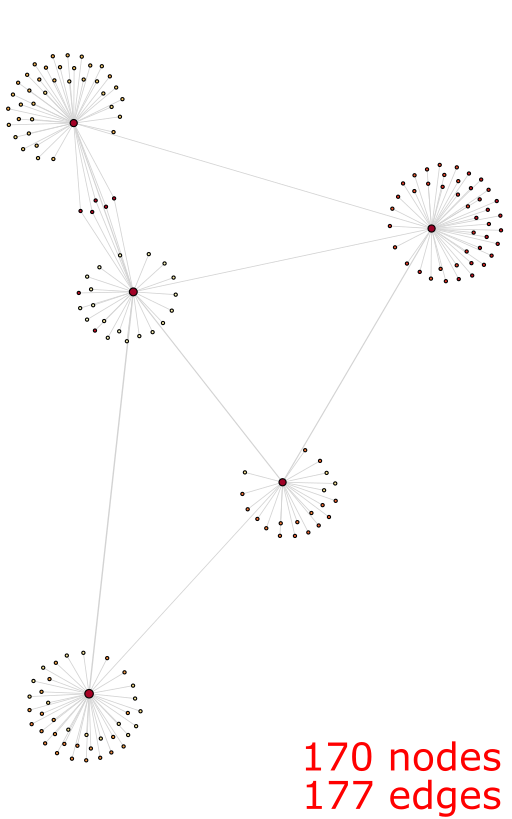}}
    \end{minipage}}
    \subfloat[\textsc{collaboration network}\label{fig:community:collaboration}]{
    \begin{minipage}[t]{0.19\linewidth}
        \centering
        \fbox{\includegraphics[width=0.4\linewidth]{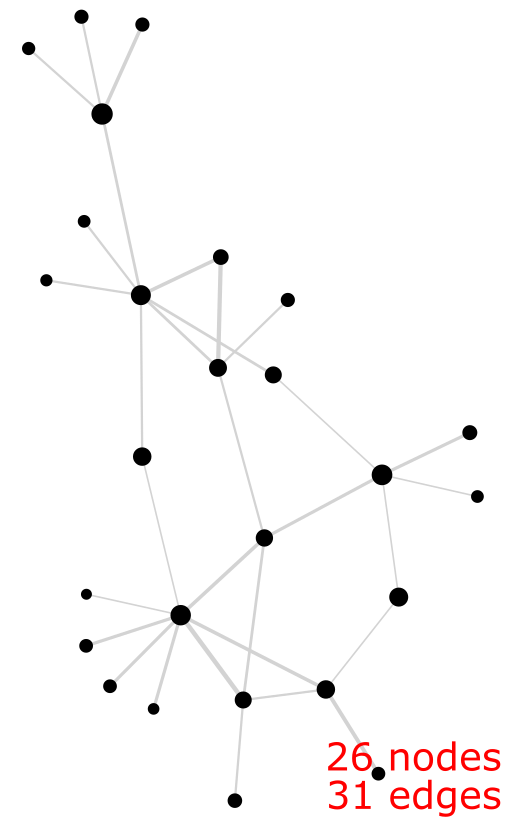}}
        \fbox{\includegraphics[width=0.4\linewidth]{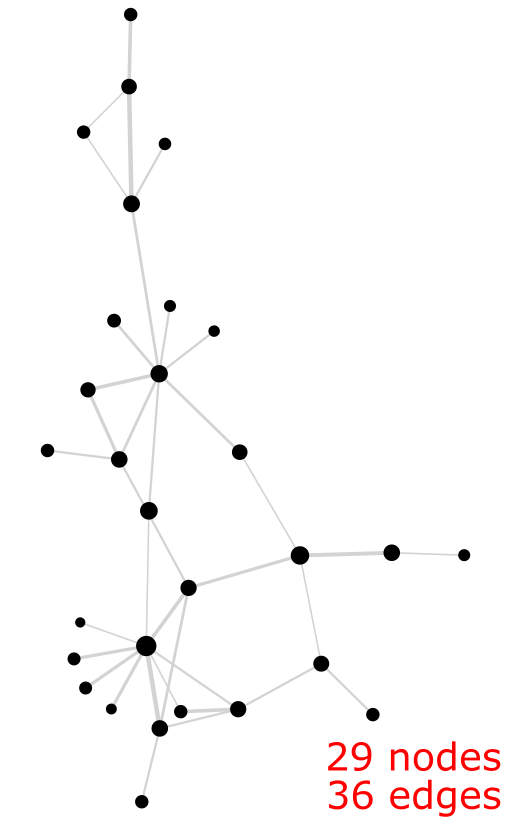}}
        
        \fbox{\includegraphics[width=0.4\linewidth]{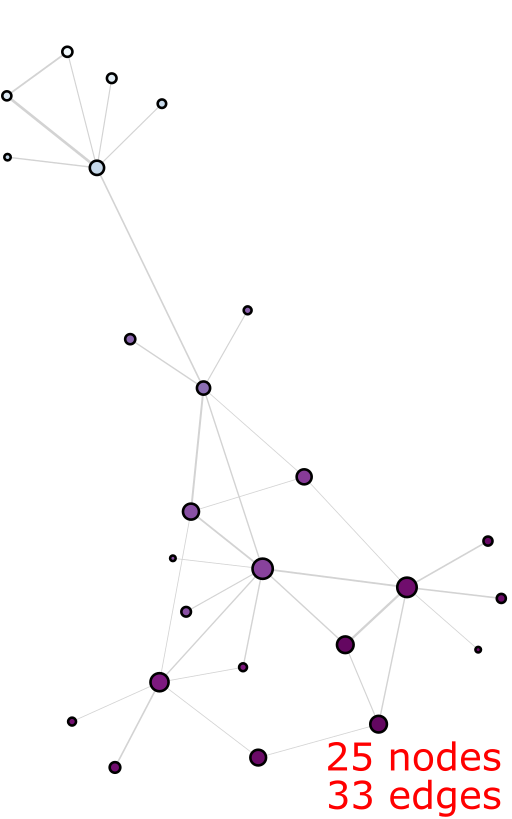}}
        \fbox{\includegraphics[width=0.4\linewidth]{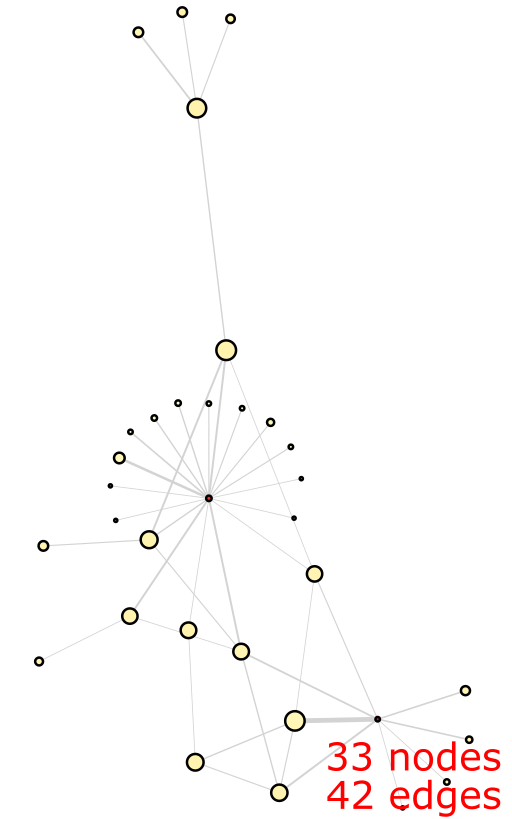}}
    \end{minipage}}
    \subfloat[\textsc{enron-email}\label{fig:community:enron}]{
    \begin{minipage}[t]{0.19\linewidth}
        \centering
        \fbox{\includegraphics[width=0.4\linewidth]{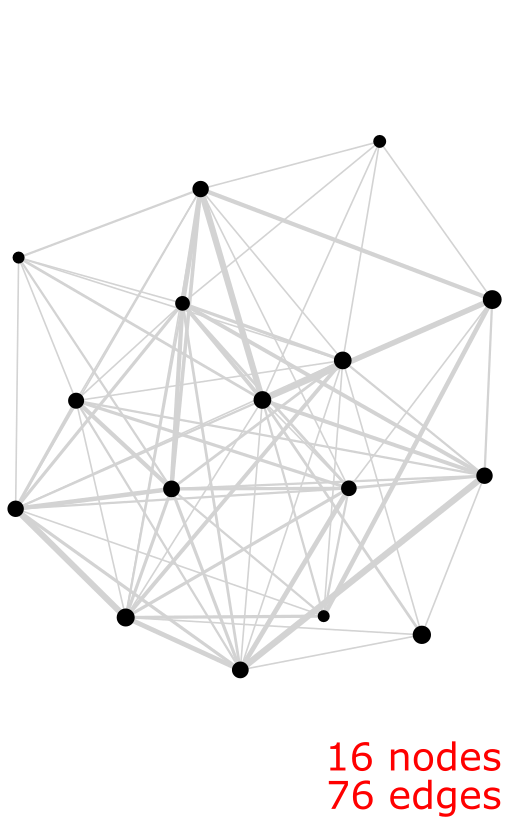}}
        \fbox{\includegraphics[width=0.4\linewidth]{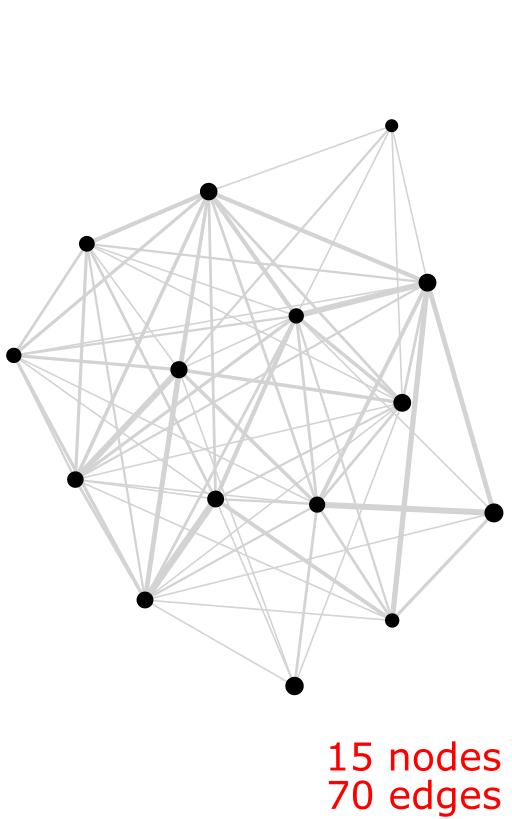}}
        
        \fbox{\includegraphics[width=0.4\linewidth]{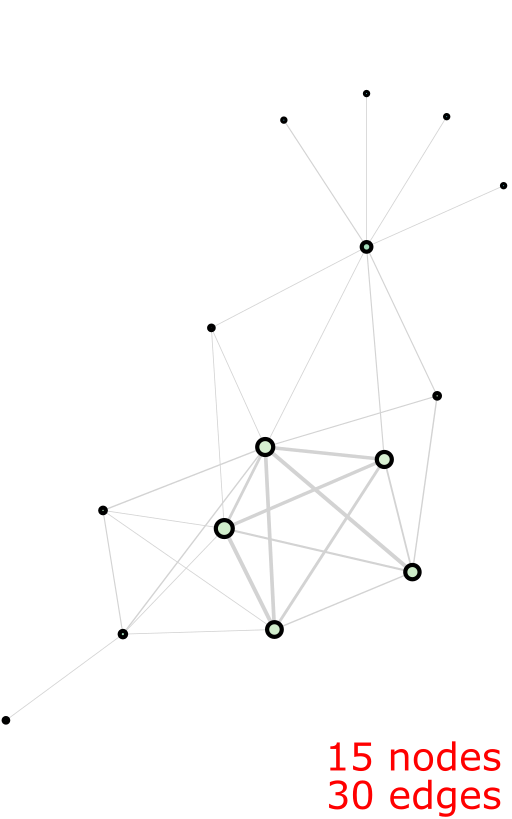}}
        \fbox{\includegraphics[width=0.4\linewidth]{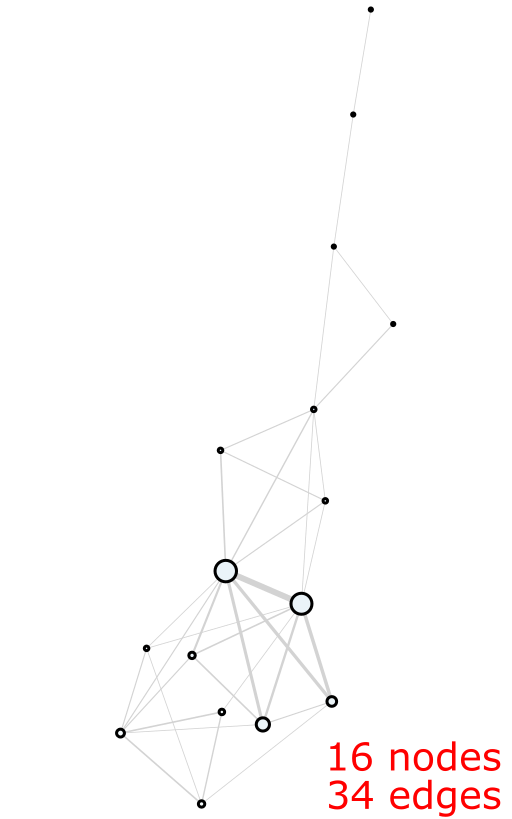}}
    \end{minipage}}
    \subfloat[\textsc{usair97}\label{fig:community:usair}]{
    \begin{minipage}[t]{0.19\linewidth}
        \centering
        \fbox{\includegraphics[width=0.4\linewidth]{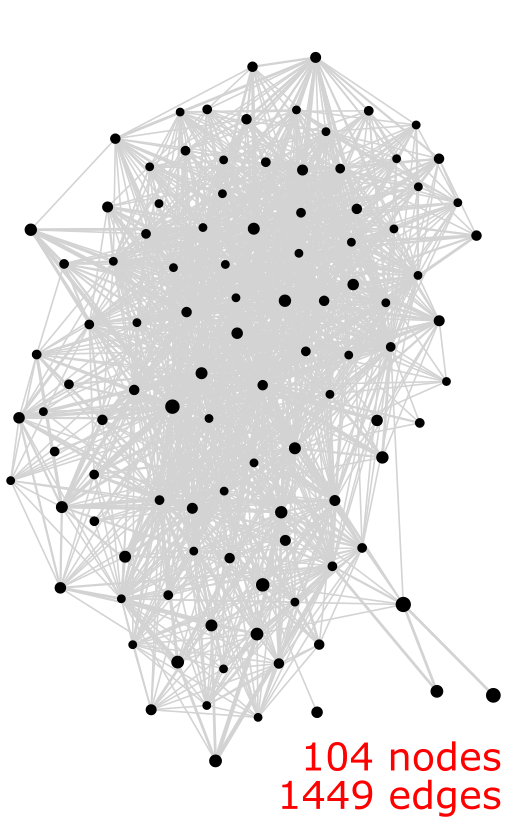}}
        \fbox{\includegraphics[width=0.4\linewidth]{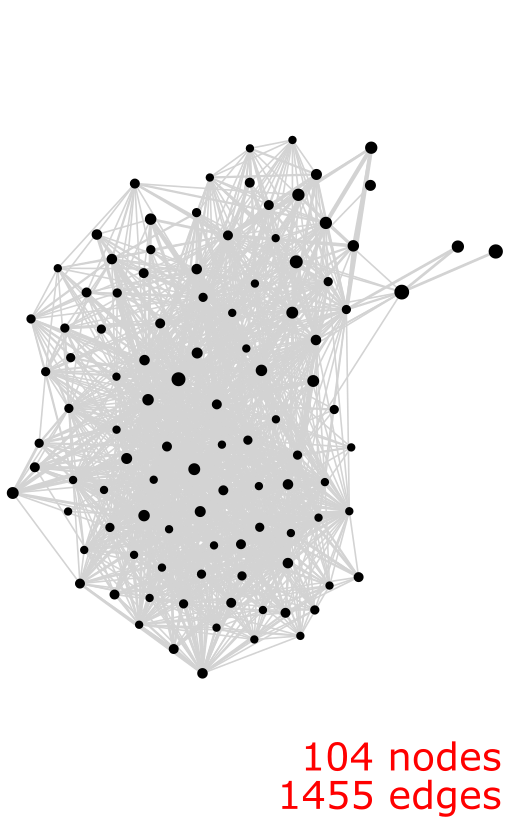}}
        
        \fbox{\includegraphics[width=0.4\linewidth]{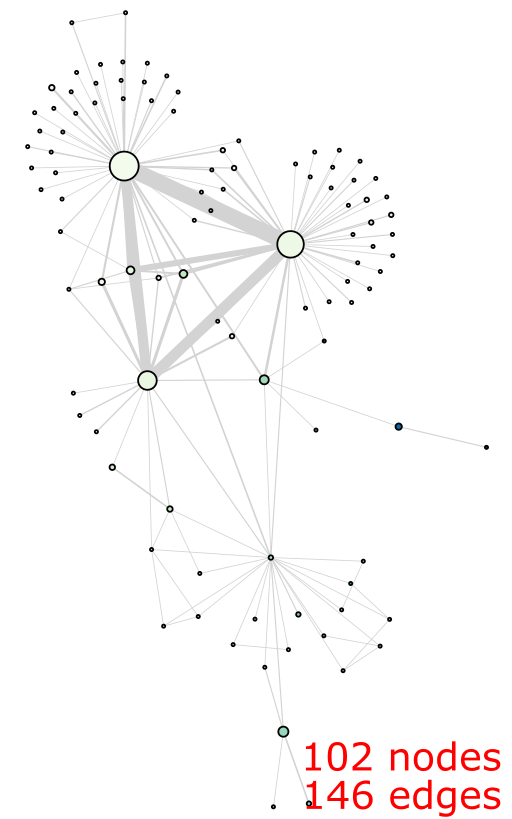}}
        \fbox{\includegraphics[width=0.4\linewidth]{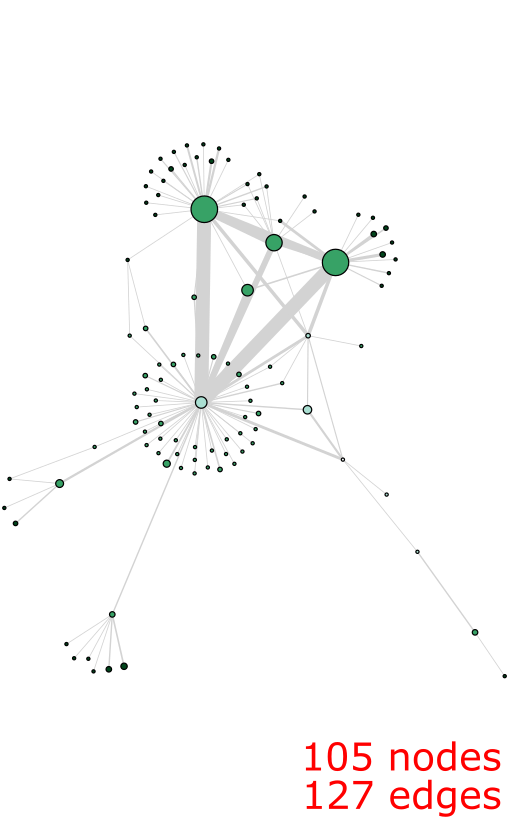}}
    \end{minipage}}
    
    \caption{Community-based skeletonization (top) compared to \mog skeletonization (bottom). Parameters were chosen so that graphs generally had a similar number of nodes. We observe that while community-based techniques sometimes preserve homology, (a-e)~\mog provides multiple perspectives on the graph, and in some cases, (d-e)~community-based approaches do not capture any structure.}
    \label{fig:community}
\end{figure*}

\subsubsection{Large Graphs}

We demonstrate the effectiveness of our approach to summarize large graphs as well. \cref{fig:large_graphs} shows results for three larges graphs, \textsc{amazon0302}, \textsc{com-amazon}, and \textsc{com-youtube}, with different numbers of cover elements $n$. 
As $n$ increases, the topological structure remains visible with more details added. 
In the case of \textsc{com-amazon}, the graph eventually becomes a hairball. 
The same would eventually happen for the other graphs as $n$ and the level-of-detail increase.

\subsubsection{Case Study: \textsc{usair97}}

\begin{figure}[!b]
    \centering
    \includegraphics[width=\linewidth]{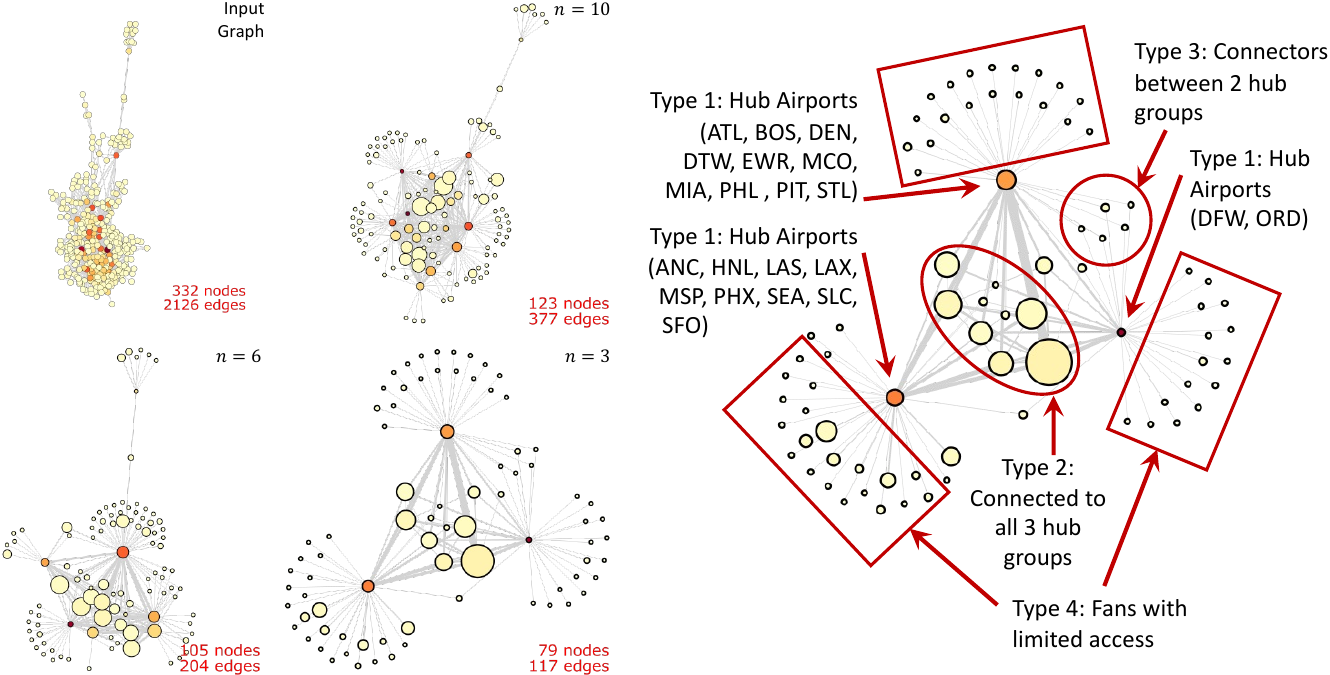}

    \vspace{-10pt}
    \subfloat[Input graph and three levels of aggregation\label{fig:usair_case:a}]{\hspace{125pt}}\hspace{0pt}
    \subfloat[Annotation of the airport group types\label{fig:usair_case:b}]{\hspace{120pt}}
    
    \caption{Case study of the \textsc{usair97} dataset using the PageRank topological lens and modularity clustering. (a)~The input graph is shown along with three \textit{mapper graphs} that have $\epsilon\!=\!0$ across different the number of cover elements. (b)~An annotated version of the $n\!=\!3$ \moag highlights the different airport groups identified.}
    \label{fig:usair_case}
\end{figure}

The \textsc{usair97} dataset is an undirected graph whose nodes are 332 airports in the US, and edges encode direct flights between the airports in 1997. We are interested in studying the properties of the network in terms of identifying which airports have high or low access to other airports in the network. Since US airlines tend to work on the hub-and-spoke model, we chose PageRank as the topological lens, since it identifies nodes of high importance (i.e., hubs). The input graph and three levels of \moag skeletonizations are shown in \cref{fig:usair_case:a}. 

Using the most aggregated \moag version (see \cref{fig:usair_case:b}), we identified several groups of airports with varying levels of access. At the highest level are hub nodes in orange (type~1), which contain airline hubs (e.g., LAX/Los Angeles, ATL/Atlanta, etc.) and popular vacation destinations (e.g., LAS/Las Vegas and MCO/Orlando). Unsurprisingly, flights from these airports can directly access many other airports, as well as connecting limited access airports. Another group with high access are the yellow nodes in the center (type~2), connected to all three hub groups. These airports can reach most destinations directly or through a single layover. The next group with medium access are the few airports that are connectors between two hub groups (type~3). These airports connect to multiple hubs and mostly require a single layover. Finally, the limited access airports are the fans that connect to hub nodes (type~4). These airports require going through at least one hub to reach any destination (besides the hub itself).

\subsection{Comparison With Community-based Skeletonization}
\label{sec:results:community}

To compare with existing approaches for skeletonization via clustering, we compare our approach with two community finding algorithms provided by NetworkX~\cite{hagberg2008exploring} that offer a resolution parameter, namely, modularity-based communities~\cite{brandes2007modularity} and Louvain community detection~\cite{blondel2008fast}. To ease comparison, graphs are constructed using the method described in \cref{sec:methods:nodes}. 
We show several examples in \cref{fig:community}. For each example, the mapper graph and community-based graph were chosen to have approximately the same number of nodes with the exception of \cref{fig:community:bn-mouse}, which had five large fans. In many cases, the community-based methods produce graphs that are similar to the \mog approach (e.g., see \cref{fig:community:bio-diseasome}). However, there are some distinctions.

\paragraph{Multiple Perspectives} Community-driven skeletonization provides only a single perspective on the graph. The ability to use different topological lenses provides \mog with the ability to provide multiple context-specific visualizations of the homological structure of an input graph. \cref{fig:community:bn-mouse} shows the most dramatic difference. However, for each example in \cref{fig:community}, the \moag reveals some variation in the structure.

\paragraph{Maintenance of Homological Features} Community-drive skeletonization often preserves some homological structures, tunnels in particular. However, these methods make no guarantees of the preservation of these features. In particular, community-based methods struggle with dense graphs where extremities are lost, and no structure is visible in the core of the graph. \cref{fig:community:enron} and \ref{fig:community:usair} are the best examples of this, where the output community-based graphs are hairballs, and the \mog graphs provide insights into the graph structure.

\begin{figure*}[!t]
    \centering

    \subfloat[Example trial\label{fig:survey_sample}]{
        \adjustbox{width=0.225\linewidth,valign=B}{
        \fbox{\includegraphics[width=0.20\linewidth]{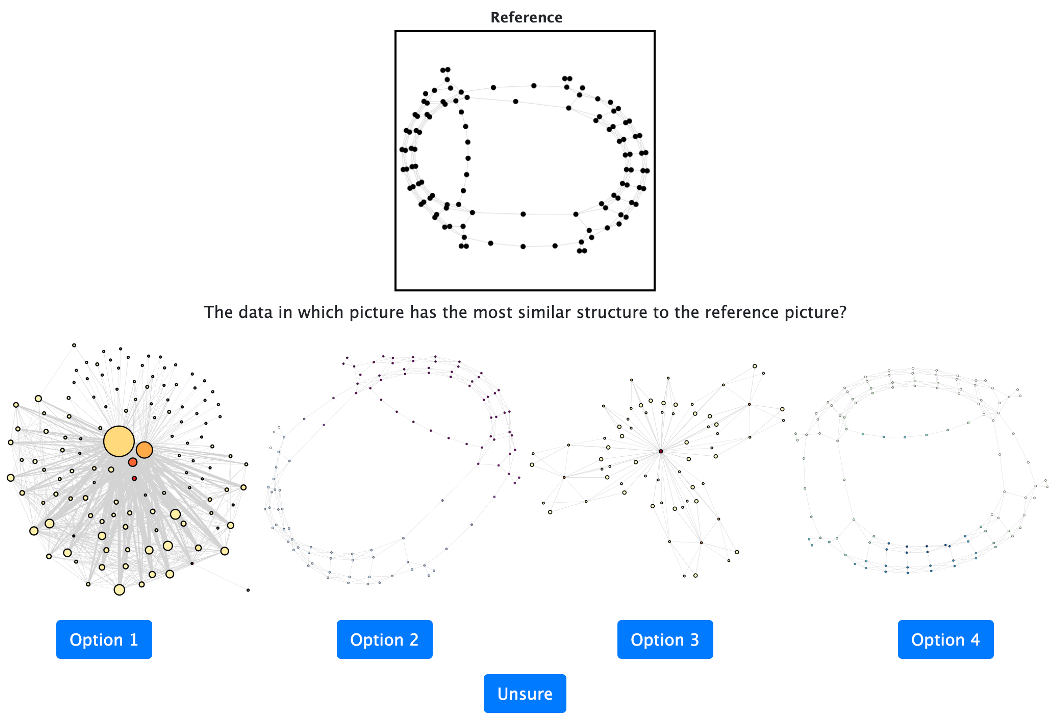}}
        }
    } 
    \subfloat[Table summarizing results\label{fig:survey_results}]{
    {\adjustbox{width=0.775\linewidth,valign=B}{
    \begin{tabular}{r|@{\hspace{3pt}}c@{\hspace{3pt}}|c|@{\hspace{3pt}}c@{\hspace{3pt}}|c|@{\hspace{3pt}}c@{\hspace{3pt}}|c|@{\hspace{3pt}}c@{\hspace{3pt}}|c|@{\hspace{3pt}}c@{\hspace{3pt}}|c|@{\hspace{3pt}}c@{\hspace{3pt}}|c|@{\hspace{3pt}}c@{\hspace{3pt}}|c|@{\hspace{3pt}}c@{\hspace{3pt}}|c|@{\hspace{3pt}}c@{\hspace{3pt}}|c|@{\hspace{3pt}}c@{\hspace{3pt}}|c}
		&	\rotatebox{90}{\textsc{barbell}}	&	\rotatebox{90}{\textsc{bcsstk}}	&	\rotatebox{90}{\textsc{bcsstk20}}	&	\rotatebox{90}{\textsc{bcsstk22}}	&	\rotatebox{90}{\textsc{bio-celegans}}	&	\rotatebox{90}{\textsc{bio-diseasome}}	&	\rotatebox{90}{\textsc{bn-mouse}}	&	\rotatebox{90}{\textsc{caltech}}	&	\rotatebox{90}{\textsc{circular ladder}}	&	\rotatebox{90}{\textsc{conn.\ caveman}}	&	\rotatebox{90}{\textsc{dorogovtsev}}	&	\rotatebox{90}{\textsc{enron-email}}	&	\rotatebox{90}{\textsc{hic 5k net 6}}	&	\rotatebox{90}{\textsc{map of science}}	&	\rotatebox{90}{\textsc{movies}}	&	\rotatebox{90}{\textsc{random lobster}}	&	\rotatebox{90}{\textsc{ring of cliques}}	&	\rotatebox{90}{\textsc{usair97}}	&	\rotatebox{90}{\textsc{watts strogatz}}	&	\rotatebox{90}{\textbf{Overall}}	\\
\hline																																									
\hline																									\arrayrulecolor{mygray}																
Correct	&	42	&	37	&	29	&	42	&	36	&	38	&	39	&	40	&	9	&	20	&	39	&	36	&	40	&	42	&	37	&	37	&	31	&	34	&	33	&	661	\\
\arrayrulecolor{mygray}\hline																																									
Incorrect	&	2	&	4	&	11	&	2	&	5	&	6	&	4	&	2	&	26	&	20	&	5	&	7	&	2	&	2	&	6	&	5	&	8	&	9	&	10	&	136	\\
\arrayrulecolor{mygray}\hline																																									
Unsure	&	0	&	2	&	4	&	1	&	4	&	0	&	0	&	1	&	7	&	5	&	2	&	2	&	2	&	1	&	1	&	1	&	6	&	2	&	2	&	43	\\
\arrayrulecolor{black}\hline																																									
Accuracy	&	95\%	&	86\%	&	66\%	&	93\%	&	80\%	&	86\%	&	91\%	&	93\%	&	21\%	&	44\%	&	85\%	&	80\%	&	91\%	&	93\%	&	84\%	&	86\%	&	69\%	&	76\%	&	73\%	&	\textbf{79\%}	\\
\arrayrulecolor{mygray}\hline																																									
$p$	&	$<.001$	&	$<.001$	&	.024	&	$<.001$	&	$<.001$	&	$<.001$	&	$<.001$	&	$<.001$	&	1.000	&	.814	&	$<.001$	&	$<.001$	&	$<.001$	&	$<.001$	&	$<.001$	&	$<.001$	&	.008	&	$<.001$	&	.001	&	\textbf{$<.001$}		
    \end{tabular}}}}
    
    \caption{(a)~an example trial for the human-subject evaluation and (b)~the results are shown.}
    \label{fig:human_study}
\end{figure*}

\begin{figure}[!b]
    \centering

    {
    \begin{minipage}[b]{0.55\linewidth}
        \centering
        \subfloat[Input Graph]{{\includegraphics[trim=0 60pt 0 90pt, clip, , height=1.75cm]{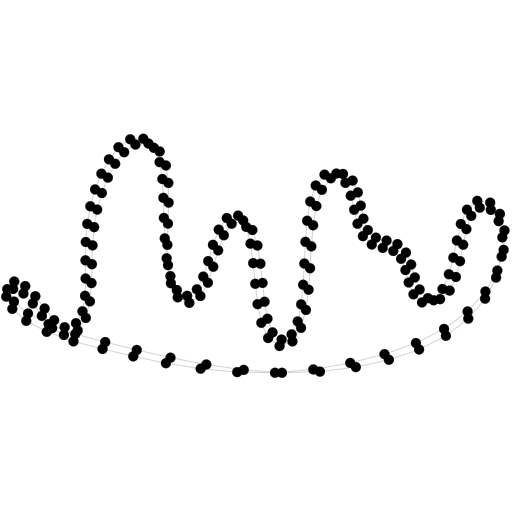}}}

        \subfloat[Den., conn.\ comp., $n\!=\!10$, $\epsilon\!=\!0$]{{\includegraphics[trim=0 15pt 0 15pt, clip, height=2cm]{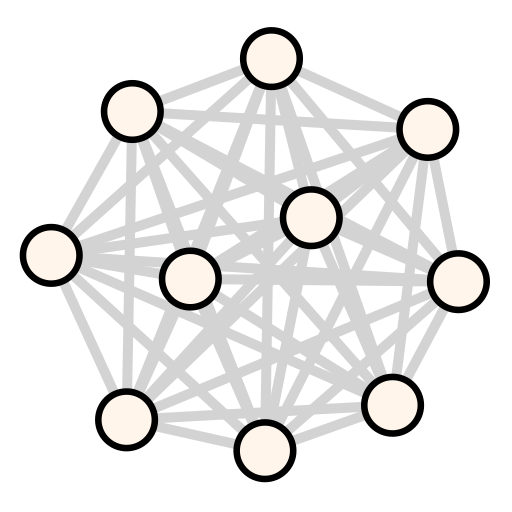}}}
        \hspace{10pt}
        \subfloat[Den., modularity, $n\!=\!10$, $\epsilon\!=\!0$]{{\includegraphics[trim=0 5pt 0 5pt, clip, height=2cm]{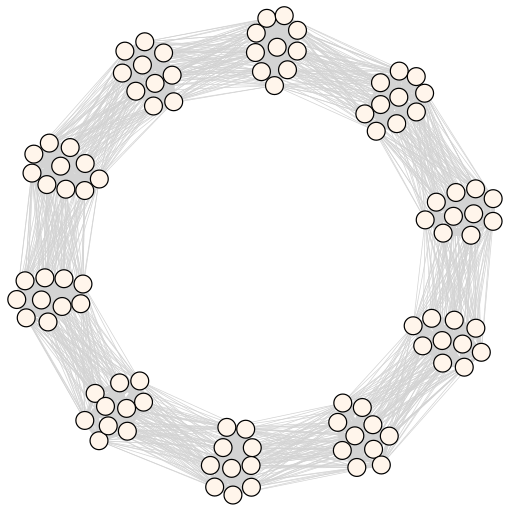}}}
    \end{minipage}}
    \hspace{1pt}
    \begin{minipage}[b]{0pt}
    \textcolor{mygray}{\rule{1pt}{3cm}}
    \vspace{0pt}
    \end{minipage}
    \hspace{-2pt}
    {
    \begin{minipage}[b]{0.35\linewidth}
        \centering
        \subfloat[Input Graph]{{\includegraphics[trim=50pt 0 60pt 0, clip, rotate=90, height=1.75cm]{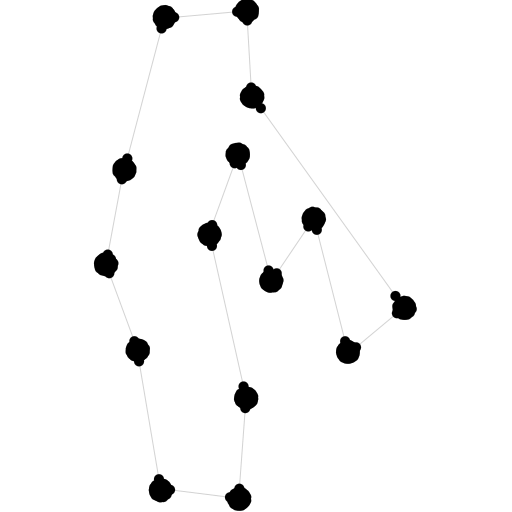}}}
        \hspace{10pt}
        \subfloat[AGD, conn.\ comp., $n\!=\!20$, $\epsilon\!=\!0$]{{\hspace{2pt}\includegraphics[height=2cm]{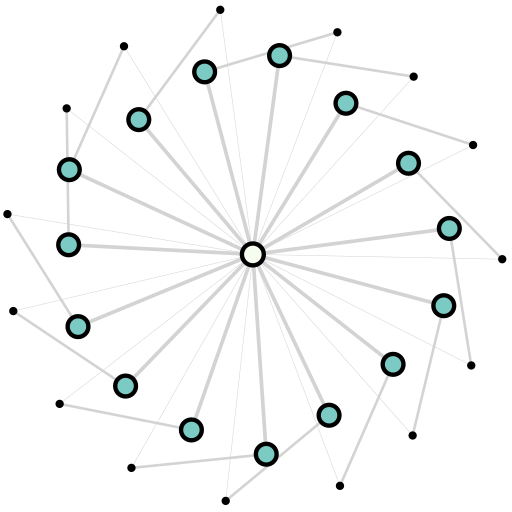}\hspace{2pt}}}
    \end{minipage}}
    
    \caption{Challenging examples from our human-subject study: (a-c)~\textsc{circular ladder}  and (d-e)~\textsc{connected caveman}.}
    \label{fig:challenge}
\end{figure}

\subsection{Human-Subject Evaluation}
\label{sec:eval:human}

To test if our approach captures the overall homology of the graph, we ran an IRB-approved human-subject evaluation on Mechanical Turk.

\subsubsection{Experiment Design}

We generated 1140 stimuli (i.e., \textit{mapper graphs}) by computing all permutations of the following parameters:

\begin{itemize}[noitemsep]
    \item Small graphs from \cref{tab:datasets} / \cref{fig:survey_results} (19 total)
    \item Number of cover elements: $n=\{6,10,20,30\}$
    \item Clustering methods: \{connected components, modularity clustering,  asynchronous label propagation\}
    \item Topological lens: \{AGD, density, eccentricity, Fiedler vector, PageRank\}
    \item Graphs were laid out using the approach in~\cite{doppalapudi2022untangling} to ensure consistency.
\end{itemize}

For each trial, subjects were exposed to one input graph and four \textit{mapper graphs} (see \cref{fig:survey_sample}): two of the \textit{mapper graphs} were generated from the input graph using different topological lenses selected at random; the other two \textit{mapper graphs} were generated from randomly selected graphs and parameters.
The subject was asked: ``Which picture has the most similar structure to the reference picture?'' The idea was that we were trying to capture if their concept of the graph shape was preserved by our approach. They could select one of the four \textit{mapper graphs} or an ``Unsure'' option.

\subsubsection{Subjects and Trials}

The study was conducted using 25 subjects. The experiment took 5-10 minutes, and subjects were compensated \$1.50 USD. Two subjects failed to complete the experiment and were excluded for a total of 23 subjects. In terms of age, subjects reported: $7\times[25,34]$; $11\times[35,44]$; $1\times[45,54]$; and $4\times[55,64]$. In terms of gender, 10 reported female, and 13 reported male. 
In terms of visualization experience, 10 reported minimal or none, 8 reported casual, and 5 reported regular or extensive experience.
The 23 subjects were each exposed to one warm-up question and 37 trials (amounting to each input graph seen twice) for a total of 851 trials. The 15-second timer expired for 11 trials, leaving a total of 840 trials for analysis.

\subsubsection{Analysis}

We predicted that the subjects would select \textit{mapper graphs} associated with the input graph with high accuracy. Since each subject was exposed to two stimuli from the input graph, the null hypothesis was that they would select one of the correct graphs $50\%$ of the time. 

\cref{fig:survey_results} summarizes the results of the survey. Accuracy is defined as  $N_{correct}/(N_{correct}+N_{incorrect}+N_{unsure})$. In the final overall column, the results show that subjects had a $79\%$ accuracy at picking one of the two correct answers, well above the $50\%$ null hypothesis. A binomial test, which is appropriate for testing whether the expected distribution of a binary outcome (correct vs.\ incorrect) matches the observed distribution, was run to determine if the difference was statistically significant with a p-value $p<0.001$, thus confirming our hypothesis.

\cref{fig:survey_results} also reports on the results for individual graphs. Participants performed well with most of the input graphs, except the \textsc{circular ladder} and the \textsc{connected caveman}. Upon further inspection, we discovered that several challenging examples were generated for these graphs (see \cref{fig:challenge}). In these cases, the combination of the topological lens, the clustering algorithm, and the level of aggregation ($n$) made the relationship difficult to identify, leading to lower scores.

\subsubsection{Ecological Validity}

Our experimental question was intentionally vague in order to elicit a response that captured subjects' instincts about what makes two graph visualizations similar. Within this narrow context, our experiment has high ecological validity.
In other words, if the visual analytics task is to determine if two graphs have the same ``structure'', our experiment showed that participants largely found our approach to satisfy that requirement. Furthermore, the layout itself, which was randomly initialized, did not significantly impact the results. Therefore, \textit{our experiment validates that our approach preserves homology and that homology-preservation is important}. However, we note that this is not the overall use case for our approach, which is multi-scale homology-preserving skeletonizations of graphs.

\section{Conclusion and Discussion}
\label{sec:discussion}

In this paper, we present a TDA-based approach for graph visualization using a construction called $\mog$.  
The approach is effective at creating skeletonizations of graphs that capture the structure across multiple scales and from multiple perspectives using different topological lenses. 

\paragraph{Parameter Selection} 
The flexibility of the \mog framework comes at a cost. 
It appears to have a number of parameters (e.g., the lens $f$, the cover parameters $n$, and $\epsilon$) that can lead to different skeletonizations of the data. 
The selection of these parameters is largely data-dependent to the point where an exploration of the parameter space may be required. 
In our case, we accomplished this using a small-multiple interface that allowed us to examine different topological lenses and numbers of cover elements. 
This interface is part of the web demo linked previously. 

For cover parameter selection, there are recent works on automatic parameter selection for mapper based on statistics~\cite{CarriereMichelOudot2018} and information criteria~\cite{ChalapathiZhouWang2021}. 
For topological lens selection, the process is not necessarily obvious or intuitive. A general guideline is that the chosen lens reflects the geometric or topological property of interest to a user. 
For instance, the PageRank lens captures node importance, where input nodes that are close to each other \emph{and} those of similar importance are grouped together to form a \moag node. As a consequence, the corresponding skeletonization encodes the distribution of input node importance.  We plan to explore (semi-)automatic parameter selection methods in the future.

\paragraph{Graph Layout} 
As pointed out earlier, the \mog technique is agnostic to the graph layout method. However, further study is needed to identify the best graph layout methods and create coherency in the layout for different parameter configurations. One possible approach would be to apply our previous method~\cite{doppalapudi2022untangling}.

\paragraph{Flexibility vs.\ Scalability} 
To achieve scalability, we have shown that limiting topological lenses to the Fiedler vector or PageRank and limiting clustering to connected components lead to highly scalable algorithms that are practical for large graphs. 
We will expand upon scalable topological lenses and clustering algorithms in the future. 

\paragraph{Relationship to Spectral Clustering} 
There is a connection between $\mog$, spectral clustering, and graph min-cut. 
The Fiedler vector $l_2$ can be used to bi-partition the graph $G$ (i.e., based on $l_2(v) > 0$ or $l_2(v) \leq 0$). 
Such a partition could also be realized by computing the \moag with $l_2$ as the lens and setting $n\!=\!2$ and $\epsilon\!=\!0$.  
Such an output not only provides a generalization of spectral clustering but also preserves the connections (which form the min-cut) between the clusters (for appropriately chosen $\epsilon>0$). 
Exploring such a connection in detail would be quite interesting.

\acknowledgments{%
This work was supported in part by grants from the NSF (IIS-1513616, DBI-1661375, IIS-2316496) and DOE (DE-SC0021015).}

\vspace{-2mm}
\bibliographystyle{abbrv-doi-narrow}
\bibliography{main-mog}

\end{document}